\documentclass[a4paper, 12pt]{article}
\pdfoutput=1
\usepackage{jheppub}

\usepackage{epsfig}
\usepackage{slashed}
\usepackage{lscape}
\usepackage{wrapfig}

\def\Dslash{\hspace{3pt}\raisebox{1pt}{$\slash$} \hspace{-9pt} D}

\def\bea{\begin{eqnarray}} \def\eea{\end{eqnarray}}
\def\be{\begin{equation}} \def\ee{\end{equation}} \def\nn{\nonumber}

\newcommand{\Lag}{\mathcal{L}}
\newcommand{\Tr}{\text{Tr}}

\newcommand{\ahat}{{\hat{a}}}

\begin{document}

\vspace*{-30mm}

\title{\boldmath Composite Dark Matter and LHC Interplay}

\author[a,b]{David Marzocca}
\author[a]{and Alfredo Urbano}
\affiliation[a]{SISSA - International School for Advanced Studies, via Bonomea 265, I-34136, Trieste, ITALY.}
\affiliation[b]{INFN sez. di Trieste, via Bonomea 265, I-34136, Trieste, ITALY.}

\vspace*{1cm}

\abstract{The actual realization of the electroweak symmetry breaking
in the context of a natural extension of the Standard Model (SM) and 
the nature of Dark Matter (DM) are two of the most compelling questions in high-energy particle physics. Composite Higgs models may provide a unified picture in which both the Higgs boson and the DM particle arise as pseudo Nambu-Goldstone bosons of a spontaneously broken global symmetry at a scale $f \sim$ TeV.
In this paper we analyze a general class of these models based on the coset $SO(6)/SO(5)$. 
Assuming the existence of  light and weakly coupled spin-1 and spin-1/2 resonances which mix linearly with the elementary SM particles,
we are able to compute the effective potential of the theory by means of  some generalized Weinberg sum rules. The properties of the Higgs boson, DM, top quark and the above resonances are thus calculable and tightly connected. We perform a wide phenomenological analysis, considering both collider physics at the LHC and astrophysical observables. 
We find that these models are tightly constrained by present experimental data, which are able to completely exclude the most natural setup with $f \simeq 800$ GeV. Upon increasing the value of $f$, an allowed region appears. In particular for $f\simeq 1.1$ TeV we find a concrete realization that predicts $m_{\rm DM}\simeq 200$ GeV for the DM mass.
This DM candidate lies close to the present sensitivity of direct detection experiments and will be ruled out -- or discovered -- in the near future.
}
\maketitle


\section{Introduction}
After a quest lasting nearly half a century, the discovery of the Higgs boson \cite{Higgs,Higgs2,Higgs3} was supposed to shed light on the mechanism triggering the electroweak symmetry breaking (EWSB)  \cite{Aad:2012tfa,Chatrchyan:2012ufa}. 
However -- as it often happens -- new discoveries prompt further and deeper questions. A light Higgs boson is unnatural in the Standard Model (SM), unless its mass is shielded from large quantum corrections. This longstanding issue of the SM is elegantly solved if the Higgs boson is protected by a new symmetry, and the most popular realization of this idea is the introduction of supersymmetry \cite{Djouadi:2005gi,Djouadi:2005gj}. 
Moreover,  some supersymmetric extensions of the SM predict the existence of a stable particle, often identified with the lightest neutralino, that can play the role of Dark Matter (DM) in the Universe \cite{Jungman:1995df}. The lack of signals of new physics first at the LEP and now at the LHC, however, has pushed these models towards a corner of their natural validity \cite{Arvanitaki:2013yja,Gherghetta:2014xea}. 

Composite Higgs models \cite{Kaplan:1983fs,Georgi:1984af,Kaplan:1983sm,Dugan:1984hq,Contino:2003ve,Agashe:2004rs} offer an alternative solution to supersymmetry based on the possibility that the Higgs boson arises as the pseudo Nambu-Goldstone boson (pNGB) of a spontaneously broken global symmetry of a new, unspecified, strongly coupled sector at the TeV scale. The minimal, phenomenologically viable, realization of this idea relies on the breaking pattern $SO(5)\to SO(4)$. Despite their undeniable theoretical complexity, Composite Higgs models provide robust and falsifiable predictions like deviations of the Higgs couplings and the presence of light (sub-TeV) top partners as a consequence of the measured value of the Higgs mass \cite{Matsedonskyi:2012ym,Redi:2012ha,Marzocca:2012zn,Pomarol:2012qf}. In ref.~\cite{Frigerio:2012uc} it has been shown that a Composite Higgs model based on  the breaking pattern $SO(6)\to SO(5)$ predicts also the existence of an extra pNGB, singlet under the SM gauge group, 
that features all the prerogatives needed to be a realistic DM candidate.
In this theoretical setup both DM and collider phenomenology are therefore tightly linked.

In this paper we realize concretely this connection making use of the \textit{Minimal Higgs Potential} hypothesis proposed in ref.~\cite{Marzocca:2012zn}. 
The key point is that the assumptions underlying this hypothesis allow to write explicitly the effective potential that involves both the Higgs and the DM particle. This effective potential, in turn, provides the possibility to compute observable quantities that can be either matched with observations -- as with top and Higgs masses -- or compared with the experimental bounds -- as with DM properties and the mass of the top partners. Equipped by this result, we will be able to subject the model to a careful analysis exploring both collider phenomenology and astrophysical implications. 

The structure of this paper is as follows. 
In section~\ref{sec:TheoreticalOverview} we present our Composite DM model. 
In section~\ref{Sec:Potential} we analyze the effective potential, while
sections~\ref{sec:phenoLHC} and \ref{sec:PhenoAstro} are devoted to the phenomenological analysis of the model.
We present our result in section~\ref{sec:results}.  Finally, we conclude in section~\ref{sec:conclusions}. In the appendices, we 
provide further details about the theoretical structure of the model.
In appendix~\ref{App:parametrizations}, we study different parametrization of the $SO(6)/SO(5)$ coset. In appendix~\ref{App:EffectivePotential}, we describe in detail the effective potential
 analyzed in section~\ref{Sec:Potential}.

\section{Composite Higgs and Dark Matter model}\label{sec:TheoreticalOverview}

In this section we present a Composite DM model in which both the Higgs doublet $H$ and the scalar singlet DM particle $\eta$ arise as composite pNGBs, characterized by the NGB \emph{decay constant} $f$ (analogous to the $f_\pi$ constant for pions in QCD), from a spontaneous symmetry breaking due to the dynamics of a new strongly coupled sector, lying at a high scale $\Lambda \sim 4 \pi f$.
The minimal scenario, considered here, is based on the $SO(6) \rightarrow SO(5)$ symmetry breaking pattern. The singlet $\eta$ is stable thanks to a parity under which
\be
	\eta \rightarrow - \eta ~.
	\label{eq:parity}
\ee
The main difference between this case and models in which $\eta$ is an elementary scalar (see, e.g., refs.~\cite{Silveira:1985rk,McDonald,Burgess}) comes from derivative interactions between $\eta$ and $H$. As we show explicitly in the next subsection, these interactions depend only on the symmetry breaking pattern and on the scale $f$. Expanding up to dimension-6 terms in $(|H|^2 , \eta^2) / f^2$, the chiral Lagrangian can be written as \cite{Frigerio:2012uc}
\be
	\Lag^{kin} \simeq |D_\mu H|^2 + \frac{1}{2} (\partial_\mu \eta)^2 + \frac{1}{2 f^2} \left( \partial_\mu |H|^2 + \frac{1}{2} \partial_\mu \eta^2 \right)^2~,
	\label{eq:ChiralLagrExpanded}
\ee
where $D_\mu H$ is the usual SM covariant derivative of the Higgs doublet.

In order to provide a mass to the SM fermions, in particular to the top quark, we assume the \emph{partial compositeness} mechanism: each SM fermion mixes with one (or more) composite vector-like fermions with the same quantum numbers \cite{Georgi:1984af,Dugan:1984hq}.
Upon integrating out the heavy fermions, the SM Yukawa interactions are generated, along with higher order interaction terms. Considering, for example, the bottom quark, up to dimension-6 terms the effective Yukawa Lagrangian can be written as
\be
	\Lag^{Yuk,b} \simeq - y_b \bar{q}_L H b_R \left( 1 - \kappa_{hb} \frac{|H|^2}{f^2} - \kappa_{\eta b} \frac{1}{2} \frac{\eta^2}{f^2} + \ldots \right) + h.c. ~,
	\label{eq:YukawaLagr}
\ee
and similarly for the other SM fermions. In our explicit model all the coefficients $\kappa_{hf} = \kappa_{\eta f} = 1$ where in general they depend on the choice of embedding of the SM fermions in (incomplete) $SO(6)$ representations and of the parametrization of the $SO(6)/SO(5)$ coset, as discussed in detail in appendix~\ref{App:parametrizations}.

These mixing terms break explicitly the global symmetry and therefore induce, at one-loop, an effective potential for the pNGBs, $V(H,\eta)_\text{eff}$. This potential presents a minimum for $H$, away from the origin, which breaks the EW symmetry to $U(1)_{em}$.
Since SM fermion masses arise via the mixing terms, the more massive the fermion, the bigger the mixing has to be.
The main contribution to the potential is thus due to the top quark mixing terms.
Another important source of explicit symmetry breaking is due to the SM EW gauge interactions.
Assuming invariance under the parity in eq.~\eqref{eq:parity}, the most general scalar potential, up to dimension 4 terms, is
\be
	V(H, \eta)_\text{eff} = \mu^2_h |H|^2 + \frac{\mu_\eta^2}{2} \eta^2 + \lambda_h |H|^4 + \frac{\lambda_\eta}{4} \eta^4 + \lambda |H|^2 \eta^2 ~,
	\label{eq:ScalarPotential}
\ee
where $\lambda$ is often dubbed \textit{Higgs portal coupling} \cite{PW}.
Assuming that $0 < - \mu^2_h < \lambda_h f^2$ and $\mu_\eta^2 - \lambda \frac{ \mu_h^2}{\lambda_h} > 0$, this potential has a minimum for
\be
	\langle H \rangle = \left( 0, \frac{v}{\sqrt{2}} \right)^t,\quad \langle \eta\rangle = 0, \quad \text{where} \quad v^2 = - \frac{\mu_h^2}{\lambda_h} \equiv \xi f^2 \simeq (246 \text{ GeV})^2 ~.
\ee
The masses of the physical fields $h$ and $\eta$, being $h$ the Higgs boson, are given by
\be
	m_h^2 = 2 \lambda_h v^2 (1 - \xi)~, \qquad m_\eta^2 = \mu_\eta^2 + \lambda v^2~,
	\label{eq:ScalarMasses}
\ee
where the $(1-\xi)$ factor in the Higgs mass is a correction due to a wave function normalization effect, see eq.~\eqref{eq:PhysFields} in the next subsection.

Following ref.~\cite{Marzocca:2012zn}, in order to render the scalar potential calculable (to be able to compute the Higgs and scalar DM masses and couplings), we assume the \emph{Minimal Higgs Potential} hypothesis, that is we assume the potential to be dominated by the contributions due to SM fields and the lighter resonances, and we impose generalized Weinberg sum rules in order to remove the quadratic and logarithmic sensitivity to the cutoff.
At one loop, the only composite states which contribute to the scalar potential are those that mix with the elementary SM particles, breaking the global $SO(6)$ symmetry with such mixings. Such states are the spin-1/2 top partners and composite spin-1 resonances, with masses of the order $m_\rho^2 \ll \Lambda^2$, which mix with the SM EW gauge bosons.

The main aim of the rest of this section is to build explicit models in order to study the allowed range of the DM mass and Higgs portal coupling in realistic cases which, in particular, correctly describe both the top and Higgs mass and which still evade the bounds from direct searches of top partners at the LHC.

\subsection{Structure and symmetries of the $SO(6)/SO(5)$ coset}

Let us review here the basic structure of next-to-minimal Composite Higgs models where the strong sector enjoys a global symmetry $SO(6) \otimes U(1)_X$\footnote{The $U(1)_X$ factor is needed in order to correctly reproduce the SM fermion hypercharges.} spontaneously broken to the subgroup $SO(5) \otimes U(1)_X$ at a scale $f$ \cite{Gripaios:2009pe,Redi:2012ha,Frigerio:2012uc}. 
Due to this spontaneous symmetry breaking, the low energy theory has 5 NGBs, which transform in the fundamental, $\bf 5$, of $SO(5)$. The custodial symmetry group is contained in the unbroken group, $SO(4) \sim SU(2)_L \otimes SU(2)_R \subset SO(5)$, and the NGBs transform as a $\bf 4 \oplus 1 \sim (2,2) \oplus (1,1)$ of the custodial group.
Here and in the following we describe the five broken $SO(6)/SO(5)$ generators as $T^{\hat{a}}$, with $\hat{a} = 1, \ldots, 5$. The 10 unbroken generators of $SO(5)$, $T^{a}$, can be divided in the 6 generators of the $SO(4)$ custodial subgroup, $T^{a_{L,R}}$ with $a_{L,R} = 1,2,3$, and the 4 generators of the $SO(5)/SO(4)$ coset, $T^\alpha$ with $\alpha = 1,\ldots, 4$ (see eq.~\eqref{eq:SO6generators} in appendix~\ref{App:parametrizations} for the explicit definition of the generators).
The SM EW gauge symmetry is identified as the subgroup $\mathcal{G}_{EW} = SU(2)_L \otimes U(1)_Y \subset SU(2)_L \otimes SU(2)_R \otimes U(1)_X$, where the hypercharge is defined as $Y = T^{3_R} + X$.

The NGBs can be described by the $\Sigma$ field
\be
	\Sigma = \frac{1}{f} \left(
	h_1, h_2, h_3, h_4, \eta,
	\sqrt{f^2 - h^2 - \eta^2 } \right)~,
	\label{eq:NGBSigma}
\ee
where $h^2 = \sum_{i=1}^4 h_i^2$ and where $h_i$ and $\eta$ live in the region $\sqrt{h^2 + \eta^2} \leq f$.\footnote{The effect of this constraint is negligible at any order in perturbation theory and therefore does not have any effect in any of the computation we perform in this work. In appendix~\ref{App:parametrizations} we will explicitly show the relations to other parametrizations used in the literature.}
The usual Higgs doublet can can be constructed as $H = \frac{1}{\sqrt{2}} ( h_1 + i h_2, h_3 + i h_4 )^t$.
In the unitary gauge $h_1(x) = h_2(x) = h_4(x) = 0$ and $h(x) \equiv h_3(x)$. See Appendix~\ref{App:parametrizations} for more details.

The chiral Lagrangian can be written in an expansion in derivatives over the cutoff. The leading term, with two derivatives, is
\be
	\Lag^{kin} = - \frac{1}{4} W_{\mu\nu}^a W^{a \mu\nu} - \frac{1}{4} B_{\mu\nu} B^{\mu\nu} + \frac{f^2}{2} (D_\mu \Sigma)^t D^\mu \Sigma~,
	\label{eq:LeadingLagr}
\ee
where $D_\mu = \partial_\mu - i \left( g_0 W^{a_L}_\mu T^{a_L} + g_0^\prime B_\mu Y \right)$ and $f > v$ is the symmetry breaking scale, that is the only parameter of the leading order chiral Lagrangian.\footnote{Our convention for the field strength is $W_{\mu\nu} = \partial_\mu W_\nu - \partial_\nu W_\mu - i g_0 [W_\mu, W_\nu]$ and $B_{\mu\nu} = \partial_\mu B_\nu - \partial_\nu B_\mu$, where $W_\mu \equiv W_\mu^{a_L} T^{a_L}. $}
The last term, in the unitary gauge, reads
\be \begin{split}
	\frac{f^2}{2} (D_\mu \Sigma)^t D^\mu \Sigma& = \frac{1}{2} \left[ (\partial_\mu h)^2 + (\partial_\mu \eta)^2 + \frac{( h \partial_\mu h + \eta \partial_\mu \eta)^2 }{f^2 - h^2 - \eta^2 } \right]  \\
		& + \frac{h^2}{8} \bigg\{ g_0^2 \left[ (W^1_\mu)^2 + (W^2_\mu)^2 \right] + (g_0^\prime B_\mu - g_0 W_\mu^3)^2 \bigg\}~.
	\label{eq:ChiralKinTerm}
\end{split} \ee
The SM gauge boson masses are given by
\be
	m_W^2 = \frac{g_0^2}{4} \langle h \rangle^2~, \qquad m_Z^2 = \frac{(g_0^2 + g_0^{\prime 2})}{4} \langle h \rangle^2~.
\ee
This fixes the EW scale $v = \langle h \rangle \equiv  f \sqrt{\xi} \simeq 246 \text{ GeV}$. Given that in the vacuum $\langle \eta \rangle = 0$, it is immediate to see that the canonically normalized fields, in this parametrization, are
\be
	h \rightarrow v + \sqrt{1- \xi} ~ h_{phys}~, \qquad \eta \rightarrow \eta_{phys}~.
	\label{eq:PhysFields}
\ee

The parity $\eta \rightarrow - \eta$, which keeps this scalar stable, corresponds to the operator
\be
	P_\eta = \text{diag}(1,1,1,1,-1,1) \in O(6)~,
\ee
and is a symmetry of the leading order chiral Lagrangian, eq.~\eqref{eq:ChiralKinTerm}. Higher derivative terms (such as the Wess-Zumino-Witten term), in general break this symmetry. As we want this scalar to be a viable DM candidate, we assume that this is a symmetry of the whole strong sector, that is we take the symmetry breaking pattern to be $O(6) \rightarrow O(5)$ \cite{Frigerio:2012uc}.

Another symmetry of eq.~\eqref{eq:ChiralKinTerm}, very relevant for the $\eta$ phenomenology, is a $SO(2)_\eta \simeq U(1)_\eta$ generated by $T^{\hat{5}}$ which rotates the fifth and sixth components of $\Sigma$ and under which $\eta$ shifts.
If the fermion mixings also respect this symmetry then $\eta$ remains an exact NGB, thus its mass and couplings from the potential vanish.

\subsection{Composite resonances Lagrangian}

Here we introduce our models, that is the Lagrangian of the spin-1 and spin-1/2 resonances which mix with the SM gauge bosons and fermions.

\subsubsection{Vector Lagrangian}

We introduce composite resonances in representations of the unbroken group $SO(5)$ using the hidden local symmetry formalism, following ref.~\cite{Contino:2011np}.
In particular, let us consider spin-1 fields in the adjoint, $\rho_\mu = \rho_\mu^a T^a \in {\bf 10}$, and in the fundamental, $a_\mu = a_\mu^{\hat{a}} T^{\hat{a}} \in {\bf 5}$.
At leading order in the number of derivatives, the Lagrangian for these fields, assumed to be lighter than the cutoff, is
\be
	\Lag^{spin-1} = - \frac{1}{4} \Tr \left( \rho_{\mu\nu}^2 \right) + \frac{f_\rho^2}{2} \Tr \left[ \left( g_\rho \rho_\mu - E_\mu \right)^2 \right] - \frac{1}{4} \Tr \left( a_{\mu\nu}^2 \right) + \frac{f_a^2}{2 \Delta^2} \Tr \left[ \left( g_a a_\mu - \Delta d_\mu \right)^2 \right] ~,
	\label{eq:spin1Lagr}
\ee
where $d_\mu$ and $E_\mu$ are the CCWZ structures \cite{Coleman1,Coleman2} defined in eq.~\eqref{eq:CCWZ} and the field strengths are defined as $\rho_{\mu\nu} = \partial_\mu \rho_\nu - \partial_\nu \rho_\mu - i g_\rho [\rho_\mu, \rho_\nu]$ and $a_{\mu\nu} = \nabla_\mu a_\nu - \nabla_\nu a_\mu$.
Let us also define the masses
\be
	m_\rho = f_\rho g_\rho~, \qquad
	m_a = f_a \frac{g_a}{\Delta} ~.
\ee
The generalization to an arbitrary number of copies is straightforward, see e.g. ref.~\cite{Marzocca:2012zn}. For simplicity we consider only the minimal case with one adjoint and one fundamental, which already allows to obtain a finite one-loop potential.

The mixing term in eq.~\eqref{eq:spin1Lagr} between $\rho_\mu$ and $E_\mu$\footnote{Expanding $E_\mu$ in the number of fields one obtains $E_\mu^a = g_0 W_\mu^{a_L} \delta^{a, a_L} + g_0^\prime B_\mu \delta^{a, 3_R} + \mathcal{O}(h^2 / f^2)~.$} induces a mixing between the SM gauge fields and the spin-1 resonances $\rho_\mu^{a_L}$ and $\rho_\mu^{3_R}$. The mass eigenvalues, before EWSB, are given by a simple rotation $W_\mu^{a_L} \rightarrow \cos \theta_g W_\mu^{a_L} + \sin \theta_g \rho_\mu^{a_L}$, $B_\mu \rightarrow \cos \theta_{g'} B_\mu + \sin \theta_{g'} \rho_\mu^{3_R}$ and similarly for $\rho_L^{a_L}$ and $\rho_\mu^{3_R}$, where $\tan \theta_g = g_0 / g_\rho$ and $\tan \theta_{g'} = g_0^\prime / g_\rho$.
The massless combinations are the physical SM EW gauge bosons while the massive ones are the spin-1 resonances. Their mass shifts, due to this mixing, at the order $\mathcal{O}(g_0^2/g_\rho^2)$.
The physical SM gauge couplings are given by $g = g_0 \cos \theta_g$, $g' = g'_0 \cos \theta_{g'}$.

\subsubsection{Fermion Lagrangian}\label{sec:fermionembedding}

In order to give mass to the SM fermions we adopt the partial compositeness scenario: the SM fields mix linearly with some fermonic operators of the composite dynamics with same quantum numbers. Assuming that such mixing terms arise from some flavor dynamics at a scale much higher than the strong dynamics scale $\Lambda$, it is reasonable to write mixing terms which transform linearly under $SO(6)$
\be
\Lag_{mix} \sim \epsilon_\psi \, \bar{\psi}_{SM} \mathcal{O}_{\Psi} + h.c. ~,
\ee
where $\mathcal{O}_{\Psi}$ belongs to some representation of $SO(6)$.
Since the SM fields are not in complete representations of $SO(6)$, such mixings will necessarily break explicitly the global symmetry. It is however useful to embed $\psi_{SM}$ in the same representation of $\mathcal{O}_{\Psi}$.
At lower energies, where the symmetry is spontaneously broken, we render explicit the NGB dependence of these terms as $\mathcal{O}_{\Psi} = U(x) \, \Psi(x)$, where $U(x)$ is the NGB matrix, see eq.~\eqref{eq:NGBmatrix}, and $\Psi(x)$ belongs to some irreducible representation of $SO(5)$.

The choice of the representation of $SO(6)$ in which to embed the SM fields is a source of model dependence, in particular the characteristics of the scalar one-loop potential and the preservation of $P_\eta$ and of $U(1)_\eta$ depend on the choice of the embedding of the third generation of quarks. It has been shown in ref.~\cite{Frigerio:2012uc} that, since $[P_\eta, T^{\hat{5}}] \neq 0$, the only way in which both symmetries can be respected by the mixing terms is if the SM fermions are embedded in representations of $SO(6)$ with vanishing $U(1)_\eta$ charge.

In the following we focus on the embedding of the SM doublets $q_L, \ell_L$ in the bi-doublet inside the $\bf 6$, with $P_\eta = +1$ and which preserves $U(1)_\eta$, and the right-handed fermions $u_R, d_R, e_R$ in the parity even singlet inside the $\bf 6$, that is its sixth component with non-zero $U(1)_\eta$ charge. The charge under  $U(1)_X$ is fixed by requiring the correct hypercharge. The embedding of the SM doublets has to be different for the mixing terms responsible for the up-type or down-type quark masses:
\be
	\xi_L^u =  \frac{1}{\sqrt{2}} \left( \begin{array}{c}
	b_L \\ -i b_L \\ t_L \\ i t_L \\ 0 \\ 0
	\end{array}\right)_{2/3}, ~~~
	\xi_R^u =  \left( \begin{array}{c}
	0 \\ 0 \\ 0 \\ 0 \\ 0 \\ t_R
	\end{array}\right)_{2/3}, ~~~
	\xi_L^d =  \frac{1}{\sqrt{2}} \left( \begin{array}{c}
	t_L \\ i t_L \\ -b_L \\ i b_L \\ 0 \\ 0
	\end{array}\right)_{-1/3}, ~~~
	\xi_R^d =  \left( \begin{array}{c}
	0 \\ 0 \\ 0 \\ 0 \\ 0 \\ b_R
	\end{array}\right)_{-1/3},
	\label{eq:SMfermEmbedding6}
\ee
where the subscript indicate the $X$ charge.\footnote{In section~\ref{sec:DirectDetection} the couplings between DM and the first two generations of quarks will be extremely important for our phenomenological analysis in the context of DM direct detection. In order to be as general as possible, therefore, we will consider also different embedding w.r.t. eq.~(\ref{eq:SMfermEmbedding6}).} We embed the SM lepton doublets and singlets in the same way as $\xi_L^d$ and $\xi_R^d$ but with $U(1)_X$ charges $X_{\ell_L} = X_{e_R} = - 1$.

Let us briefly comment on the case in which the right handed top quark is embedded in a $\bf 15$ of $SO(6)$, in order to preserve the $U(1)_\eta$ symmetry. In this case the breaking of this symmetry, 
and therefore the contribution to the $\eta$ potential, comes only from the bottom quark, assuming its right chirality is embedded in the $\bf 6$.
Since the bottom mixings to the composite sector are much smaller than those of the top, we expect that in this case the singlet is much lighter, $m_\eta \lesssim \mathcal{O}(10)$ GeV. From the expression of the DM mass in eq.~\eqref{eq:ScalarMasses}, assuming $\mu_\eta^2 > 0$, this implies that also the coupling $\lambda$ is generically small: $\lambda \lesssim 10
^{-3}$. In this case the bound from the Higgs invisible width is able to exclude such a framework for any value of $\xi \gtrsim 0.05$. For this reason, we will not further consider this possibility in the rest of this paper.

Let us now focus on the fermion partners responsible to give mass to the top quark, since the mixing terms with these fermions provide the leading contributions to the effective potential. We assume that the right-handed top is an elementary state, as all the other SM fermions.
Following the logic of ref.~\cite{Marzocca:2012zn}, we introduce $N_F$ vector-like composite fermions in the fundamental, $F \in \textbf{5}$ with $X = \frac{2}{3}$ (each contains two doublets $F_{1/6} \in (\textbf{2}, \frac{1}{6})$, $F_{7/6} \in (\textbf{2}, \frac{7}{6})$ and one singlet $F_{5} \in (\textbf{1}, \frac{2}{3})$ under $SU(2)_L \times U(1)_Y$), and $N_S$ vector-like singlets, $S \in \textbf{1}$, of $SO(5)$, with $X = Y = \frac{2}{3}$. We embed the SM fermions in the \textbf{6} of $SO(6)$. The leading Lagrangian for the top sector, relevant for the computation of the one-loop effective potential, is given by
\begin{eqnarray}\label{eq:LGeneralFermions}
	\mathcal{L}_{f}  &=&   \bar q_L i \Dslash q_L +\bar t_R i \Dslash t_R + \sum_{i=1}^{N_S} \bar{S}_i (i \slashed \nabla - m_{iS}) S_i  + \sum_{j=1}^{N_F} \bar F_j (i \slashed \nabla - m_{jF}) F_j \\
	&&  + \sum_{i=1}^{N_S} \left( \epsilon_{tS}^i \bar \xi_R P_L U S_i + \epsilon_{qS}^i \bar \xi_L P_R U S_i \right) + \sum_{j=1}^{N_F} \left( \epsilon_{tF}^j \bar \xi_R P_L U F_j + \epsilon_{qF}^j \bar \xi_L P_R U F_j \right) + h.c.~, \nonumber
\end{eqnarray}
where $P_{L,R}= \frac{1 \mp \gamma^5}{2}$ are chirality projectors and
\be
\nabla_{\mu} = \partial_\mu - i E_\mu - i q_X g^\prime_0 B_\mu~.
\ee

In general, with our field content, at the same order in the expansion in derivatives it is possible to write other invariants which do not involve the elementary fields. For this reason they do not contribute at one loop to the effective potential. The most general couplings at leading order are (see ref.~\cite{Marzocca:2012zn})
\be \begin{split}
	\Lag_{int} =  \sum_{\eta = L,R} &\left[
	k_{ij}^{V,\eta} \bar{F}_i \gamma^\mu (g_\rho \rho_\mu - E_\mu) P_\eta F_j  \right. \\
& \left. + k_{ij}^{A,\eta} \bar{S}_i \gamma^\mu a_\mu P_\eta F_j  + k_{ij}^{d,\eta} \bar{S}_i \gamma^\mu d_\mu P_\eta F_j  + h.c. 	\right]~.
	\label{eq:GenericExtraFermLagr}
\end{split} \ee
The last term in eq.~(\ref{eq:GenericExtraFermLagr}), in particular, can play an important role in the phenomenology of single production processes of top partners \cite{DeSimone:2012fs,Azatov:2013hya} and in the fermion contributions to EW precision tests \cite{Grojean:2013qca}. However, since they do not influence the scalar potential at one-loop, we neglect the terms in eq.~(\ref{eq:GenericExtraFermLagr}) in the following.

\section{Analysis of the potential and parameter scans}
\label{Sec:Potential}

The mixing terms between the elementary SM states and the heavy composite resonances, introduced in the previous section, break explicitly the $SO(6)$ symmetry. At one loop they generate a Coleman-Weinberg effective potential for the pNGBs $h$ and $\eta$.
In general, after renormalization, the field-dependent terms of the one-loop effective potential are scale dependent which would imply the need of fixing some boundary conditions and therefore a lack of predictability. Using the cutoff regularization, this issue can be seen as quadratic and logarithmic divergences in the computation of the one-loop potential (see appedix~\ref{App:EffectivePotential} for more details).

In order to cure the UV behavior of the potential and cancel this UV sensitivity (i.e. the scale and scheme dependence), some generalized Weinberg sum rules are imposed \cite{Marzocca:2012zn,Pomarol:2012qf} in both the gauge and fermion sectors.
Once these Weinberg sum rules are enforced, it is possible to expand the potential in powers of $h$ and $\eta$, in order to extract the coefficients of eq.~\eqref{eq:ScalarPotential}.\footnote{A well known fact is that the quartic terms in this expansion suffer from a spurious infrared divergence which is due to the fact that the SM particles are massless in the $\langle h \rangle \rightarrow 0$ limit, therefore the potential contains terms proportional to $h^4 \log h^2/f^2$, which do not allow a Taylor expansion around $h = 0$.
In the following analytic studies we simply cutoff this divergence with the $W$ or top mass (depending on the sector we are considering), however in the numerical analysis we always consider the full potential in which case there is no infrared divergence. For a more complete discussion on this issue, in the context of $SO(5)/SO(4)$ models, we refer to appendix A of ref.~\cite{Marzocca:2013fza}.}
In this section we present the main results of this approach, focusing the discussion on the analysis of the effective potential. Further technical details are collected in appendix~\ref{App:EffectivePotential}. Analytical approximations and full numerical results are explicitly computed using two benchmark values for the parameter $\xi = v^2/f^2$, namely $\xi = 0.1$, corresponding to $f\simeq 800$ GeV, and 
$\xi = 0.05$, corresponding to $f\simeq 1.1$ TeV.

\subsection{Vector contribution}
\label{Sec:GaugePotential}

The gauge sector, described by the Lagrangian of eq.~\eqref{eq:spin1Lagr}, contributes to the potential only via the $h^2$ dependence, therefore only to the $\mu_h^2$ and $\lambda_h$ coefficients of eq.~\eqref{eq:ScalarPotential}.
In general, this contribution is quadratically divergent, see appendix~\ref{App:GaugePotential} for the details.
We require the cancellation of this quadratic divergence by imposing the sum rule
\be
	(\text{WSR } 1)_{gauge}~: \qquad \frac{f^2}{2} + f_a^2 - f_\rho^2 = 0~,
	\label{eq:WSR1gauge}
\ee
while the logarithmic divergence is removed requiring
\be
	(\text{WSR } 2)_{gauge}~: \qquad f_a^2 m_a^2 = f_\rho^2 m_\rho^2~.
	\label{eq:WSR2gauge}
\ee
We use these two sum rules to express $f_a$ and $m_a$ in terms of the other parameters; note that this fixes all the parameters of the $a_\mu$ fields relevant for the effective potential, since only the combination $g_a^2 / \Delta$ enters in the potential.
The sum rule of eq.~\eqref{eq:WSR1gauge} requires a bound $f_\rho > f / \sqrt{2}$, that is compatible with the \emph{partial UV completion} (PUVC) criterion introduced in ref.~\cite{Contino:2011np} which predicts $f_\rho \sim f$.

In order to obtain a simple analytic expression for the gauge contribution to the potential let us take $g^\prime = 0$, $f_\rho = f$ and expand for $g^2 \ll 1$. We obtain
\be
	(\mu_h^2)^{g} \simeq \frac{9 g^2 f^2 m_\rho^2}{32 \pi^2} \log 2~, \qquad
	(\lambda_h)^{g} \simeq - \frac{9 g^4 f^4 }{256 \pi^2} \left( \log \frac{32 m_\rho^2}{m_W^2} - 5 \right)~.
	\label{eq:PotentialCoeffGauge}
\ee

\subsection{Fermion contribution}
\label{Sec:FermionPotential}

In general, the fermion sector contributes to all the coefficients of the potential in eq.~\eqref{eq:ScalarPotential}.
As in the gauge sector, also in this case the potential is generically quadratically sensitive to the cutoff, see appendix~\ref{App:FermionPotential} for the derivation of the potential.
To cure this divergence we impose the sum rules
\be
\text{(WSR 1)}_{ferm}~:\quad  \left\{
\begin{split}
	 \sum_{j=1}^{N_F} |\epsilon_{qF}^j|^2 =& \sum_{i=1}^{N_S} |\epsilon_{qS}^i|^2~, \\
	\sum_{j=1}^{N_F} |\epsilon_{tF}^j|^2 =& \sum_{i=1}^{N_S} |\epsilon_{tS}^i|^2~.
\end{split} \right.
\ee
In order to cancel the residual logarithmic divergence we further require
\be
\text{(WSR 2)}_{ferm}~:\quad  \left\{
\begin{split}
	 \sum_{j=1}^{N_F} m_{jF}^2 |\epsilon_{qF}^j|^2 =& \sum_{i=1}^{N_S} m_{iS}^2 |\epsilon_{qS}^i|^2~, \\
	 \sum_{j=1}^{N_F} m_{jF}^2 |\epsilon_{tF}^j|^2 =& \sum_{i=1}^{N_S} m_{iS}^2 |\epsilon_{tS}^i|^2~.
\end{split} \right.
\ee

The rest of the section is devoted to analyze in more detail two specific models. First we consider the minimal scenario which allows to enforce both sum rules and to reproduce the top mass, that is with only one fundamental $F$ and one singlet $S$.
Then we study the next-to-minimal scenario, in which we add a second singlet, since it allows more freedom in exploring the parameter space of these composite Higgs models.

\subsubsection{Minimal case: $N_F = N_S = 1$}\label{sec:minimalcase}

In this minimal model it is straightforward to obtain the mass spectrum of the top partners before EWSB from the Lagrangian of eq.~\eqref{eq:LGeneralFermions}. The SM top is massless at this level, the singlet $S$ gets a mass $M_S^2 = m_S^2 + |\epsilon_{tS}|^2$, the doublet $F_{1/6}$ has a mass $M_{F_{1/6}}^2 = m_F^2 + |\epsilon_{qF}|^2$ while the other doublet, $F_{7/6}$, and the other singlet, $F_{5}$, are degenerate with a mass $M_{F_{7/6}} = M_{F_5} = m_F$.
After EWSB the fermions with same electric charge mix and these masses shift by an amount of the order $\mathcal{O}(v \epsilon / m)$. From eq.~\eqref{eq:GenericTopMass} we obtain the top mass, at leading order for small $\xi$, \cite{Marzocca:2012zn}
\be
	M_{top} \simeq \frac{| \epsilon_{qF} \epsilon_{tS} | }{\sqrt{2} M_{F_{1/6}} M_{S}} \left| m_S \frac{\epsilon_{tF}}{\epsilon_{tS}} + m_F \frac{\epsilon_{qS}}{\epsilon_{qF}}  \right| \sqrt{\xi} ~.
	\label{eq:topMassMinMod}
\ee
In this minimal setup, the first sum rule is solved by imposing
\be
	|\epsilon_{qF}|^2 = |\epsilon_{qS}|^2 \equiv \epsilon_Q^2  \qquad \text{and} \qquad
	|\epsilon_{tF}|^2 = |\epsilon_{tS}|^2 \equiv \epsilon_T^2~.
\ee
The second sum rule further fixes
\be
	\quad m_F = m_S = m~,
\ee
where we used the field basis where the masses are real and positive.
Assuming for simplicity that the mixing parameters are real, the only solution (up to field redefinition) for which the potential does not vanish is
\be
	\epsilon_{qF} = \epsilon_{qS} = \epsilon_Q~, \qquad
	\epsilon_{tF} = \epsilon_{tS} = \epsilon_T~.
\ee
In this case, it turns out that
\be
	\frac{(\mu_\eta^2)^f }{f^2}= \lambda_\eta^f = 0~, \qquad
	\lambda^f = \lambda = - \frac{(\mu_h^2)^f}{f^2}~.
	\label{eq:ConstraintsMinMod}
\ee
Since $\mu_\eta^2$ does not receive any contribution neither from the gauge sector nor from the fermion sector, it vanishes and therefore the singlet will be light (its mass is $\xi$-suppressed, as the Higgs mass, eq.~\eqref{eq:ScalarMasses}).

In this simple model it is straightforward to obtain exact analytic formulae for these coefficients, however in order to get an understanding of the behavior of this model it is useful to make some approximations. For example assuming big mixings, that is $m^2 \ll M_{F_{1/6}}^2, M_{S}^2$, we get $M_{top}^2 \simeq 2 m^2 \xi$ and
\be
	\lambda = \lambda^f = -  \frac{(\mu_h^2)^f}{f^2} \simeq \frac{1}{2} \lambda_h^f \simeq \frac{N_c M_{top}^2}{4 \pi^2 v^2} \frac{M_{F_{1/6}}^2 M_{S}^2 }{f^2 (M_{F_{1/6}}^2 - M_{S}^2)} \log \frac{M_{F_{1/6}}^2}{M_{S}^2}~,
	\label{eq:EstimateParamFermMinMod1}
\ee
which is evidently always positive. The top mass fixes $m = M_{F_{7/6}} \sim 350$ GeV which, as we show in section~\ref{sec:TopPartners}, is experimentally excluded, therefore this region is disfavored.
In the opposite limit, that is $\epsilon_Q^2, \epsilon_T^2 \ll m^2$, we obtain $M_{top}^2 \simeq 2 \xi \epsilon_Q^2 \epsilon_T^2 / m^2 $ and
\be
	\lambda = \lambda^f = - \frac{(\mu_h^2)^f}{f^2} \simeq \frac{1}{2} \lambda_h^f \simeq \frac{N_c M_{top}^2}{4 \pi^2 v^2} \frac{m^2}{f^2}~.
	\label{eq:EstimateParamFermMinMod2}
\ee
In this case, the scale of the top partner masses $m$ has to be smaller than $\sim 1.5 f \simeq 1.2~(1.6)$ TeV for $\xi = 0.1~(0.05)$, in order to reproduce the correct Higgs mass.
We have checked numerically that, indeed, the relation $\lambda^f  \simeq \frac{1}{2} \lambda_h^f $ holds, up to $\mathcal{O}(20\%)$ corrections, in all the parameter space. This fact, using eq.~\eqref{eq:ScalarMasses} and the fact that the gauge contribution to $\lambda_h$ is always negligible, allows us to conclude that in this model, for a given $\xi$, the Higgs mass fixes both the DM mass and portal coupling
\be
	m_\eta \simeq \frac{1}{2} m_h \simeq 63 \text{ GeV}~, \quad \text{and} \quad
	\lambda = \frac{m_\eta^2}{v^2} \simeq \frac{1}{4} \frac{m_h^2}{v^2}\simeq 0.065~.
	\label{eq:RelationsMinimalModel}
\ee

Let us finally discuss how $\xi$ can be tuned to realistic values, in particular our benchmark values $\xi = 0.1, 0.05$. From the relation $- \frac{(\mu_h^2)^f}{f^2} \simeq \frac{1}{2} \lambda_h^f$ and eq.~\eqref{eq:ScalarMasses} we get
\be
	\xi \simeq \frac{1}{2} - \frac{(\mu_h^2)^g}{m_h^2} 2 \xi~,
	\label{eq:XiEstimateMinModel}
\ee
where we neglected the gauge contribution to $\lambda_h$ since it is always negligible with respect to the fermionic one. The gauge contribution to $\mu_h^2$ is therefore necessary in order to reduce $\xi$. Eq.~\eqref{eq:PotentialCoeffGauge} allows to fix the composite vector mass as a function of the Higgs mass (for a given value of $f_\rho / f$, which has been set to 1 in this example)
\be
	m_\rho \sim \sqrt{\frac{2}{\log 2}} \frac{\pi}{3} \frac{m_h}{m_W} \frac{v}{\sqrt{\xi}} \simeq 2 \text{ TeV ~~~~~(for $\xi = 0.1$)}~.
	\label{eq:VectorMassMinMod}
\ee
From eq.~\eqref{eq:XiEstimateMinModel} we see that, in absence of the gauge contribution, the natural value of $\xi$ would be $\sim 0.5$. Therefore, we can estimate the amount of tuning needed to get a smaller value with the simple relation
\be
	\Delta \sim \frac{1}{2\xi}~,
	\label{eq:TuningMinMod}
\ee
that is, a $\sim 20\%$ tuning for $\xi = 0.1$.
Such a low amount of tuning in this model is due to the fact that the extreme simplicity of the model after imposing the Weinberg sum rules fixes $- \frac{(\mu_h^2)^f}{f^2}$ to be of the same order (actually, a factor of 2 smaller) of $\lambda_h$, see eqs.~(\ref{eq:EstimateParamFermMinMod1},~\ref{eq:EstimateParamFermMinMod2}).
This and the relations in eq.~\eqref{eq:ConstraintsMinMod} are non-generic features of these kind of models: in general the mass term in the potential is expected to be generated at quadratic order in the mixings while the self-coupling term only at quartic order, so that $\left| \frac{(\mu_h^2)^f}{f^2 \lambda_h^f} \right|$ would be naturally much bigger than 1 and therefore the needed amount of tuning much larger.
For this reason, in order to assess with more generality the viability of these DM model, in the next section we study also a non-minimal model, in which this more generic feature is indeed present.

\begin{figure*}[t]
\begin{center}
\fbox{\footnotesize $N_F = 1$, $N_S = 1$}
\end{center}
\vspace{-0.5cm}
\hspace{-0.5cm}
\minipage{0.5\textwidth}
  \includegraphics[width=\linewidth]{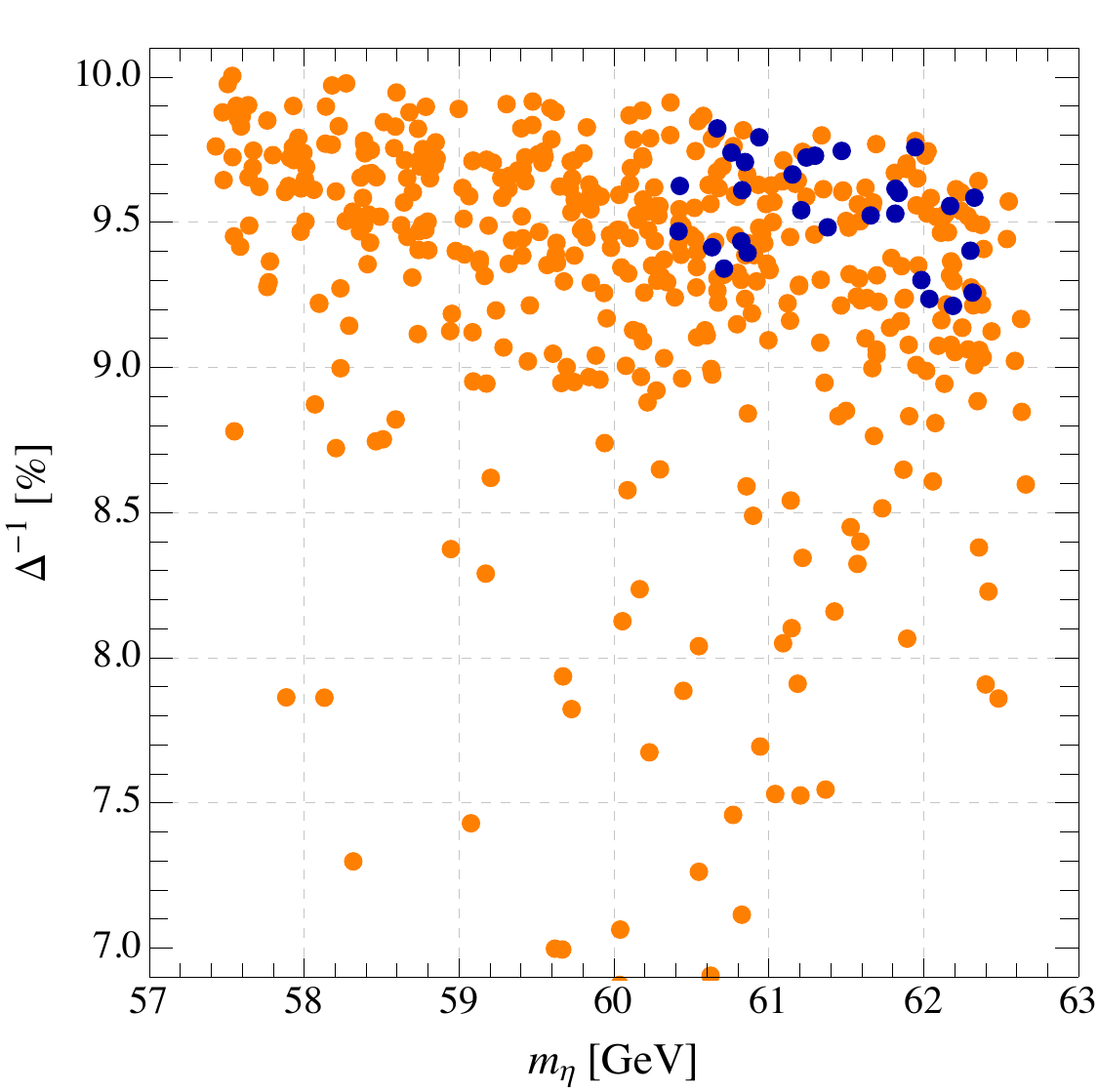}
\endminipage \hspace{0.5cm}
\minipage{0.5\textwidth}
  \includegraphics[width=\linewidth]{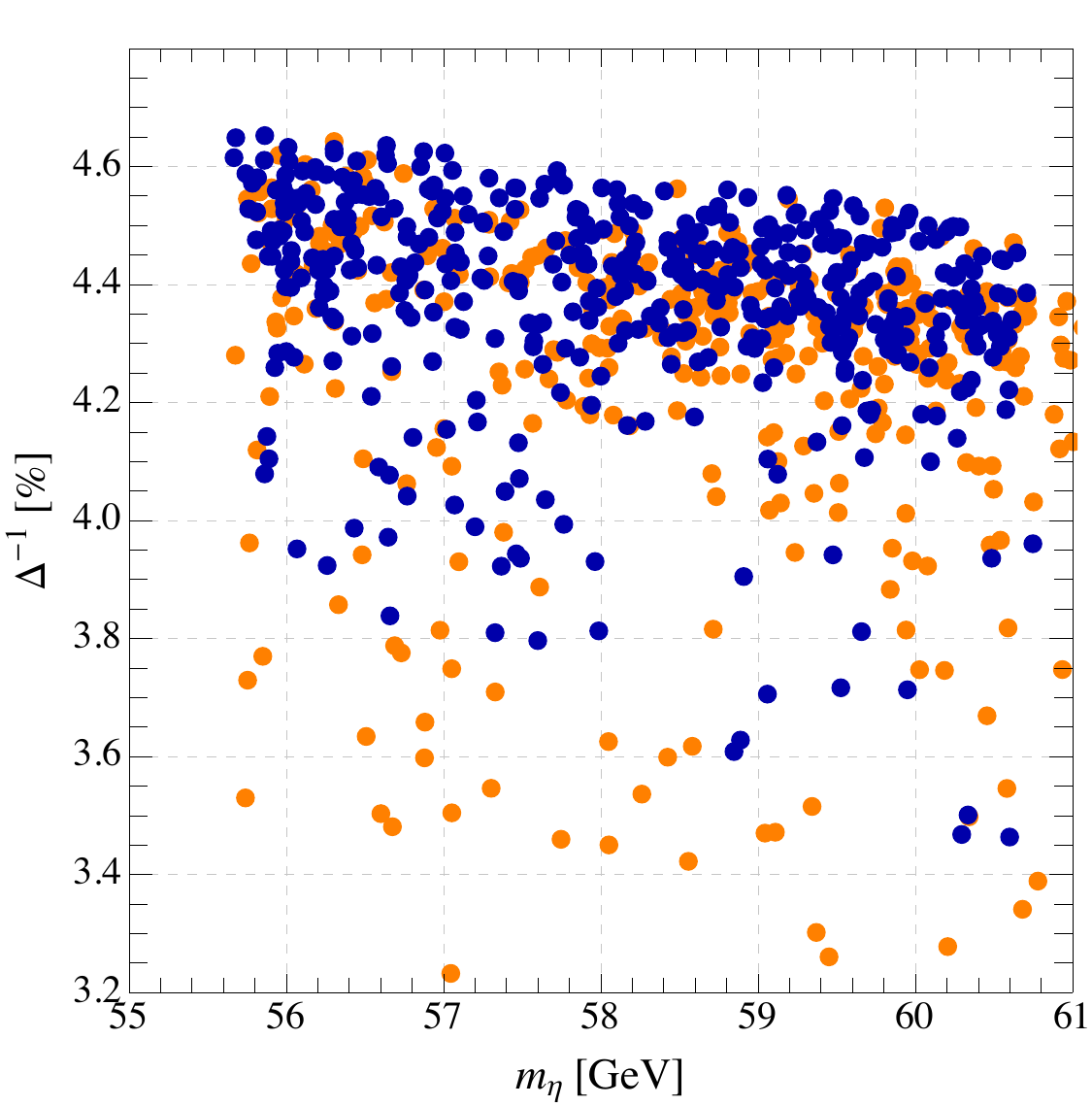}
\endminipage
 \caption{
\textit{Here we show the distribution of the fine-tuning $\Delta$, computed summing in quadrature the logarithmic derivatives of $\xi$ with respect to all the free parameters of the model after imposing the Weinberg sum rules, versus $m_\eta$. The left plot is for $\xi = 0.1$ while the right one is for $\xi = 0.05$. All the points here reproduce the correct top and Higgs masses. The blue points pass the direct searches bounds described in section~\ref{sec:TopPartners}, the orange ones do not.}}
 \label{fig:ScanResultsNS1}
\end{figure*}

To verify the conclusions obtained by our analytic study, we performed a numerical parameter scan of the model, extracting randomly the parameters $f_\rho \in [\frac{1}{\sqrt{2}} f ,2 f]$, $\epsilon_T \in [0.2 f, 6 f]$, $m \in [0, 6 f]$ and obtaining $\epsilon_Q$ by requiring the correct top mass at the TeV scale $M_{top}(1 \text{ TeV}) \simeq 155$ GeV. The vector mass $m_\rho$ finally has been fixed by requiring the desired value of $\xi$ (we took as benchmark points $\xi = 0.1$ and $\xi = 0.05$). After computing the full potential with the chosen parameters, we selected only the points with a Higgs mass between $120$ GeV and $130$ GeV.\footnote{This loose interval has been chosen in order to obtain a sufficient number of points from the scan and because a $\mathcal{O}(5)$ GeV deviation in $m_h$ does not have a significant relevance in our models. Moreover, we expect some small correction to $m_h^2$ to arise from the bottom quark mixing, which we didn't include in the scan.} As can be seen from fig.~\ref{fig:ScanResultsNS1}, our scan confirms the analytical estimations presented above, in particular the relation in eq.~\eqref{eq:RelationsMinimalModel}, within a few percent deviation.
For each point of the scan we computed the fine tuning in $\xi$ adding in quadrature the logarithmic derivatives of $\xi$ with respect to all the free parameters of the model after fixing the Weinberg sum rules (that is $c_i \in \{ f_\rho / f, m_\rho, m, \epsilon_T, \epsilon_Q \}$),
\be
	\Delta = \sqrt{\sum_i \left( \frac{\partial \log \xi}{\partial \log c_i}\right)^2}~,
\ee
 and found that $\Delta^{-1} \simeq 10\%$ for $\xi=0.1$ and $\Delta^{-1} \simeq 5\%$ for $\xi=0.05$, confirming the estimate of eq.~\eqref{eq:TuningMinMod}.

\subsubsection{Next-to-minimal case: $N_F =1$, $N_S = 2$}\label{sec:NTMmodel}

Let us now move to discuss the next-to-minimal scenario with one fundamental and two fermionic singlets.
Also in this model, the mass spectrum before EWSB can be easily obtained from eq.~\eqref{eq:LGeneralFermions}. The mass of the fields in the fundamental is the same as in the previous model, while the two singlets now have a mass
\be
	M_{S_{1,2}}^2 = \frac{1}{2} \left\{ \widetilde m^2 \mp \sqrt{\widetilde m^2 - 4\left[m_{1S}^2 m_{2S}^2 + (\epsilon_{tS}^1)^2 m_{2S}^2 + (\epsilon_{tS}^2)^2 m_{1S}^2\right]} \right\}~,
\ee
where we defined $\widetilde m^2 \equiv m_{1S}^2 + m_{2S}^2 + (\epsilon_{tS}^1)^2 + (\epsilon_{tS}^2)^2$. In the limit where $m_{2S}$ is much bigger than the other masses, these two expressions reduce to $M_{S_{X=1,2}}^2 \simeq m_{XS}^2 + (\epsilon_{tS}^X)^2$.
From eq.~\eqref{eq:GenericTopMass} we get the top mass, at leading order in $\xi \ll 1$
\be
	\hspace{-0.5cm} M_{top} \simeq \frac{\sqrt{\xi} \epsilon_{qF}^1 \epsilon_{tS}^1 \epsilon_{tS}^2  \left| \frac{m_{1S} m_{2S} \epsilon_{tF}^1}{\epsilon_{tS}^1 \epsilon_{tS}^2} + \frac{m_F}{\epsilon_{qF}} \left( \frac{m_{1S} \epsilon_{qS}^2}{\epsilon_{tS}^1} +  \frac{m_{2S} \epsilon_{qS}^1}{\epsilon_{tS}^2} \right)\right|}{\sqrt{2} M_{F_{1/6}} \sqrt{ (M_{S_2}^2 + M_{S_2}^2)^2 - (M_{S_2}^2 - M_{S_2}^2)^2 }}~.
\ee

In this case the most general solution to the first sum rule is (assuming real mixings)
\be \text{(WSR }1)_{ferm}:~\quad\left\{ \begin{split}
	\epsilon_{qF} = \epsilon_Q~, \quad & \epsilon_{qS}^1=\epsilon_Q \cos \theta~, \quad  \epsilon_{qS}^2=\epsilon_Q \sin \theta~, \\
	\epsilon_{tF} = \epsilon_T~, \quad & \epsilon_{tS}^1=\epsilon_T \cos \phi~, \quad  \epsilon_{tS}^2=\epsilon_T \sin \phi~.
	\label{eq:NModelWSR1ferm}
\end{split}  \right. \ee
After imposing this, the second sum rule becomes
\be \text{(WSR }2)_{ferm}:~\quad\left\{ \begin{split}
	m_F^2 &= m_{1S}^2 \cos^2 \theta + m_{2S}^2 \sin^2 \theta~, \\
	m_F^2 &= m_{1S}^2 \cos^2 \phi + m_{2S}^2 \sin^2 \phi~.
\end{split} \right. \ee
Solving these two conditions in terms of $m_{2S}$ and $\phi$, up to arbitrary signs, we get
\be \text{(WSR }2)_{ferm}:~ \quad \left\{ \begin{split}
	m_{2S} & = \frac{1}{\sin \theta} \sqrt{m_F^2 - m_{1S}^2 \cos^2 \theta}~, \\
	\sin \phi & = \sin \theta~.
	\label{eq:NModelWSR2ferm}
\end{split} \right. \ee
Without loss of generality we take $m_{2S} > m_{1S}$. This and eq.~\eqref{eq:NModelWSR2ferm} imply that the relation $m_F^2  > m_{1S}^2$ has to be satisfied.

In this model, from our numerical parameter scans, we find two characteristic regions depending on the values of $m_F$ and $\sin \theta$.
In the limit of small $m_F$, that is of big mixing terms, the DM quadratic term $\mu_\eta^2$ goes to zero, so the DM mass is expected to be of the order of the Higgs mass, and, like in the minimal model, the other coefficients are related by $\mathcal{O}(1)$ factors:
\be
	\lambda = \lambda^f = - \frac{(\mu_h^2)^f}{f^2} \simeq \frac{1}{2} \lambda_h^f \simeq \frac{N_c m_F^2}{8 \pi^2 f^4} (9 + 7 |\sin \theta|) \frac{\epsilon_Q^2 \epsilon_T^2 }{\epsilon_Q^2 - \epsilon_T^2 } \log \frac{\epsilon_Q^2}{\epsilon_T^2}~,
	\label{eq:EstimateParamFermNMinMod}
\ee
where we fixed $m_{1S} = m_F / 2$ in order to respect the bound from the second sum rule and to simplify the expression. In this region this model behaves like the minimal model discussed in the previous section, in particular we expect the DM mass to be $m_\eta \sim 63$ GeV and the coupling $\lambda \sim 6 \times 10^{-2}$, eq.~\eqref{eq:RelationsMinimalModel}. A similar result is obtained by expanding for small mixings $\epsilon_Q$ and $\epsilon_T$ (in order to obtain simple analytic expressions) and going in the $\sin \theta \rightarrow 1$ limit, due to a term proportional to $\log \sin^2 \theta$ in the leading term in $\mu^2_h$ and $\mu^2_\eta$, as in eq.~\eqref{eq:EstimateParamFermNMinMod2}. In this case we exactly reproduce the relations of eq.~\eqref{eq:EstimateParamFermMinMod2}, and therefore the same conclusions apply.

A different region is reached (always in an expansion for small mixings) in the limit of big $m_F \gg f$ and small $\sin \theta \ll 1$, that is with a hierarchy $m_{2S} \gg m_F \gg m_{1S} \sim f$. In this case we obtain
\be \begin{split}
	(\mu_h^2)^f \simeq& ~ - \frac{N_c}{8 \pi^2} \frac{m_F^2 (\epsilon_Q^2 - 2 \epsilon_T^2)}{f^2} \log \frac{1}{\sin^2 \theta}~, \\
	\mu_\eta^2 \simeq& ~ \frac{N_c}{4 \pi^2} \frac{m_F^2 \epsilon_T^2}{f^2} \log \frac{1}{\sin^2 \theta} ~, \\
	\lambda_h^f \simeq& ~ \frac{N_c}{16 \pi^2 f^4} \left[ -2(\epsilon_Q^2 - 2 \epsilon_T^2)^2 + (\epsilon_Q^4 + 4 \epsilon_T^4) \log \frac{m_F^2}{m_S^2} \right]~, \\
	\lambda \simeq& ~ \frac{N_c}{4 \pi^2} \frac{\epsilon_T^2}{f^4} \left( \epsilon_Q^2 - 2 \epsilon_T^2 + \epsilon_T^2 \log \frac{m_F^2}{m_S^2} \right)~.
	\label{eq:EstimateParamFermNMinMod2}
\end{split} \ee
In this case the DM mass can be arbitrarily high (for big $m_F$ and small $\sin \theta$), while in order to obtain the correct EW scale, that is to suppress $(\mu_h^2)^f$, it is necessary to tune $\epsilon_Q^2 \sim 2 \epsilon_T^2$. If this tuning is avoided here, then the gauge contribution to $\mu_h^2$ has to provide the necessary cancellation, which will imply higher values of the vector mass $m_\rho$ than the case in eq.~\eqref{eq:VectorMassMinMod}. In both cases, we expect the tuning in this region to be higher than in the cases examined previously, for which the expected tuning is as in eq.~\eqref{eq:TuningMinMod}.
Taking $\epsilon_Q^2 \sim 2 \epsilon_T^2$, from the expression for $\lambda_h$ in eq.~\eqref{eq:EstimateParamFermNMinMod2} we can fix $\epsilon_T$ by requiring the correct Higgs mass and then substitute this in the formula for $\lambda$. We obtain
\be\label{eq:Lambda2}
	\lambda \simeq \frac{m_h^2}{4 v^2} \simeq 0.065 ~,
\ee
which is the same value we obtained in the minimal model.

\begin{figure}[!htb!]
\vspace{-0.8cm}
\begin{center}
\fbox{\footnotesize $N_F = 1$, $N_S = 2$, $\xi=0.1$} \\
\hspace*{-0.65cm} 
\begin{minipage}{0.5\linewidth}
\begin{center}
	\includegraphics[width=\linewidth]{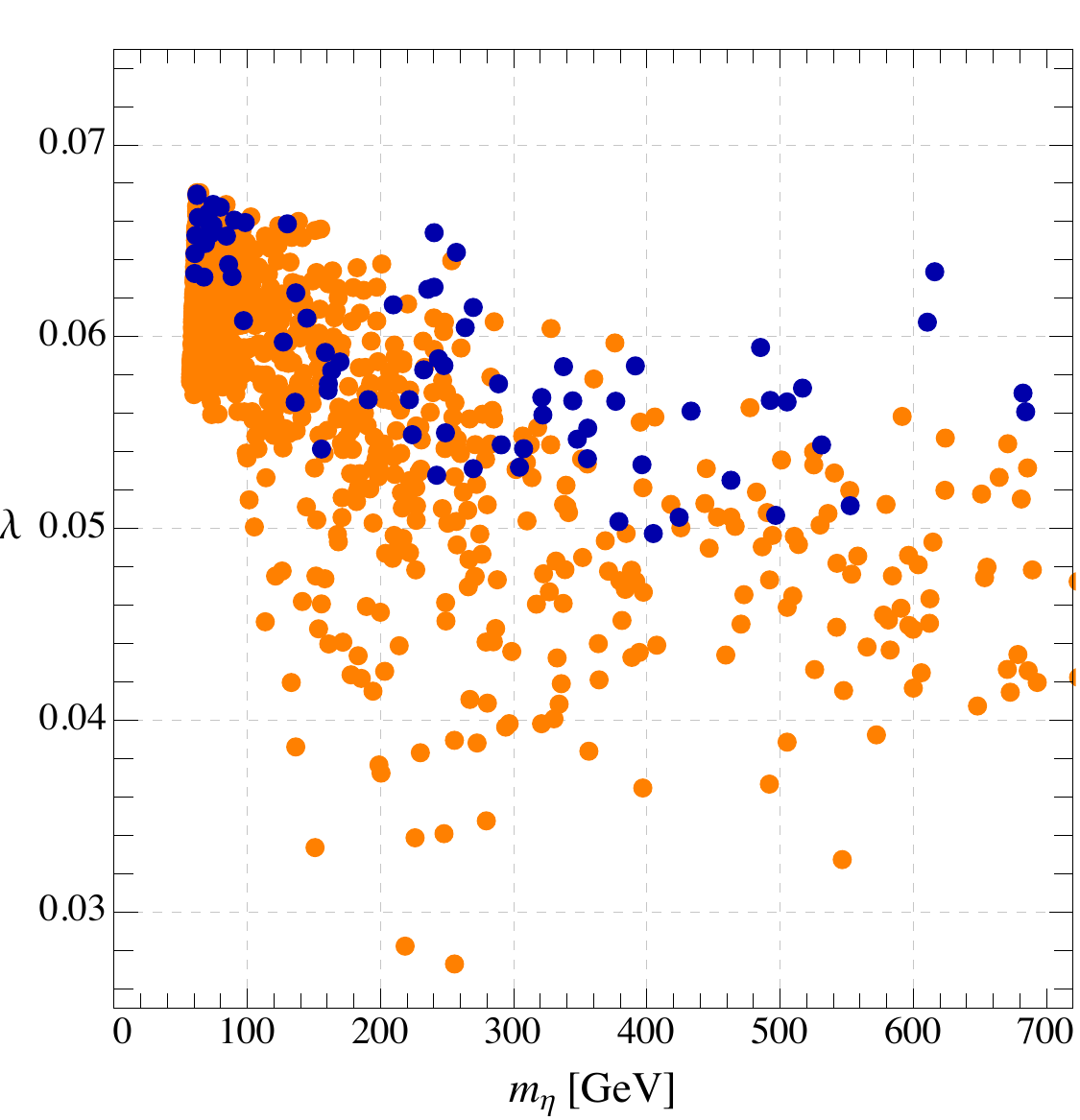}
\end{center}
\end{minipage}
\begin{minipage}{0.5\linewidth}
\begin{center}
	\includegraphics[width=\linewidth]{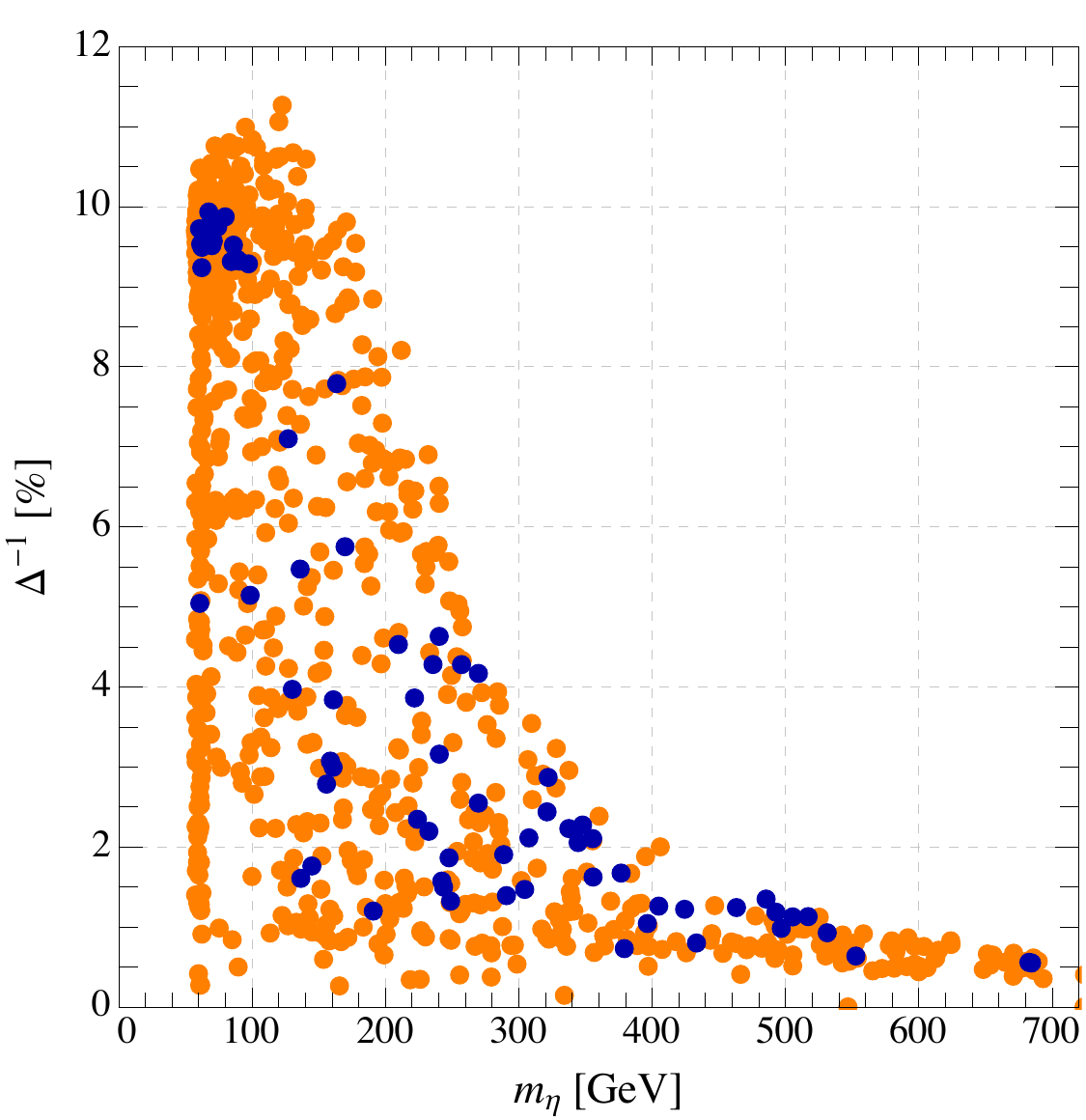}
\end{center}
\end{minipage}
\\[0.1cm]
\fbox{\footnotesize $N_F = 1$, $N_S = 2$, $\xi=0.05$} \\
\hspace*{-0.65cm} 
\begin{minipage}{0.5\linewidth}
\begin{center}
	\includegraphics[width=\linewidth]{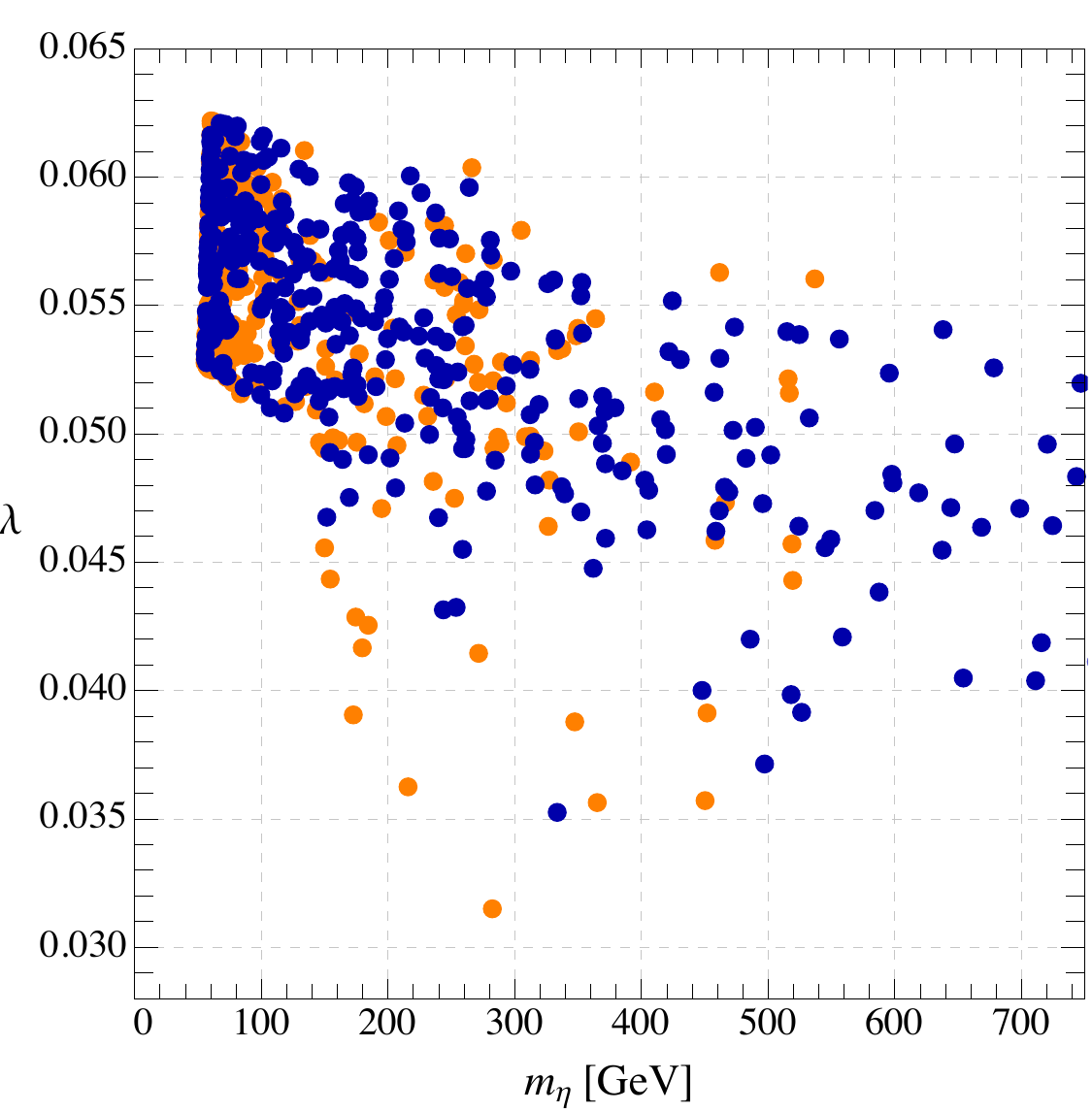}
\end{center}
\end{minipage}
\begin{minipage}{0.5\linewidth}
\begin{center}
	\includegraphics[width=\linewidth]{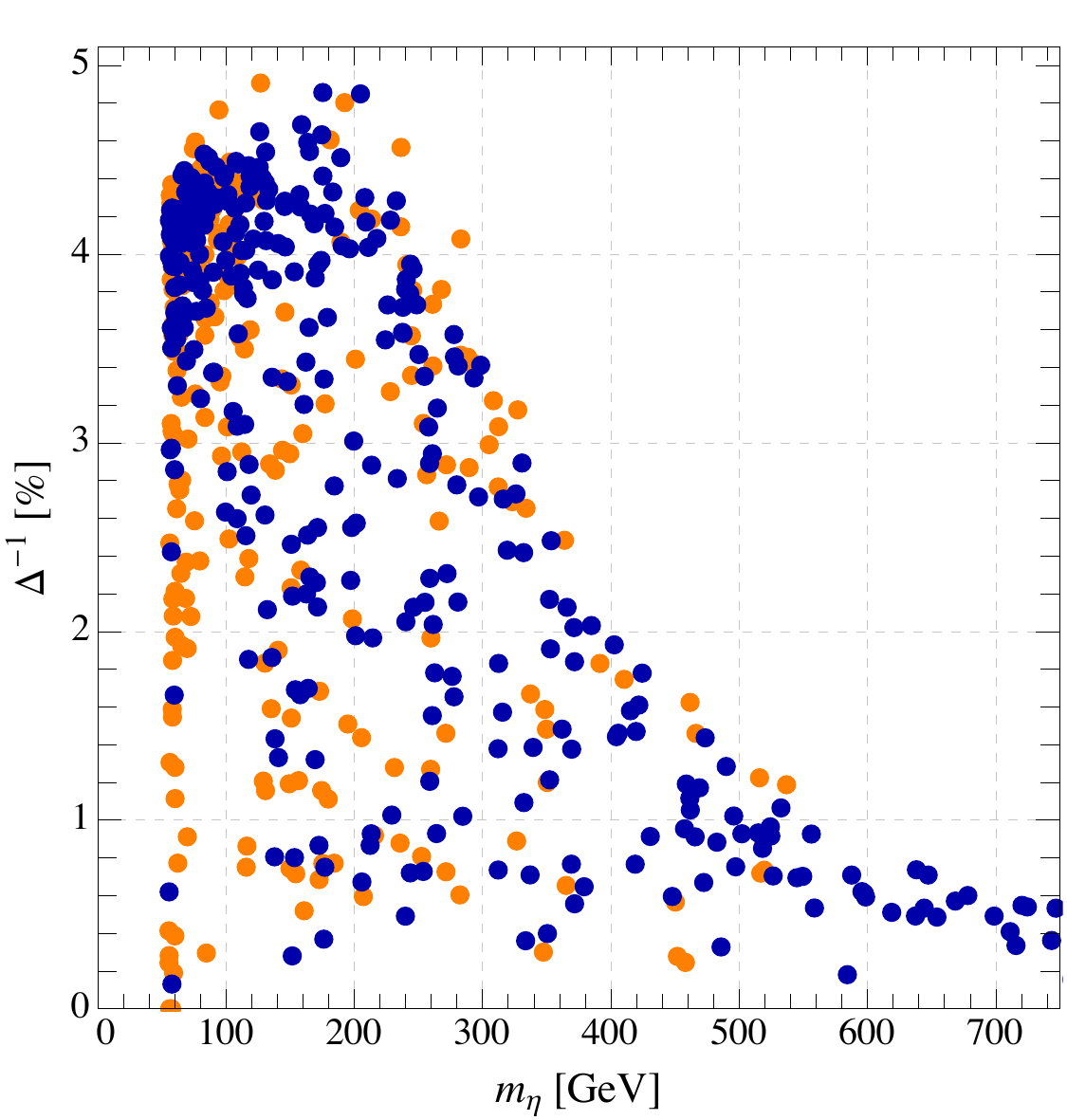}
\end{center}
\end{minipage}
\end{center}
\vspace*{-0.2cm}
\caption{\label{fig:ScanResultsNS2}
\textit{In the left column we show the distribution of the points obtained from the scan of the next-to-minimal model in the $(m_\eta, \lambda)$ plane, while in the right column we show the distribution of the fine-tuning $\Delta$, computed summing in quadrature the logarithmic derivatives of $\xi$ with respect to all the parameters of the model, versus $m_\eta$. The upper row is for $\xi = 0.1$ while the lower one for $\xi = 0.05$. All the points here reproduce the correct top and Higgs masses. The blue points pass the direct searches bounds described in section~\ref{sec:TopPartners}, the orange ones do not.}}
\end{figure}

Also in this case we performed a numerical parameter scan of the model, extracting randomly $f_\rho \in [\frac{1}{\sqrt{2}} f,2 f]$, $\epsilon_T \in [0.2 f, 6 f]$, $m_S \in [0, 8 f]$, $m_F \in [m_S, 8 f]$, $\theta \in [0, \frac{\pi}{2}]$ and obtaining $\epsilon_Q$ by requiring the correct top mass at the TeV scale $M_{top}(1 \text{ TeV}) \simeq 155$ GeV. As in the minimal model, the vector mass $m_\rho$ has been fixed by requiring $\xi = 0.1$ (or $0.05$) and we selected only the points with a Higgs mass between $120$~GeV and $130$~GeV.
From these scans we observe that, even when relaxing the tuning condition $\epsilon_Q^2 \sim 2 \epsilon_T^2$, the value of the coupling $\lambda$ remains always of the same order of magnitude, that is in the range $3 \times 10^{-2} \lesssim \lambda \lesssim 7 \times 10^{-2}$, while the DM mass can vary from $m_\eta \sim m_h/2$ up to $m_\eta \sim \mathcal{O}(700)$ GeV, see figure~\ref{fig:ScanResultsNS2}.

Computing the fine-tuning as presented in the minimal model, we find that for $m_\eta \lesssim 200$ GeV most of the points present $\Delta^{-1} \sim \xi$ with a tail of points with $\Delta^{-1} \lesssim 0.5\%$, as can be seen in the right panels of figure~\ref{fig:ScanResultsNS2}. Increasing $m_\eta$ the fine-tuning increases: for $m_\eta \simeq 600$ GeV we have $0.5 \% \lesssim \Delta^{-1} \lesssim 1\%$.

\subsection*{Relaxing the second Weinberg sum rules}

In order to assess the generality of our prediction for $\lambda \sim 6 \times 10^{-2}$, which we obtain both in the minimal and in the next-to-minimal models presented above, we also consider a generalization of the next-to-minimal model in which we impose only eq.~\eqref{eq:NModelWSR1ferm}, relaxing the second Weinberg sum rules of eq.~\eqref{eq:NModelWSR2ferm}.
As discussed before, and in more detail in appendix~\ref{App:EffectivePotential}, this renders the effective potential incalculable. In particular, relaxing the second sum rules leaves a logarithmic divergence (i.e. a scale dependence) in $\mu_h^2$ and $\mu_\eta^2$. On the other hand, the quartic couplings $\lambda$, $\lambda_h$ and $\lambda_\eta$ are still scale-independent and therefore calculable.
As a consequence, both $\xi$ and $m_\eta^2$ can not be explicitly computed in this case but need to be fixed as boundary conditions.

\begin{figure*}[t]
\centering
\fbox{\footnotesize $N_F = 1$, $N_S = 2$, $\xi = 0.1$, only first WSR} \\
\hspace{-0.3cm}
\minipage{0.5\textwidth}
  \includegraphics[width=0.96\linewidth]{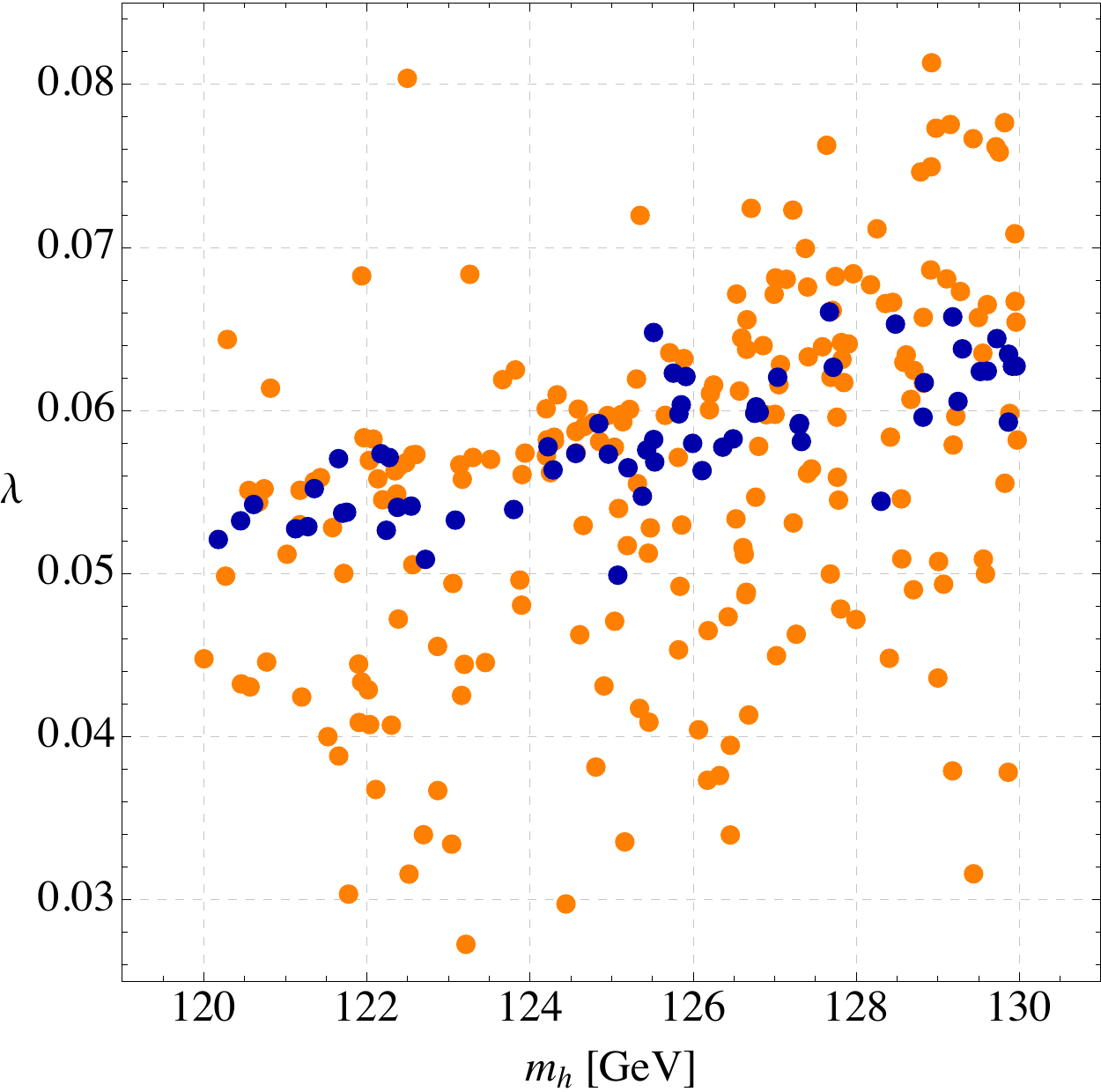}
\endminipage
\minipage{0.5\textwidth}
  \includegraphics[width=0.98\linewidth]{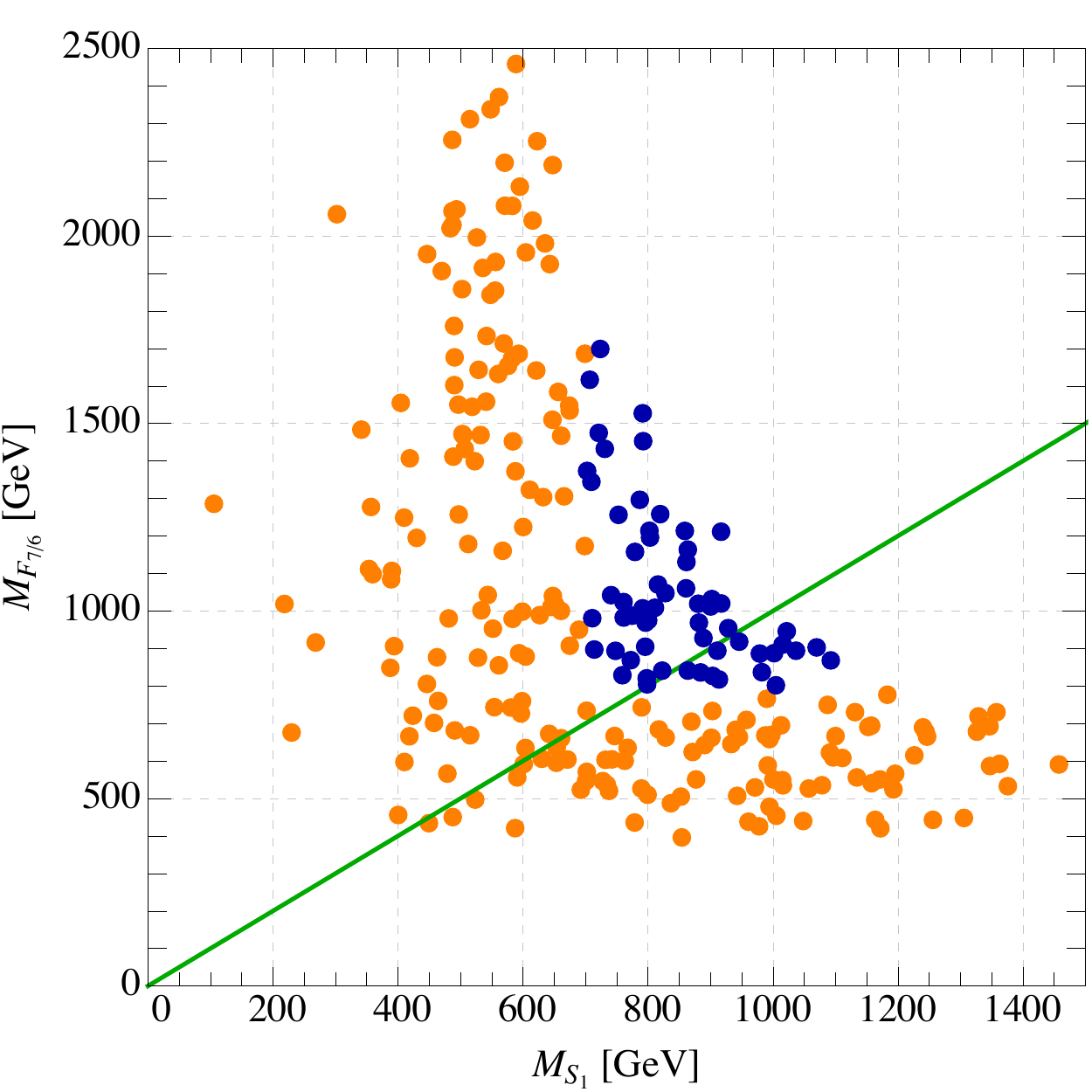}
\endminipage
 \caption{\textit{In the left plot we show the points obtained from the parameter scan in the model with $N_S=2,~N_F=1$ relaxing the second Weinberg sum rules, in the ($m_h, \lambda$) plane. In the right one we show the lightest top partner masses, the green line is a reference line for $M_{F_{7/6}} = M_{S_1}$. The blue points pass the direct searches bounds described in section~\ref{sec:TopPartners}, the orange ones do not.}}
 \label{fig:lambda_mh_logdiv}
\end{figure*}

Since we are mostly interested in the range of $\lambda$ given the measured Higgs mass, we performed a parameter scan of this model fixing $\xi = 0.1$ and extracting randomly $\epsilon_T \in [0.2 f, 6 f]$, $m_{1S} , m_F \in [0, 8 f]$, $m_{2S} \in [m_{1S}, 8 f]$, $\theta \in [0, \frac{\pi}{2}]$, $\phi \in [0, \frac{\pi}{2}]$ and obtaining $\epsilon_Q$ by requiring the correct $M_{top}.$\footnote{We took into consideration only the fermion sector, since the gauge contribution to the Higgs mass is always negligible due to the $g^4$ factor as well as a numerical suppression, see eq.~\eqref{eq:PotentialCoeffGauge}.} For each point we computed $\lambda$ and $m_h$ and selected only the points with $m_h$ between $120$ GeV and $130$ GeV.
As shown in the left panel of figure~\ref{fig:lambda_mh_logdiv}, we obtain that $\lambda$ ranges from $\sim 3\times 10^{-2}$ and $\sim 8 \times 10^{-2}$, with the distribution of the points peaked near $\lambda \sim 6 \times 10^{-2}$, thus confirming the range obtained in the cases where both Weinberg sum rules were being imposed. The DM mass $m_\eta$, not being calculable, is in this case a free parameter.

\section{Phenomenological analysis -- part I: LHC}\label{sec:phenoLHC}
In this section we analyze the constraints placed on the parameter space of our Composite DM model by the LHC. In section~\ref{sec:InvisibleWidth} we discuss the bound on the invisible Higgs  decay width, while in section~\ref{sec:TopPartners} we consider  direct searches of composite resonances.

\subsection{Invisible Higgs decay width}\label{sec:InvisibleWidth}

If $m_{\eta}<m_{h}/2$, the Higgs boson can decay invisibly into two DM particles. The invisible decay width 
corresponding to this process is given by \cite{Frigerio:2012uc}
\begin{equation}\label{eq:InvisibleWidth}
\Gamma_{\rm inv}(h\to \eta\eta) = 
\frac{v^2}{32\pi m_h}\left(
\frac{m_h^2 \xi}{v^2\sqrt{1-\xi}} - 2\lambda \sqrt{1-\xi}
\right)^2\sqrt{1-\frac{4m_{\eta}^2}{m_h^2}}~\theta(m_h - 2m_{\eta})~.
\end{equation}
In addition to the invisible Higgs decay width in eq.~(\ref{eq:InvisibleWidth}), composite Higgs models also predict   $\mathcal{O}(\xi)$ deviations of the tree level Higgs couplings to gauge bosons and fermions w.r.t. their SM values \cite{Giudice:2007fh,Montull:2013mla}. 
In particular in our model we have
\begin{equation}\label{eq:deviations}
g_{hVV} =g_{hVV}^{\rm SM}\,\,\sqrt{1-\xi}~,~~~~~~~~~~
g_{hf\bar{f}} =g_{hf\bar{f}}^{\rm SM}\,\,\frac{1-2\xi}{\sqrt{1-\xi}}~,
\end{equation}
with $V=W,Z$, see table~\ref{table:CouplingsParam}.
It should be noted here that the $\xi$-dependence in the modified coupling of the Higgs with  EW gauge bosons is model-independent,\footnote{In general the couplings depend on the chosen parametrization of the coset, only when computing physical observables this parametrization-dependence is removed. See appendix~\ref{App:parametrizations} for a detailed discussion of this issue.} whereas the coupling with fermions is modified according to the representation of $SO(6)$ in which the SM fermions are embedded. Following the discussion in section~\ref{sec:fermionembedding}, the result in eq.~(\ref{eq:deviations}) refers to the embedding of SM fermions in the fundamental $\textbf{6}$ of $SO(6)$.\footnote{See ref.~\cite{Urbano:2013aoa} for a special case, based on the non-compact global symmetry $SO(4,1)$, in which $g_{hVV} =g_{hVV}^{\rm SM}\,\,\sqrt{1 + \xi}$.}
Loop-induced couplings -- i.e. Higgs couplings to gluons, photons and $Z\gamma$ -- are also modified as an indirect consequence of eq.~(\ref{eq:deviations}).
For instance the Higgs coupling to gluons, whose value sets the Higgs production cross-section via gluon fusion, is dominated by the top triangle loop and modified according to $g_{hgg}\approx g_{hgg}^{\rm SM}\,\,(1-2\xi)/\sqrt{1-\xi}$. 

\begin{figure*}[!t]
\hspace{-0.2cm}
\minipage{0.5\textwidth}
  \includegraphics[width=\linewidth]{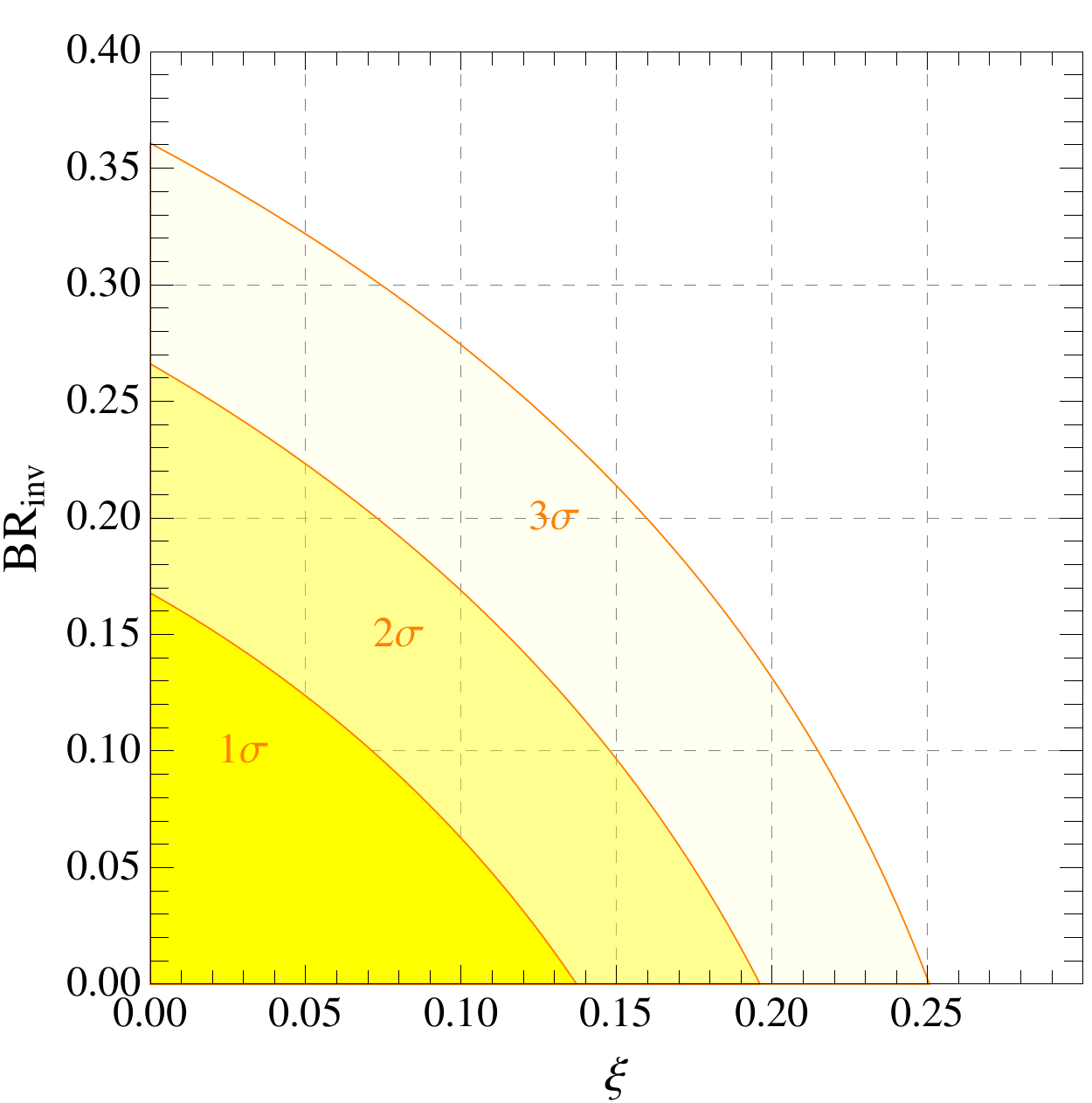}
\endminipage
\hspace{0.2cm}
\minipage{0.5\textwidth}
  \includegraphics[width=\linewidth]{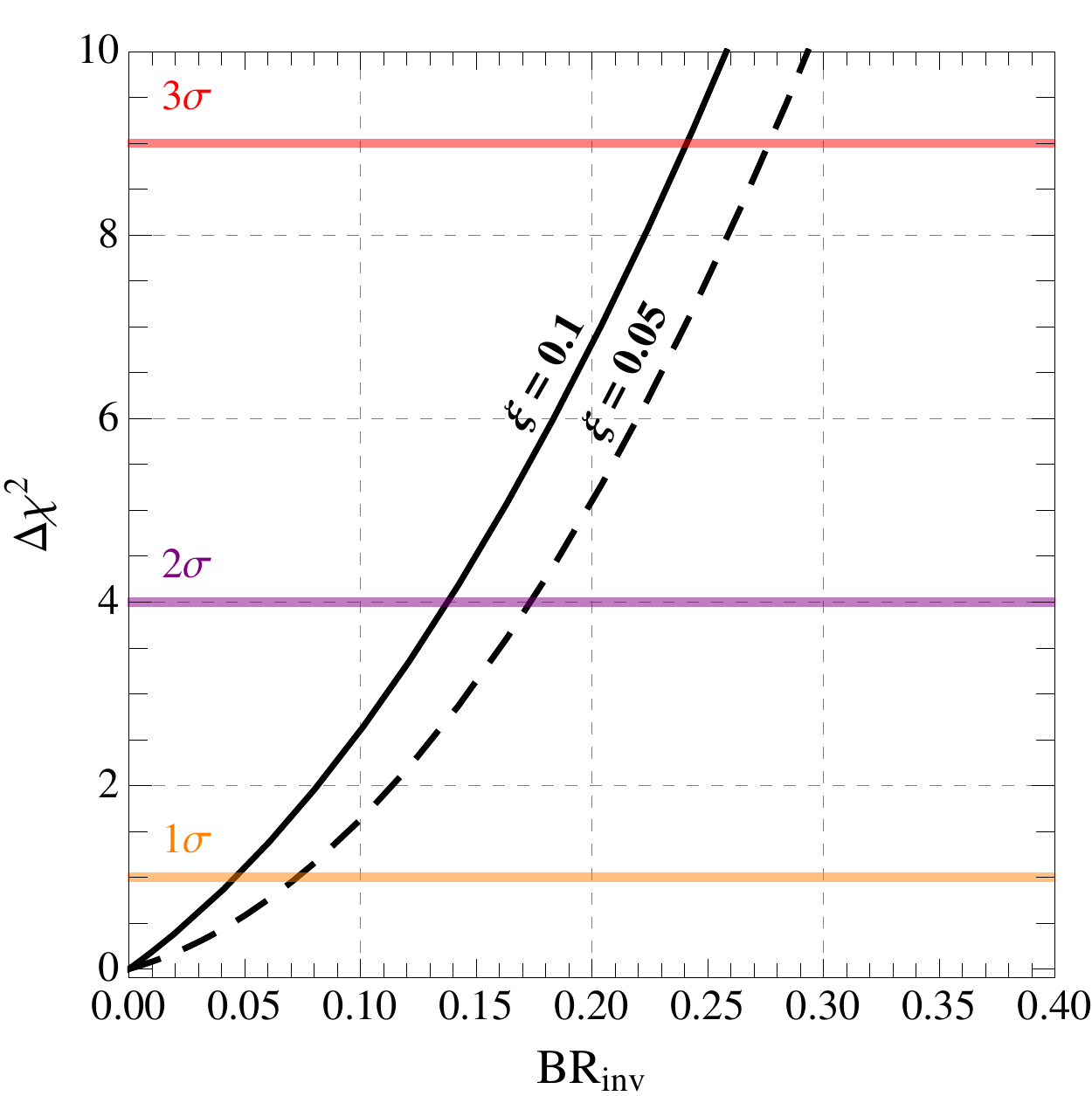}
\endminipage
 \caption{\textit{
Results of the $\chi$-square fit obtained considering all the Higgs searches under investigation at the LHC and the TeVatron (see ref.~\cite{Falkowski:2013dza} for details).
 In the left panel we show the 1$\sigma$, 2$\sigma$ and 3$\sigma$  confidence regions 
obtained considering a two-dimensional fit of the data as a function of the invisible branching ratio and the parameter $\xi$. In the right panel we show the
$\Delta\chi^2=\chi^2-\chi^2_{\rm min}$ distribution together with
 the corresponding 1$\sigma$, 2$\sigma$ and 3$\sigma$ confidence contours as a function of the invisible branching ratio for a fixed value of $\xi$, where $\chi^2_{\text{min}} = 52~(51)$ for $\xi = 0.1~(0.05)$.
 }}\label{fig:InvisibleBR}
\end{figure*}

The proprieties of the Higgs boson, and in particular its couplings to each of the SM 
gauge bosons and fermions, are currently under investigation at the LHC.
The couplings are measured by the ATLAS \cite{ATLAS_couplings} and CMS \cite{CMS_couplings} experiments considering the channels
$h\to \gamma\gamma$, $h\to ZZ^*$ (with $ZZ^*\to 4l,2l2\nu,2l2q,2l2\tau$), $h\to WW^*$ (with $WW^*\to l\nu l\nu,l\nu qq$), $h\to b\bar{b}$ and $h\to \tau^+\tau^-$ (with both leptonic and hadronic $\tau$-decays).
 The invisible decay width of the Higgs boson
 is strongly constrained by the fact that 
the rates associated to the channels listed above are compatible with
the predictions of the SM \cite{Aad:2014iia,CMS_invisible}. 
In our analysis we perform a combined 
fit of all the data related to the Higgs searches under investigation at the LHC and the TeVatron taking into account 
both the modified Higgs couplings in eq.~(\ref{eq:deviations}) and the invisible decay width  in eq.~(\ref{eq:InvisibleWidth}). The latter is rephrased in terms of the following invisible branching ratio
\begin{equation}\label{eq:BRinv}
{\rm BR}_{\rm inv} \equiv \frac{\Gamma_{\rm inv}(h\to \eta\eta)}
{\Gamma_{\rm SM}^{~\xi}+\Gamma_{\rm inv}(h\to \eta\eta)}~,
\end{equation}
where $\Gamma_{\rm SM}^{~\xi}$ is the decay width of the Higgs boson into SM particles
obtained including the deviations of the Higgs couplings in eq.~(\ref{eq:deviations}).
We perform a $\chi$-square fit following ref.~\cite{Falkowski:2013dza} (see also refs.~\cite{Giardino:2013bma,Carmi:2012in,Azatov:2012bz,Montull:2012ik,Espinosa:2012im,Ellis:2013lra} for similar analysis) and we present our results in figure~\ref{fig:InvisibleBR}.
In the left panel of figure~\ref{fig:InvisibleBR} 
 we show the result of a two-dimensional fit considering as free parameters both ${\rm BR}_{\rm inv}$ and $\xi$.
Notice that larger values of ${\rm BR}_{\rm inv}$ are allowed only if combined with small values of $\xi$. The reason is that a high value of $\xi$ suppresses the Higgs production cross-section via gluon fusion, as immediately follows from the modified coupling $g_{hgg}$ previously discussed. This suppression, in turn, gives a tighter bound on the invisible branching fraction since, intuitively, less Higgses than expected are produced \cite{Falkowski:2013dza}.
 In the right panel of figure~\ref{fig:InvisibleBR} we restrict our analysis to a one-dimensional fit
 obtained considering 
 as  free parameter only the invisible branching ratio, while we fix the parameter $\xi$ to the two benchmark values $\xi =0.1$ and $\xi = 0.05$. For $\xi=0.1$ ($\xi=0.05$)
 we find that ${\rm BR}_{\rm inv}>0.24$ (${\rm BR}_{\rm inv}>0.275$) is excluded at 3$\sigma$ level. 
 Writing explicitly ${\rm BR}_{\rm inv}$ as a function of the DM mass and the Higgs portal coupling -- using eqs.~(\ref{eq:InvisibleWidth},~\ref{eq:BRinv}) -- it is possible to draw an exclusion curve in the plane $(m_{\eta},\lambda)$. We will show this bound in section~\ref{sec:results}, together with all the other phenomenological constraints that we will derive in the following sections.

\subsection{Direct searches of composite resonances}
\label{sec:TopPartners}

\begin{figure}[!htb!]
\vspace{-0.1cm}
\begin{center}
\fbox{\footnotesize $N_F = 1$, $N_S = 1$} \\
\hspace*{-0.65cm} 
\begin{minipage}{0.5\linewidth}
\begin{center}
	\includegraphics[width=\linewidth]{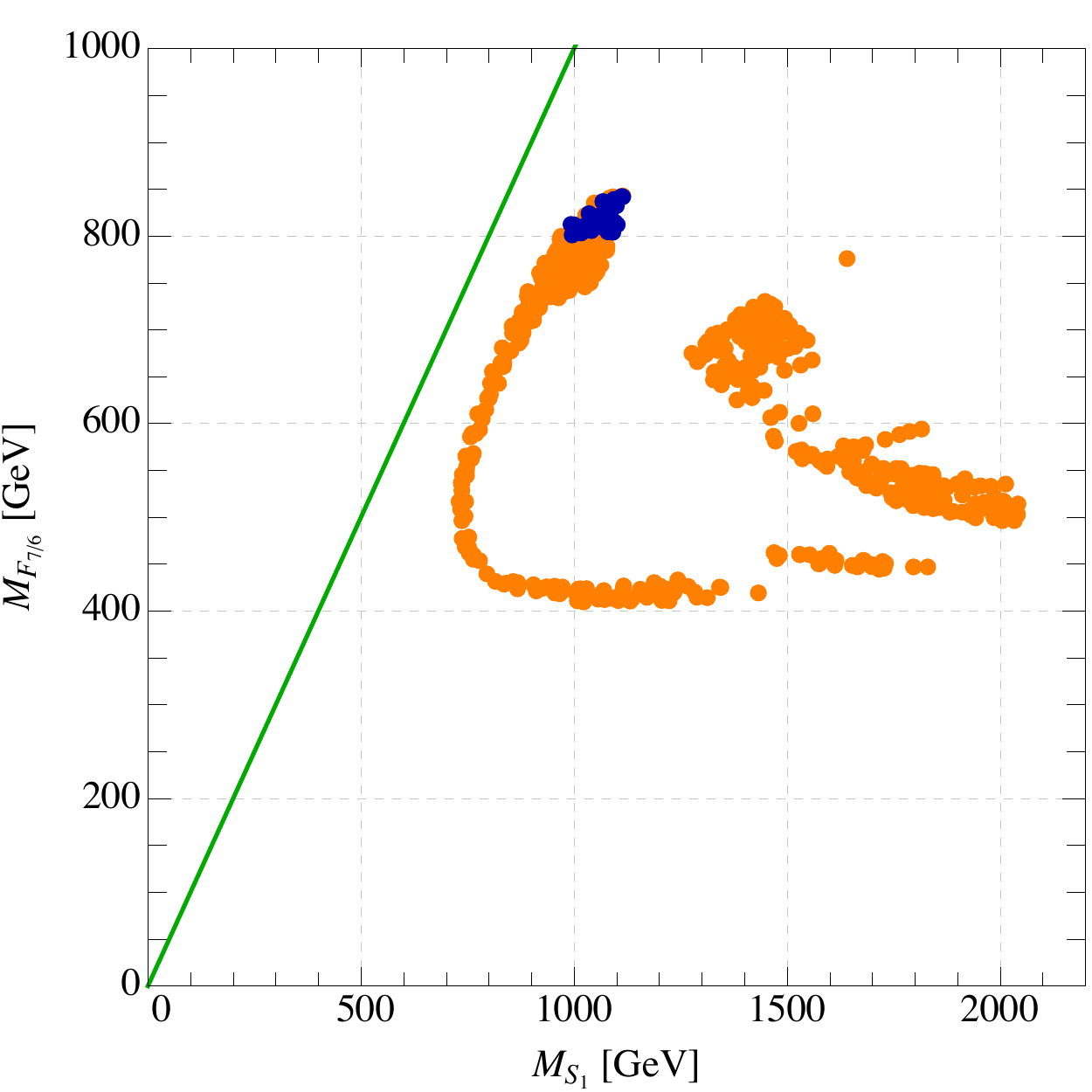}
\end{center}
\end{minipage}
\begin{minipage}{0.5\linewidth}
\begin{center}
	\includegraphics[width=\linewidth]{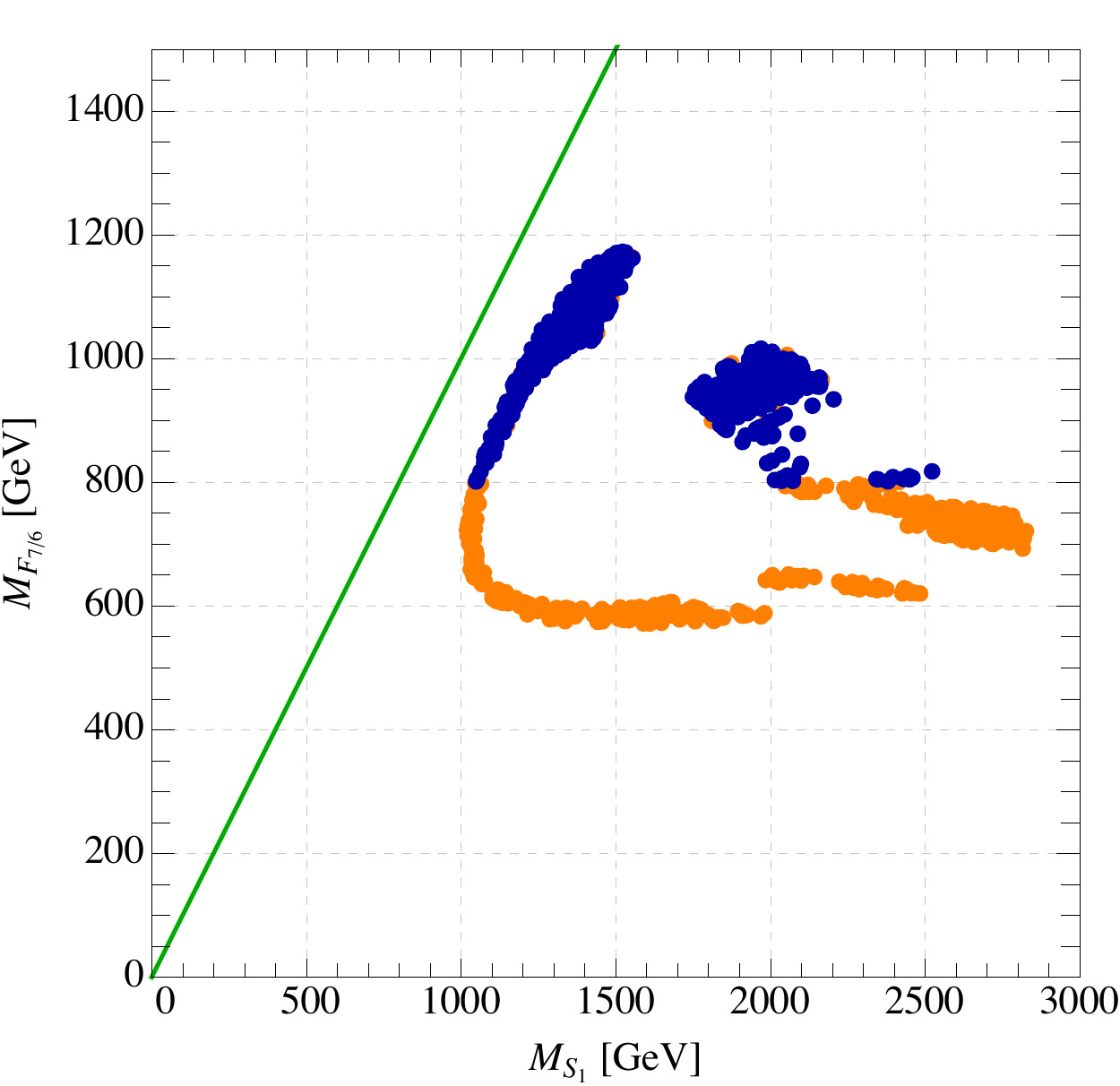}
\end{center}
\end{minipage}
\\[0.1cm]
\fbox{\footnotesize $N_F = 1$, $N_S = 2$} \\
\hspace*{-0.65cm} 
\begin{minipage}{0.5\linewidth}
\begin{center}
	\includegraphics[width=\linewidth]{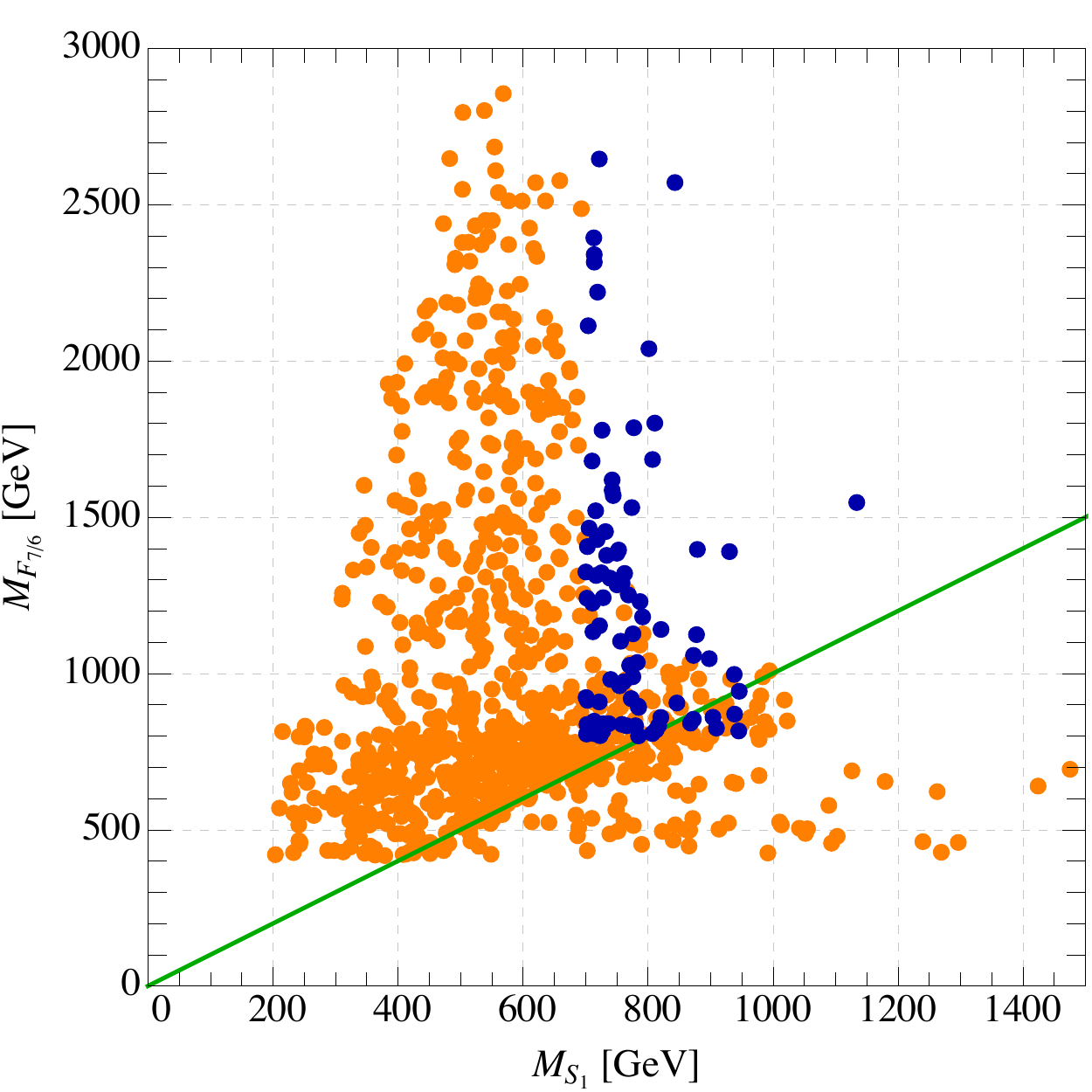}
\end{center}
\end{minipage}
\begin{minipage}{0.5\linewidth}
\begin{center}
	\includegraphics[width=\linewidth]{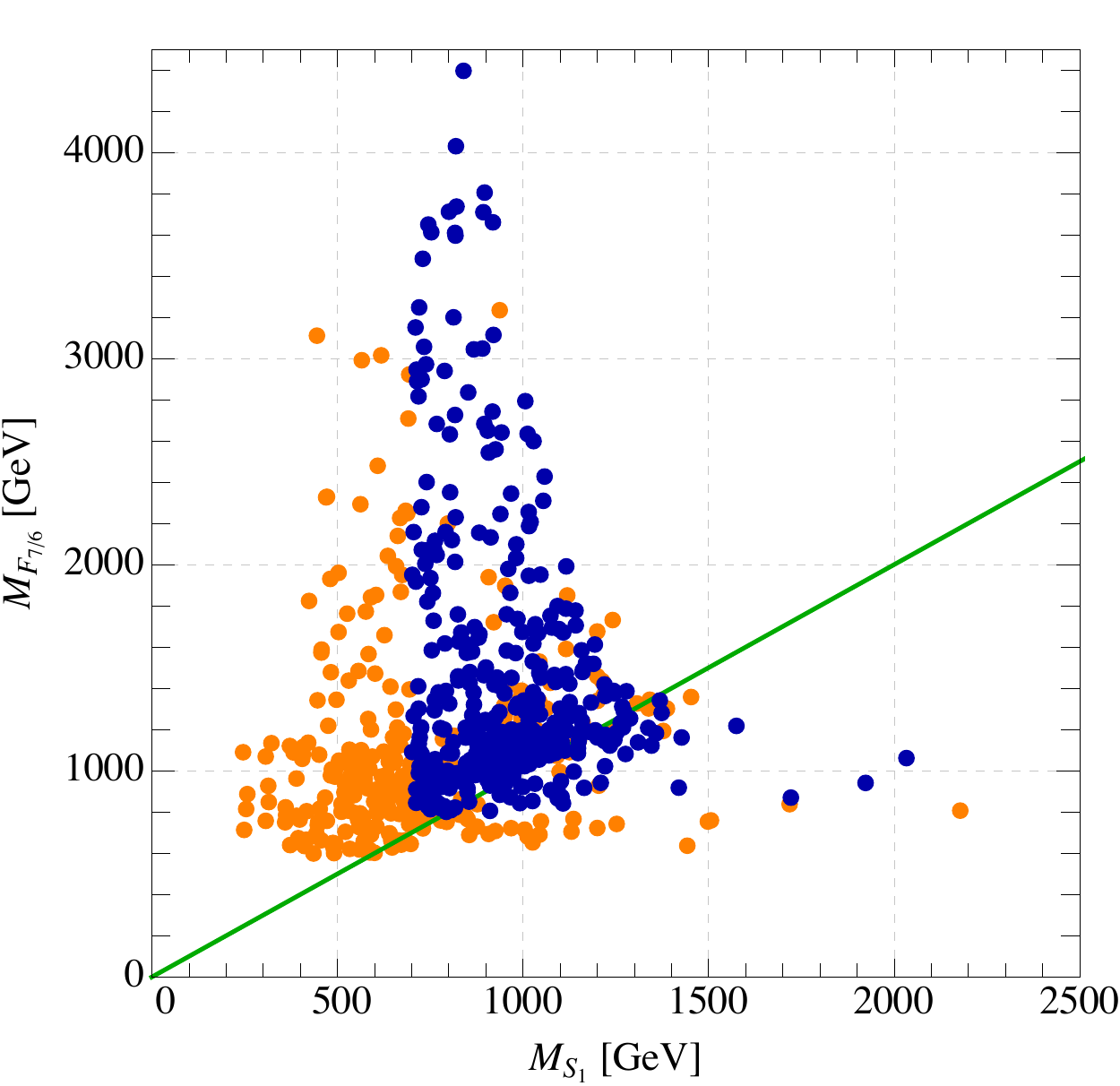}
\end{center}
\end{minipage}
\end{center}
\vspace*{-0.2cm}
\caption{\label{fig:topPartners}
\textit{In the upper (lower) row we show the lightest top partner masses (before EWSB) in the minimal (next-to-minimal) model with for $\xi = 0.1~[0.05]$ in the left [right] plot. The points reproduce the correct top and Higgs masses, up to a $\sim 5$ GeV tolerance on $m_h$. The blue points pass the selection while the orange ones are excluded by direct searches of top partners and vector resonances, eqs.~(\ref{eq:ExpBoundVectors},~\ref{eq:ExpBoundTopPartners}). The green line is a reference for $M_{F_{7/6}} = M_{S_1}$.}}
\end{figure}

In this section we focus on constraints from the LHC on the composite resonances present in our models, discussed in sections~\ref{sec:minimalcase},~\ref{sec:NTMmodel}. It is already well established that, in the context of composite pseudo-Nambu Goldstone Higgs models with partial compositeness, the measured value of the Higgs mass requires the presence of top-partners with a mass below the TeV scale \cite{Matsedonskyi:2012ym,Redi:2012ha,Marzocca:2012zn,Pomarol:2012qf}.
The parameter scans we performed for our models and which we presented in the previous section confirm this fact, as can be seen from figure~\ref{fig:topPartners} (see also the right panel of figure~\ref{fig:lambda_mh_logdiv}).
Moreover, in the minimal model and in some regions of the second model, the spin-1 resonances are expected to be near the $\sim 2$ TeV scale \eqref{eq:VectorMassMinMod}.

The present experimental bounds on spin-1 resonances and, more importantly, on spin-1/2 top partners are already able to rule out a relevant part of the parameter space of our models.\footnote{In this work we decided to focus on bounds from direct searches and not consider constraints from EW precision tests. Even though the latter, in particular those from the oblique $S$ and $T$ parameters and from $Z b \bar{b}$ coupling deviations, can in principle provide similar bounds as direct searches, they suffer from a larger model dependence and, in the case of strongly coupled models, some lack of predictability. For example, even though vector resonances contribute to $S$ at tree level, the IR one-loop contribution to the oblique parameters due to the deviation in the Higgs couplings to the SM gauge bosons and the loop contribution from composite fermions are both very important and all have to be taken into account. In particular it has been shown \cite{Grojean:2013qca} that some of the couplings in eq.~\eqref{eq:GenericExtraFermLagr}, which do not contribute to the effective potential, can instead give important contributions to $S$ and $T$. In addition, the bounds from direct searches have already reached a similar sensitivity to those from indirect constraints.}

Ref.~\cite{Pappadopulo:2014qza} recently studied the bounds from direct searches at the LHC of spin-1 resonances introducing a simplified model with a triplet of $SU(2)_L$ and presenting the bound in the $(g_\rho, m_\rho)$ plane.
Our model presents a more complicated spectrum of vector resonances: the adjoint of $SO(5)$ ($\rho_\mu^a$), with masses of the order $m_\rho$, contains a $\bf (3,1)\oplus (1,3) \oplus (2,2)$ of $SU(2)_L \otimes SU(2)_R$ and the fundamental of $SO(5)$ ($a_\mu^{\hat{a}}$), with mass $m_a$, contains $\bf (2,2) \oplus (1,1)$. In order to obtain experimental bounds on these states it would be necessary to perform a complete collider study of the model, including also possible chain decays involving composite fermions through the interactions of eq.~\eqref{eq:GenericExtraFermLagr}, see ref.~\cite{Vignaroli:2014bpa} for a recent phenomenological analysis of this issue.
Since this is well beyond the purpose of this work we take at face value, as an approximate reference value of the experimental bound on these states, the result of ref.~\cite{Pappadopulo:2014qza}. Fixing the two benchmark values of $\xi = 0.1,~0.05$ and taking for simplicity $f_\rho = f$, so that $m_\rho \simeq g_{\rho} f = g_{\rho} \frac{v}{\sqrt{\xi}}$, we get that the allowed region is approximately
\be
	m_\rho \gtrsim 1.8~(2.2) \mbox{ TeV} \quad \text{for } \quad \xi = 0.1~(0.05)~.
	\label{eq:ExpBoundVectors}
\ee
This is comparable with the bound one can extract from the tree-level contribution of the spin-1 resonances to the $\hat{S}$ parameter \cite{Peskin:1990zt,Barbieri:2004qk} of eq.~\eqref{eq:Sparam}, assuming no correlation with other contributions. From the constraint $ \hat{S} \lesssim 2 \times 10^{-3}$ \cite{gfitter} one obtains a bound of $m_\rho \gtrsim 1.8~(2.4)$ TeV for $f_\rho = f / \sqrt{2}$ $(= 2 f)$.

Both ATLAS and CMS collaborations are providing bounds on pair produced top partners, studying different decay modes. The relevant searches for our models are those for colored vector-like fermions, $X$, with electric charge $Q = 5/3$ decaying in $W^+ t$ with $\text{BR}(X \rightarrow W^+ t) = 100\%$ \cite{Chatrchyan:2013wfa,ATLAS:2012130} and for vector-like top partners $T^\prime$ with $Q = 2/3$ decaying into $b W^+$, $t Z$ and $t h$ \cite{Chatrchyan:2013uxa,ATLAS:2013060,ATLAS:2013018}.
The $Q=5/3$ fermion decays with unity probability to $t W^+$ when it is the lightest and masses $M_X < 800$ GeV are excluded at 95$\%$ C.L. by CMS \cite{Chatrchyan:2013wfa}.
The branching ratios of the $T^\prime$ in the three channels listed before are instead model-dependent and the 95$\%$ C.L. bound given in ref.~\cite{Chatrchyan:2013uxa} varies from $\sim 680$ GeV up to $\sim 780$ GeV.  Applying the Equivalence Theorem gives a reference value, for the singlet branching ratios, of $\text{BR}(T^\prime \rightarrow W^+ b) \simeq 2 \text{BR}(T^\prime \rightarrow Z t) \simeq 2 \text{BR}(T^\prime \rightarrow h t) \simeq 50 \%$ \cite{DeSimone:2012fs}, in which case the bound is $\sim 700$ GeV.
These analysis are always performed under the assumption that only one new state is present at low energy while the others are much heavier. This assumption is very strong and seldom realized in concrete models, including our case.
For these reasons a complete analysis of the experimental results in order to adapt them to the realistic case would be needed, but is beyond the purpose of the present work. 

Let us classify the parameter space of our models in three broad regions depending on the mass of the doublet which includes the exotic $Q=5/3$ fermion, $M_{7/6}$, and the mass of the lightest of the two $SO(5)$ singlets, $M_{S_1}$. The first region is defined as $M_{S_1} \ll M_{7/6}$ (light singlet) in which case we expect that the bound on the singlet $T^\prime$ to be approximately valid since all other states are heavier.
In the opposite case, $M_{7/6} \ll M_{S_1} $, the $Y=7/6$ doublet is the lightest but, as we described in the previous section, up to EWSB effects it is degenerate with the singlet in the fifth component of the fundamental of $SO(5)$, $F_5$, and all these three states have an equal mass $m_F$. Mixing effects after EWSB will slightly lift this degeneracy, leaving only the $Q=5/3$ state exactly with the mass $m_F$. 
Since the experimental bound on this state is the strongest, we still expect that it will put the strongest constraint on this region. Even though the precise value of the bound may differ from the one in the simplified model with only one resonance, for our purposes we take that as a reference value. The same argument applies also in the region where $M_{7/6} \sim M_{S_1}$.
Therefore, as a first approximation we adopt the following constraints:
\be
	M_{F_{7/6}} \gtrsim 800 \text{ GeV}~, \qquad
	M_{S_{1}} \gtrsim 700 \text{ GeV}~.
	\label{eq:ExpBoundTopPartners}
\ee
In figure~\ref{fig:topPartners} we present the results of the parameter scans we performed for the two models (the minimal in the upper row, the next-to-minimal in the lower one) showing the points which reproduce the correct top and Higgs masses, as well as the desired value of $\xi$, in the plane $(M_{S_1}, M_{F_{7/6}})$. The blue (orange) points are those which pass (do not pass) the bounds of eqs.~(\ref{eq:ExpBoundVectors},~\ref{eq:ExpBoundTopPartners}) while the green is a reference for the two regions specified before.
We see that the models with lower tuning, $\xi = 0.1$, are already on the verge to be excluded by direct searches and also for $\xi = 0.05$ the bounds cut a sizable part of the parameter space of the models.

\section{Phenomenological analysis -- part II: astrophysics}\label{sec:PhenoAstro}

In this section we analyze all the relevant bounds placed on the parameter space
of our Composite DM model  by the most constraining DM searches currently ongoing in high-energy astrophysics.
In section~\ref{sec:RelicDensity} we discuss the DM relic abundance, while in section~\ref{sec:DirectDetection} we analyze the result of the LUX experiment in the context of direct detection of DM particles. In section~\ref{sec:IndirectDetection} 
we study indirect detection experiments, focusing in particular on the measurement of the antiproton energy spectrum.

\subsection{Relic density}\label{sec:RelicDensity}

In this paper we assume a standard cosmological scenario in which DM is a weakly-interacting cold thermal relic. According to this paradigm, 
in the early Universe DM particles are kept in thermal equilibrium through their interactions with other species populating the thermal bath. In full generality this means that processes converting heavy particles into lighter ones and vice-versa occur at the same rate. As the Universe expands and cools, however,  the conditions to support this delicate equilibrium no longer exist because of two main reasons: 
on the one hand the thermal kinetic energy of lighter particles is no longer sufficient to produce heavier particles, on the other one the expansion of the Universe dilutes the number density of the latter in such a way that their annihilation processes become less and less frequent.
Eventually, heavier particles ``freeze-out'' and their number density, no longer affected by interaction processes, remains constant.
Considering the freeze-out of DM particles, 
the evolution of their number density $n(x)$ during the expansion of the Universe, being $x\equiv m_{\eta}/T$ where $T$ is the temperature,
 is quantitatively described using a Boltzmann equation. In terms of the yield ${\rm Y}(x)=n(x)/s(x)$, where $s(x)$ is the entropy density, this equation reads 
  \begin{equation}\label{eq:BoltzEq}
  \frac{d{\rm Y}}{dx}=-Z(x)\left[{\rm Y}^2(x) - {\rm Y}_{\rm eq}^2(x)\right]~,
  \end{equation}
  where
  \begin{equation}
  Z(x) \equiv \sqrt{\frac{\pi}{45}}\frac{m_{\eta}M_{\rm PL}}{x^2}\sqrt{g_*(T)}
   \langle \sigma v_{\rm rel} \rangle(x)~,
  \end{equation}
$M_{\rm PL}=1.22\times10^{19}$ GeV is the Planck mass and $g_*(T)$ is the number of relativistic degrees of freedom.
  The thermally averaged annihilation cross-section is given by
  \begin{equation}\label{eq:Thermal}
  \langle \sigma v_{\rm rel}\rangle(x) =
  \int_{4m_{\eta}^2}^{\infty}ds\, \frac{s\sqrt{s-4m_{\eta}^2}
  K_1(\sqrt{s}/T)}{16Tm_{\eta}^4K_2^2(m_{\eta}/T)}\,\sigma v_{\rm rel}(s)~,
  \end{equation}
 where $s$ is the center of mass energy squared, $K_{\alpha=1,2}$ are the modified Bessel functions of second kind and $\sigma v_{\rm rel}(s)$ is the total annihilation cross-section times relative velocity of two DM particles.
At the equilibrium
  \begin{equation}\label{eq:EquilibriumYield}
  {\rm Y}_{\rm eq}(x) = \frac{45}{4\pi^4}\frac{x^2}{h_{\rm eff}(T)}K_2(x)~,
  \end{equation}
  where $h_{\rm eff}(T)$ is the effective entropy.\footnote{Solving numerically the Boltzmann equation, we keep the temperature dependence both in $g_*(T)$ and $h_{\rm eff}(T)$ (see ref.~\cite{Huang:2013apa}).} The integration of the Boltzmann equation gives the yield today, ${\rm Y}_0$, which is related to the DM relic density through
  \begin{equation}\label{eq:RDtheory}
  \Omega_{\eta} h^2 = \frac{2.74\times 10^8 m_{\eta} {\rm Y}_0}{{\rm GeV}}~,
  \end{equation}
  where $\Omega_{\eta}\equiv \rho_{\eta}/\rho_{\rm c}$ is the ratio between the energy density of DM and the critical energy density of the Universe and $h\equiv H_{0}/(100~{\rm km}/{\rm s}/{\rm Mpc})$ is the reduced value of the present Hubble parameter.
We solved numerically the Boltzmann equation in eq.~(\ref{eq:BoltzEq}), requiring to reproduce the value observed by the Planck collaboration, $\Omega_{\rm DM}h^2= 0.1199 \pm 0.0027$ ($68\%$ C.L.) \cite{Ade:2013zuv}. 

In our analysis we included the annihilation processes $\eta\eta \to \bar{f}f$,  $\eta\eta \to W^+W^-$,  $\eta\eta \to ZZ$, $\eta\eta \to hh$. 
 The relevant SM fermions entering in the computation are the bottom and the top quark. Moreover, below the kinematical threshold for the annihilation into two on-shell gauge bosons, we also include the three-body processes  $\eta\eta \to WW^*$,  $\eta\eta \to ZZ^*$. Given the great precision reached by the measurement of the relic abundance, in fact, the inclusion of these radiative effects is mandatory in order to obtain an accurate matching \cite{Cline:2012hg}.\footnote{See refs.~\cite{Ciafaloni:2013hya,Boudjema:2014gza} for a more general discussion about the role of radiative corrections for  the computation of the relic abundance.}
Let us now discuss the results of our analysis from a more quantitative point of view.
 \begin{figure*}[!htb!]
\minipage{0.5\textwidth}
  \includegraphics[width=\linewidth]{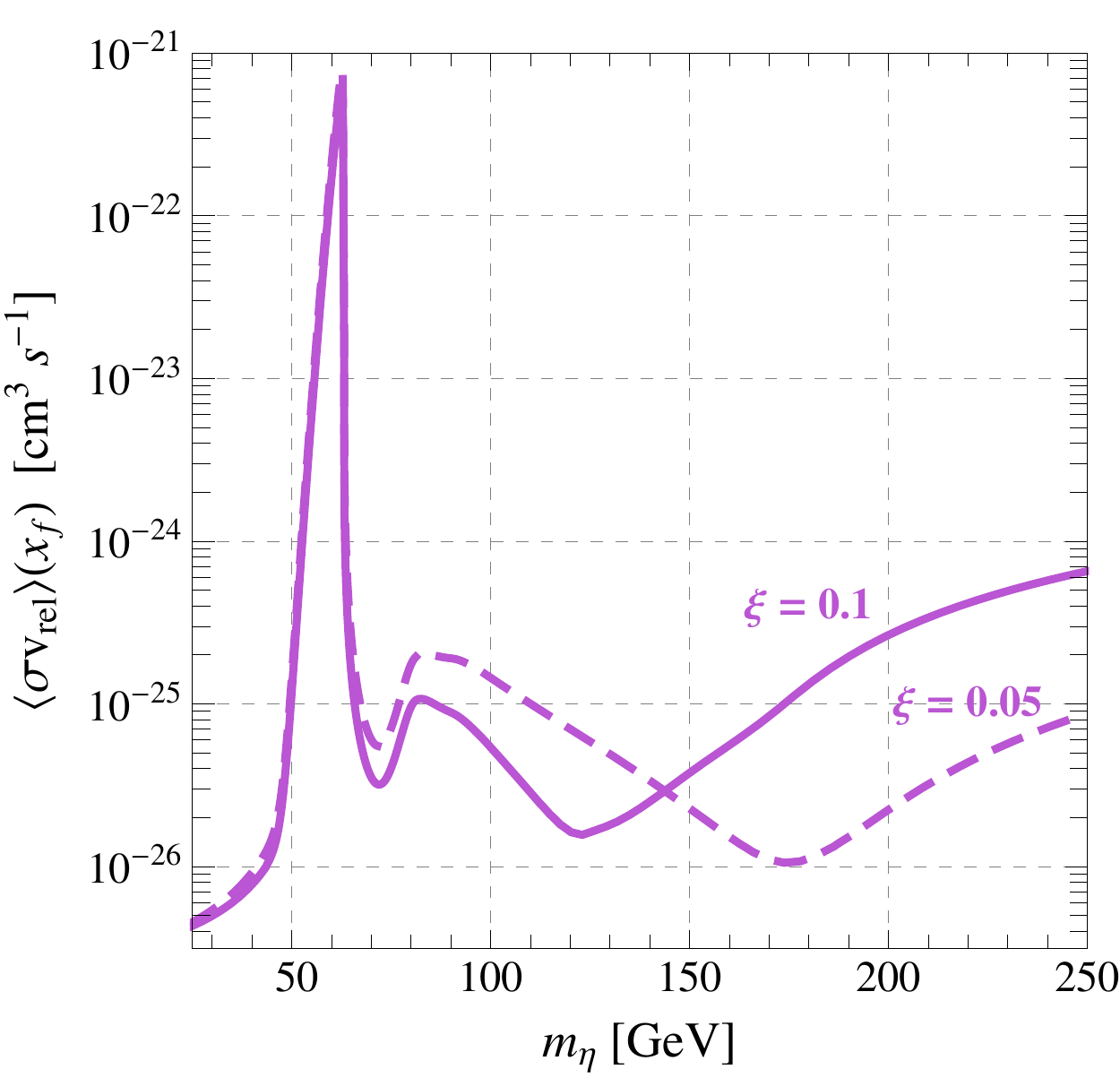}
\endminipage
\minipage{0.5\textwidth}
\includegraphics[width=\linewidth]{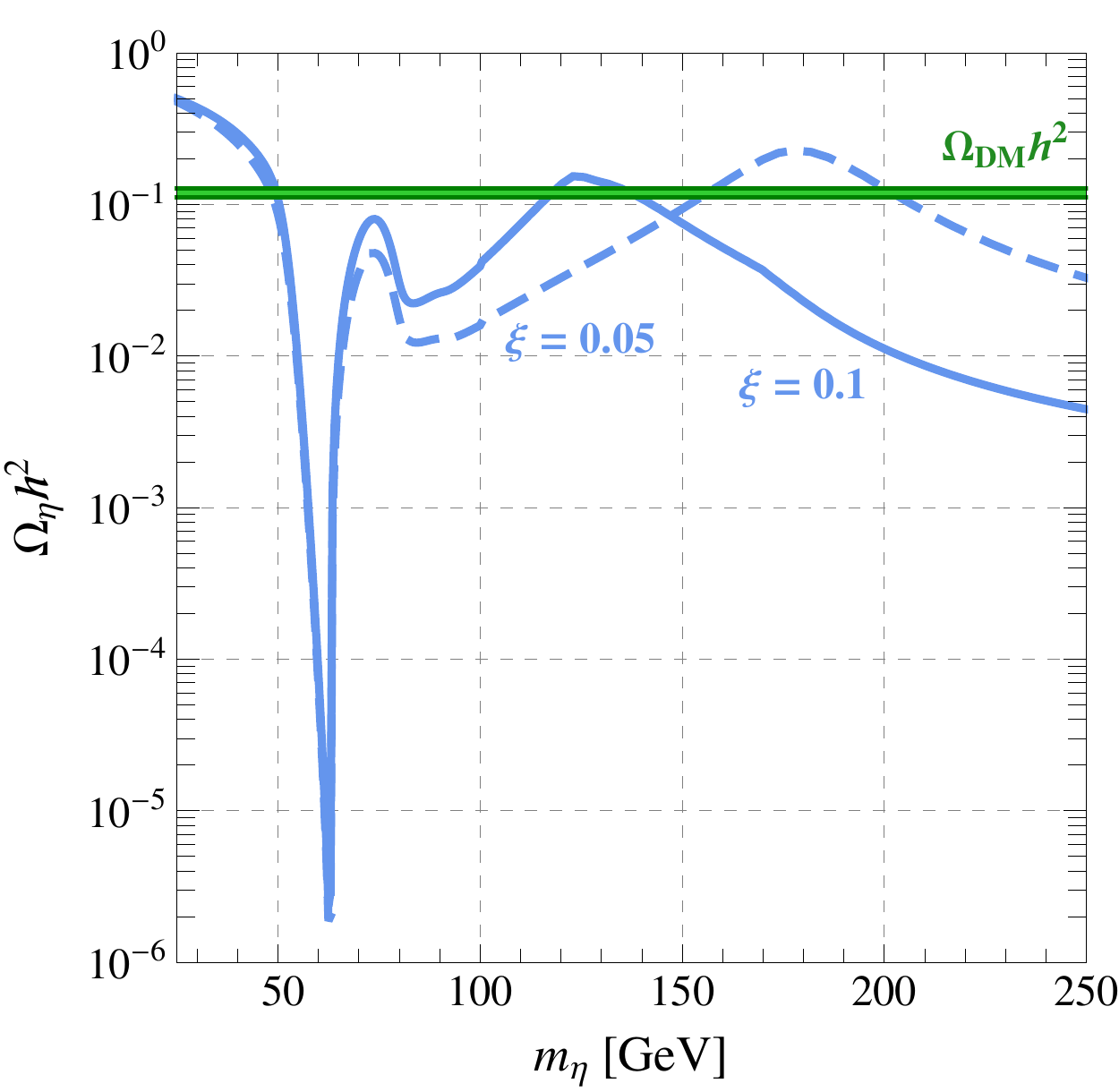}
\endminipage
 \caption{\textit{
Left panel: thermally averaged annihilation cross-section $\langle \sigma v_{\rm rel}\rangle(x)$
evaluated at the typical freeze-out temperature for a weakly-interacting DM particle, namely $T_f =m_{\eta}/x_f$ with $x_f=20$. Right panel: DM relic density $\Omega_{\eta} h^2$ in eq.~(\ref{eq:RDtheory}) compared with the 3$\sigma$ interval measured by the Planck collaboration (green band).
We show two different values $\xi =0.1$ (solid line) and $\xi =0.05$ (dashed line) while we fix $\lambda = 0.065$ as suggested by eqs.~(\ref{eq:RelationsMinimalModel},~\ref{eq:Lambda2}).
 }}\label{fig:RelicDensityFixedLambda}
\end{figure*}

In the left panel of figure~\ref{fig:RelicDensityFixedLambda} we plot, as a function of the DM mass $m_{\eta}$, 
the thermally averaged annihilation cross-section at the freeze-out epoch, i.e. assuming $x_f = 20$, for the benchmark values $\xi=0.1$ and $\xi=0.05$. We take $\lambda = 0.065$, as suggested by eqs.~(\ref{eq:RelationsMinimalModel},~\ref{eq:Lambda2}). Going from small to large values for the DM mass $m_{\eta}$ it is possible to recognize the Higgs resonance ($m_{\eta}\approx 63$ GeV), the two-body threshold for annihilation into two on-shell W bosons ($m_{\eta}\approx 80$ GeV) and the effect of the momentum-dependent  interactions of the chiral Lagrangian in eq.~(\ref{eq:ChiralLagrExpanded}). The latter, growing proportionally to the square of the total energy in the c.o.m., become important for large values of the DM mass enhancing the annihilation cross-section. 
Finally, notice that the dip around $130$ GeV for $\xi = 0.1$ ($180$ GeV for  $\xi = 0.05$) corresponds to the value of $m_{\eta}$ that solves the equation 
$s-2\lambda\xi(1-\xi)/v^2=0$, with $s=4m_{\eta}^2/(1-v_{\rm rel}^2/4)$ and $v_{\rm rel}\approx 1/2$ at the freeze-out. This condition corresponds to an accidental cancellation between the derivative and the $\lambda$ contribution to the $\eta$-$\eta$-$h$
vertex (see appendix~\ref{App:parametrizations} and ref.~\cite{Frigerio:2012uc}).
 
In the right panel of figure~\ref{fig:RelicDensityFixedLambda} we plot, as a function of the DM mass $m_{\eta}$, the value of the relic density in eq.~(\ref{eq:RDtheory})
compared with the 3$\sigma$ interval measured by the Planck collaboration. 
As before, we take 
$\xi=0.1$ and $\xi=0.05$, with $\lambda = 0.065$. At the qualitative level the result can be understood bearing in mind that
a na\"{\i}ve but useful approximated solution of the Boltzmann equation is given by
\begin{equation}
\frac{\Omega_{\eta} h^2}{0.1199}\simeq
\frac{3\times 10^{-26}~{\rm cm^3s^{-1}}}{\langle \sigma v_{\rm rel}\rangle(x_f)}~.
\end{equation}
As a consequence the relic abundance retraces, upside down, the same contour   of the thermally 
averaged annihilation cross-section.

In section~\ref{sec:results} we will present our numerical results for the computation of the relic density 
from a more general viewpoint as contour plot  in the plane $(m_{\eta},\lambda)$. In this way we will be able to 
compare the region of the parameter space in which the model can reproduce the observed value of the relic
abundance with the other constraints analyzed in the rest of this paper. 

\subsection{Direct detection}\label{sec:DirectDetection}

Direct detection of DM can occur through elastic scattering  between an incident DM particle and a nucleus at rest inside a detector beneath the surface of the Earth. Direct detection experiments aim to measure, as fingerprints of these interactions,
the nuclear recoil energy.
The LUX experiment \cite{LUXwebsite} has recently reported the most stringent limit on the spin-independent DM-nucleon 
elastic cross-section $\sigma_{\rm SI}$ \cite{Akerib:2013tjd}.

In our model the spin-independent DM-nucleon 
elastic cross-section is generated by two types of diagrams. On the one hand, 
the $\eta$-$\eta$-$h$ vertex in the chiral Lagrangian in eq.~(\ref{eq:ChiralLagrExpanded}) generates a tree-level contribution via the exchange in the t-channel of the Higgs boson which, in turn, couples to quarks and gluons inside the nucleon.
On the other one, the Yukawa Lagrangian in eq.~(\ref{eq:YukawaLagr}) generates an effective  operator proportional to $(m_q/f^2)\eta^2\bar{q}q$, thus leading to a contact interaction between DM and quarks. Note that in both cases we have a scalar-mediated interaction with quarks, i.e. the interactions involving quarks are always proportional to the scalar operator $m_q\bar{q}q$.  In full generality, the spin-independent DM-nucleon 
elastic cross-section mediated by scalar interactions can
be parametrized as follows 
\begin{equation}\label{eq:SI}
\sigma_{\rm SI} = \frac{1}{\pi}
\left(\frac{m_{\rm N}}{m_{\eta} + m_{\rm N}}\right)^2
\frac{[Zf_{p} + (A - Z)f_n]^2}{A^2}~,
\end{equation}
where $m_{\rm N}=(m_n + m_p)/2=938.95$ MeV is the nucleon mass while $Z$ and $A-Z$ are the number of protons and neutrons inside the nucleus, with $Z=54$ and $A=130$ for a nucleus of Xenon.
In eq.~(\ref{eq:SI}) $f_p$  and $f_n$ describe the coupling between DM and, respectively, protons and neutrons. They are given by
\begin{equation}
f_{n,p} = 
\sum_{q = u,d,s}f_{T_q}^{(n,p)}a_q m_{n,p} + \frac{2}{27}\,f_{T_G}
\sum_{q=c,b,t}a_q m_{n,p}~,
\end{equation}
where for the nuclear matrix elements we take \cite{Berlin:2014tja,Junnarkar:2013ac} $f_{T_u}^{(n)}=0.026$, $f_{T_d}^{(n)}=0.020$, $f_{T_u}^{(p)}=0.020$, $f_{T_d}^{(p)}=0.026$,  $f_{T_s}^{(n,p)}=0.043$, and 
$f_{T_G} = 1-f_{T_u}^{(n,p)}-f_{T_d}^{(n,p)}-f_{T_s}^{(n,p)}=0.911$. The coefficients $a_q$
describe the effective interactions between DM and quarks, normalized as $\mathcal{L}^{\rm DD}_{\eta} \supset \sum_q a_q m_q \eta^2 \bar{q}q$.
In order to write down explicitly these coefficients in our model, 
we need to specify the contact interactions between DM and the first two generations of quarks.
 \begin{figure*}[!t]
\minipage{0.5\textwidth}
  \includegraphics[width=\linewidth]{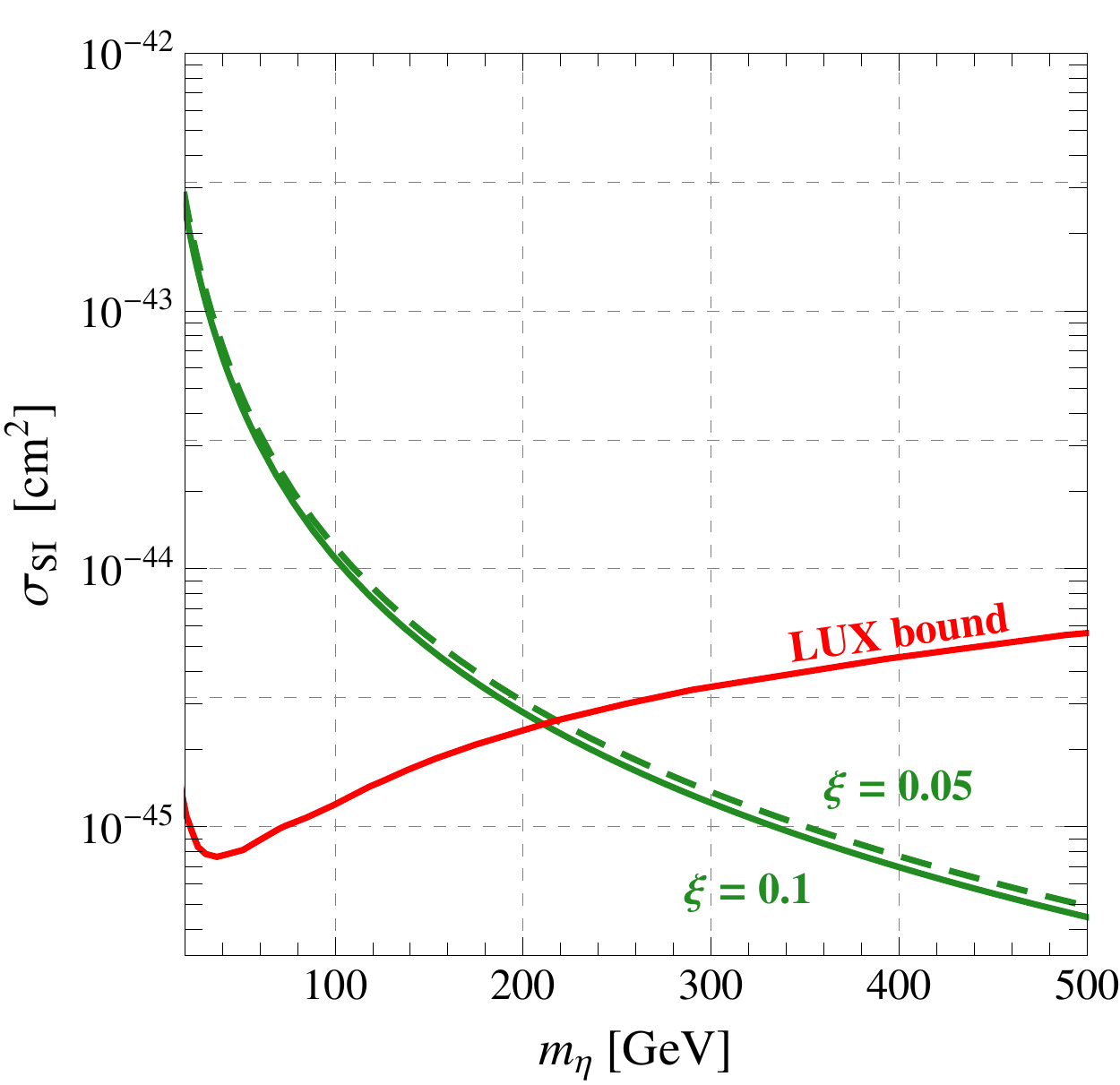}
\endminipage
\minipage{0.5\textwidth}
  \includegraphics[width=\linewidth]{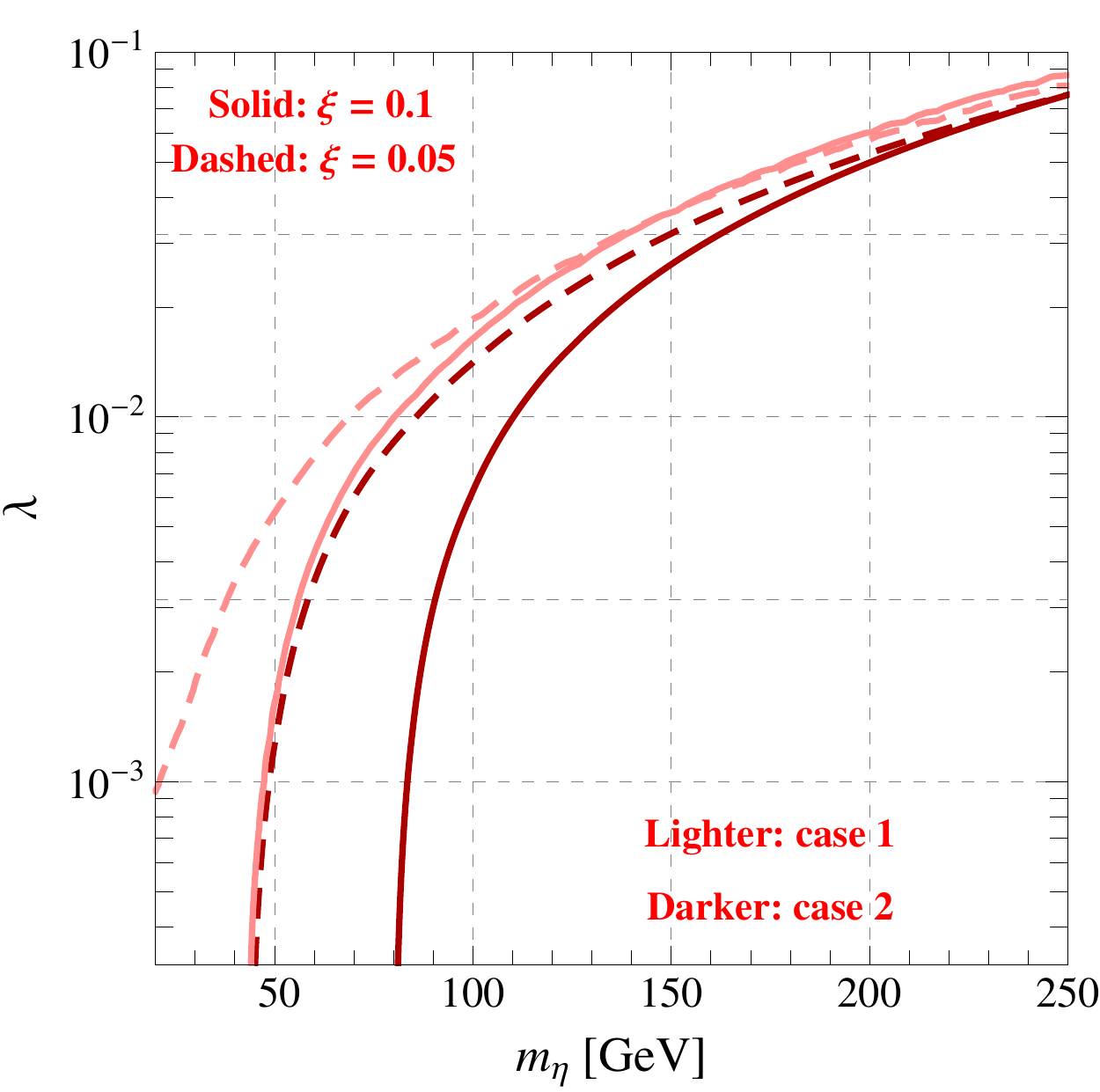}
\endminipage
 \caption{\textit{
 Left panel: comparison between the spin-independent elastic cross-section  $\sigma_{\rm SI}$ in 
 eq.~(\ref{eq:SI})
 and the bound extracted by the LUX experiment (the region above the red line is excluded). We plot the value of
  $\sigma_{\rm SI}$ corresponding to $\lambda = 0.065$  limited to case 1 in eq.~(\ref{eq:DDCase1}), with $\xi = 0.1$ (green solid line) and $\xi = 0.05$ (green dashed line). Right panel: region of the parameter space $(m_{\eta}, \lambda)$ excluded by the LUX experiment. We show the corresponding bound for 
  $\xi = 0.1$ (red solid line) and $\xi = 0.05$ (red dashed line), considering both case 1 in 
  eq.~(\ref{eq:DDCase1}) (lighter red) and case 2 in 
  eq.~(\ref{eq:DDCase2}) (darker red).
 }}\label{fig:DD}
\end{figure*}
Since the computation of the spin-independent elastic cross-section
is the only place in which these interactions play an important phenomenological role,
we decided to distinguish between two cases
\begin{eqnarray}
{\rm Case~1:}~~~~~~&&a_{q=u,d,c,s} = \frac{\lambda(1-2\xi)}{m_h^2}~,~~~
a_{q=t,b} = \frac{\lambda(1-2\xi)}{m_h^2} + \frac{\xi}{2(1-\xi)v^2}~,\label{eq:DDCase1}\\
{\rm Case~2:}~~~~~~&&a_{q=u,d,c,s,t,b} =  \frac{\lambda(1-2\xi)}{m_h^2} + \frac{\xi}{2(1-\xi)v^2}~.\label{eq:DDCase2}
\end{eqnarray}
In the first case -- eq.~(\ref{eq:DDCase1}) -- we set to zero the contact interaction between $\eta$ and all the quarks belonging to the first two generations.
This setup can be easily realized, for instance, considering the embedding of the right handed 
quarks of the first two generations into the $\textbf{15}$ of $SO(6)$.
 The only non-zero contribution to $a_{q=u,d,c,s}$, as a consequence, 
arises from the t-channel exchange of the Higgs boson. This contribution has been computed 
neglecting the square of the momentum transferred, $q^2$, both in the t-channel 
Higgs propagator and in the derivative interaction arising from the chiral Lagrangian in eq.~(\ref{eq:ChiralLagrExpanded}). This approximation is justified by the fact that
in the elastic scattering  we have $-q^2/m_h^2$, $-q^2/f^2\ll 1$, with
 $q^2 = -2m_{\rm Xe}E_{\rm re}$ where the mass of a nucleus of Xenon is $m_{\rm Xe}= 121$ GeV while for the typical kinetic recoil energy one has $E_{\rm re}\sim$ few keV. 
 The coefficients $a_{q=t,b}$ receive, in addition to the term generated by the t-channel exchange of the Higgs, an extra contact interaction from the Yukawa Lagrangian in eq.~(\ref{eq:YukawaLagr}); according to the discussion in section~\ref{sec:fermionembedding},
 this contribution has been computed assuming the embedding of the bottom and top quark into the fundamental representation $\textbf{6}$ of $SO(6)$.
 In the second case -- eq.~(\ref{eq:DDCase2}) -- we assumed non-zero
  contact interactions also for the quarks belonging to the first two generations, adopting the same embedding into the $\textbf{6}$ of $SO(6)$ characterizing the top-bottom sector.
We show our results in figure~\ref{fig:DD}.  In the left panel we compare the spin-independent elastic cross-section computed in our model with the bound set by the LUX experiment. Following our choice of benchmark values, we plot $\sigma_{\rm SI}$ for $\lambda = 0.065$ and for $\xi = 0.1$, $\xi = 0.05$. Moreover, for definiteness, 
we show only the setup corresponding to eq.~(\ref{eq:DDCase1}).
The bound of LUX turns out to be very stringent, and only values of DM mass larger than $200$ GeV are allowed. 
The two lines for $\xi = 0.1$ and $\xi = 0.05$ are almost indistinguishable. The difference between these two values, in fact, starts to be significant when $\lambda(1-2\xi)/m_h^2 < \xi/2(1-\xi)v^2$, i.e. for $\lambda \lesssim 10^{-2}$. In the right panel of  figure~\ref{fig:DD} we illustrate 
the difference between case 1 and case 2 in eqs.~(\ref{eq:DDCase1},~\ref{eq:DDCase2})
showing the bound of the LUX experiment in the parameter space $(m_{\eta}, \lambda)$,
both for $\xi = 0.1$ and $\xi = 0.05$. For small values of $\lambda$, i.e. $\lambda \lesssim 10^{-2}$, the role of the additional contact interactions  in case 2 starts to be significant, 
 pushing the excluded region towards larger values of DM mass if compared with those allowed in case 1. For $m_{\eta}\gtrsim 150$ GeV, where the LUX bound can exclude only large values of $\lambda\gtrsim 10^{-2}$ in order to compensate the $m_{\eta}^{-2}$ suppression in $\sigma_{\rm SI}$, the difference between case 1 and case 2 is less relevant. 
 
 In section~\ref{sec:results} we will use the result in the right panel of figure~\ref{fig:DD} 
 in order to combine the bound of LUX with all the other phenomenological constraints under investigation in our analysis.

\subsection{Indirect detection}\label{sec:IndirectDetection}

DM annihilation into lighter SM particles in the halo of the Milky Way galaxy copiously produces stable particles -- e.g. photons, 
positrons, antiprotons and neutrinos  --
 giving rise, in principle, to detectable signals on Earth \cite{Bertone:2010zza,Cirelli:2010xx}. 
The major task that has to be addressed in order to detect such signal is to understand, for each of the stable species mentioned above, the contribution of the astrophysical background, mostly originated from the interactions of ultra high-energy cosmic rays of extragalactic origin with the interstellar medium in the Galaxy. 
In this context, the measurement of the antiproton flux plays a central role for three main reasons: \textit{i}) among all stable particles that may be produced from DM annihilation, the ratio between the DM signal and the astrophysical background is largest in the antiproton channel, \textit{ii}) the theoretical prediction for the astrophysical background  -- i.e. secondary
production of antiprotons from primary cosmic rays protons interacting
with gas and dust in the Galaxy -- is moderately under control, relying on a strict analogy with 
the analysis carried for heavier nuclei, like the measurement of the boron-to-carbon ratio \cite{Donato:2001ms,DiBernardo:2009ku},
\textit{iii}) simple arguments, based on kinematics, show that background and signal should have completely different spectral features -- i.e. a spectrum suppressed at small energies and peaked around few GeV for the background versus a broader spectrum for the DM signal \cite{Profumo:2013yn}.
The balloon-borne experiment BESS \cite{Matsunaga:1998he,Maeno:2000qx,Asaoka:2001fv} and the 
space-based experiment PAMELA \cite{Adriani:2010rc} have measured with good precision the antiproton energy spectrum
in the energy range from $0.1$ GeV up to about 
$180$ GeV. A further improvement is expected when the antiproton data collected by the  AMS-02 experiment will be released \cite{AMSwebsite}. The measured rate agrees well with standard
background estimate; this result, as a consequence, can be used to set limits on the yield of
antiprotons from exotic sources like
DM annihilation.

In our analysis we closely followed the approach outlined in ref.~\cite{Evoli:2011id} and further re-examined
in ref.~\cite{MeToAppear} in the context of scalar Higgs portal models (see also 
refs.~\cite{Fornengo:2013xda,Cirelli:2013hv} for related analysis). In a nutshell this approach 
is based on a careful scrutiny of the uncertainties associated with the astrophysical background.
Five different models for the propagation of charged cosmic rays in the Galaxy 
have been constructed 
by using different assumptions -- i.e. different rigidities for the diffusion 
coefficient, different thickness for the Galactic halo and the possibility to have strong convection -- and requiring 
to fit the recently updated boron-to-carbon and proton data \cite{Adriani:2011cu}. 
Once one of these propagation models is chosen, it can be used to compute the antiproton flux,
 testing the background plus DM hypothesis  versus the background prediction. Strong bounds
  on the 
 DM thermally averaged annihilation cross-section times relative velocity can be extracted using this strategy. 
 Let us now describe in more detail our approach.
First we computed the antiproton energy spectrum produced by DM annihilation -- i.e. the number of 
 antiprotons per each annihilation process -- according to
 \begin{equation}\label{eq:Spectrum}
\left. \frac{dN}{dE}\right|_{\overline{p}} =
\sum_{f}{\rm BR}_f \times \left. \frac{dN}{dE}\right|^f_{\overline{p}}~,
 \end{equation}
 where the sum runs over all the possible final states $\eta\eta \to f$  
 that are kinematically allowed for a given value of DM mass $m_{\eta}$. In addition to two-body final states,
 we included the three-body annihilation processes $\eta\eta \to WW^*$, $ZZ^*$ below the kinematical threshold 
 for the annihilation into two on-shell gauge bosons. In eq.~(\ref{eq:Spectrum})  $\left.dN/dE\right|^f_{\overline{p}}$
 is the number of antiprotons per each annihilation into the finale state $\eta\eta \to f$ whose branching ratio is given by 
  ${\rm BR}_f$. We obtained these energy spectra using the Monte Carlo event generator PYTHIA 8.1 \cite{PYTHIAwebsite} including the 
  effects of three-body final states as described in refs.~\cite{Ciafaloni:2010ti,Ciafaloni:2011sa}. 
  The number of antiproton per unit energy, time and volume produced by DM annihilation is therefore given by the following source term
  \begin{equation}
  Q_{\bar{p}} = \frac{1}{2}\left[\frac{\rho_{\rm DM}(r)}{m_{\eta}}\right]^2
  \langle \sigma v_{\rm rel}\rangle_{0} 
  \left. \frac{dN}{dE}\right|_{\overline{p}}~,
  \end{equation}
  where $\langle \sigma v_{\rm rel}\rangle_{0}$ is the thermally averaged annihilation cross-section times relative velocity describing DM annihilation today. Concerning the DM halo profile $\rho_{\rm DM}(r)$ we adopted three different possibilities, namely the Einasto \cite{Navarro:2008kc,Graham:2005xx},  NFW \cite{Navarro:1995iw} and Isothermal \cite{Burkert:1995yz} profiles.
  Using the public code DRAGON \cite{DRAGONwebsite,Evoli:2008dv}, we then propagated the antiprotons produced by DM annihilation considering for definiteness  two  different propagation models among those described in
  refs.~\cite{Evoli:2011id}, i.e. the KOL and CON propagation models.
  The former -- more constraining -- assumes Kolmogorov turbulence, while the latter -- less constraining -- includes convective effects (see ref.~\cite{Evoli:2011id} for a more detailed discussion). The comparison between these two different choices should give an idea of the uncertainties affecting the propagation of charged particles in the Galaxy.\footnote{It is worth noticing that models based on a thin diffusion zone (i.e. the THN model in ref.~\cite{Evoli:2011id}) give bounds that in general are less constraining if compared with those obtained using the CON model. These models, however, are disfavored by recent studies on synchrotron emission, radio maps and low energy positron
spectrum \cite{DiBernardo:2012zu}. For this reason we do not consider in our analysis this possibility.}  Finally, comparing the DM antiproton signal with the background generated using the same propagation models, we were able to extract 
   exclusion curves for  $\langle \sigma v_{\rm rel}\rangle_{0}$. In particular, we required that the total (background $+$ signal) antiproton flux does not exceed the measured flux \cite{Adriani:2010rc} at any energy by more than 3$\sigma$.\footnote{In addition to the measurement of the absolute antiproton flux, the PAMELA collaboration has reported in ref.~\cite{Adriani:2008zq} the measurement of the antiproton-to-proton flux ratio. However, we do not use these data in our analysis. The reason
   is that ref.~\cite{Evoli:2011id} already used proton data in the definition of the propagation models. 
   If we use the antiproton-to-proton ratio in order to extract our bound, then we will inconsistently use the same proton data twice: one for the definition of the propagation model (thus without the inclusion of any exotic component in addition to the background contribution), the other one for the fit of the DM signal (thus including an exotic component in addition to the background contribution).}
   \begin{figure*}[!t]
\minipage{0.5\textwidth}
  \includegraphics[width=\linewidth]{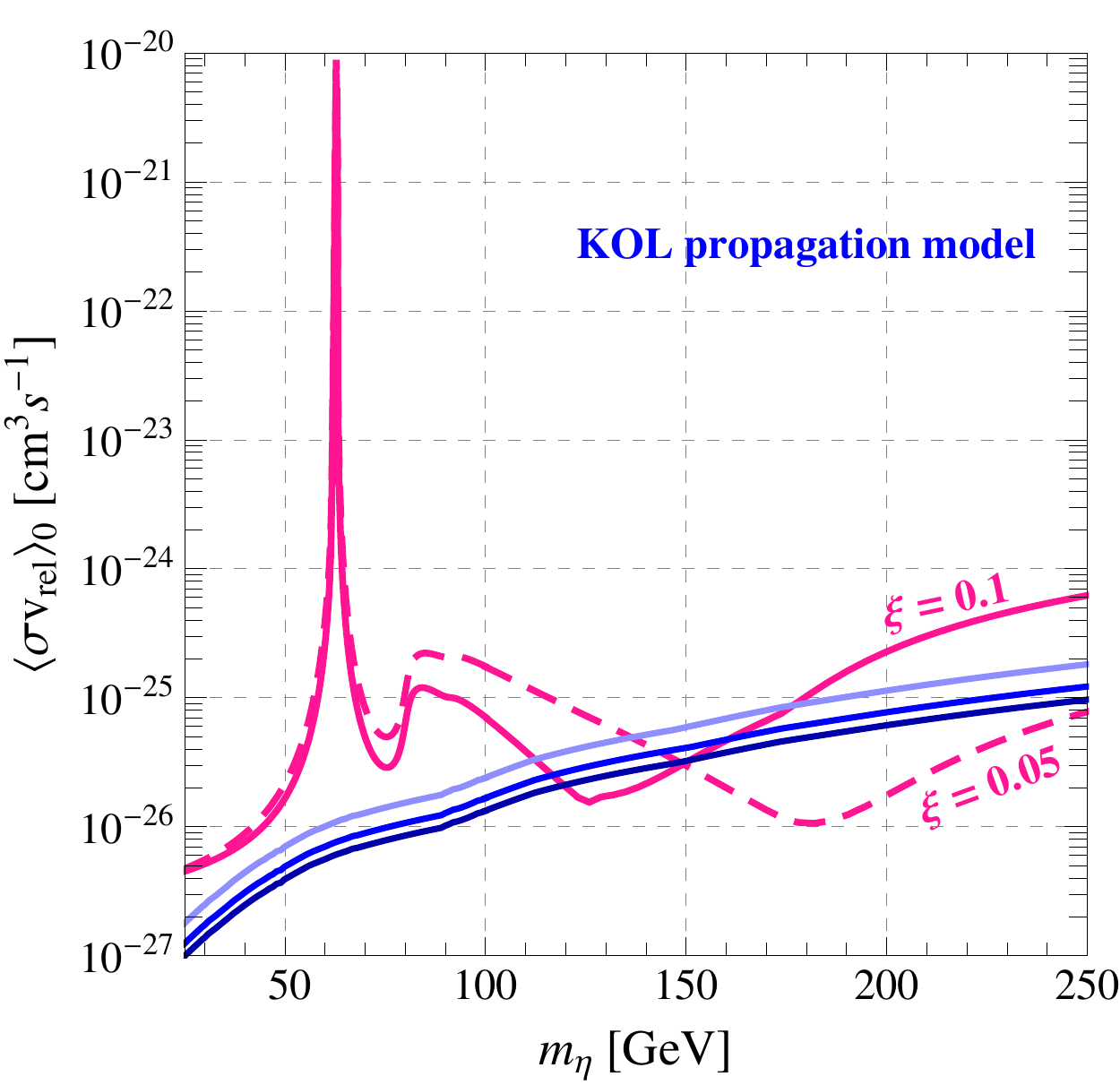}
\endminipage
\minipage{0.5\textwidth}
  \includegraphics[width=\linewidth]{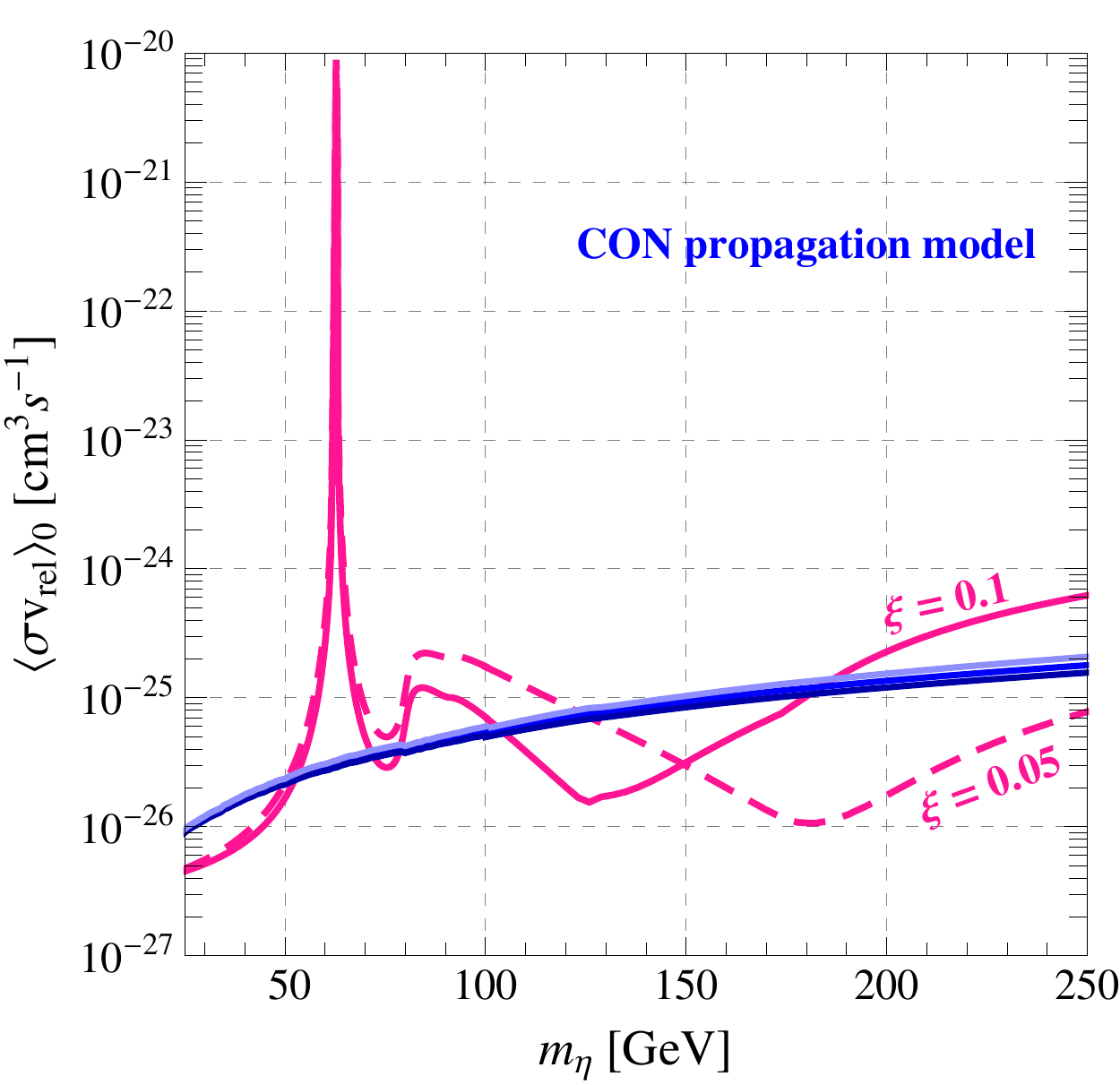}
\endminipage
 \caption{\textit{
 Bounds on the thermally averaged annihilation cross-section times relative velocity $\langle \sigma v_{\rm rel}\rangle_{0}$
 obtained using the antiproton flux measured by the PAMELA experiment. The region above the blue lines is excluded at 3$\sigma$ level.
 We show the bounds obtained using two different  models for the propagation of charged cosmic rays in the Galaxy, namely the KOL (left panel) and CON (right panel) propagation models \cite{Evoli:2011id,MeToAppear}. In both cases we plot three lines corresponding to different DM density profiles, namely -- from bottom to top -- 
 Einasto (darker blue), NFW (blue), Isothermal (lighter blue).
  We also show the value of $\langle \sigma v_{\rm rel}\rangle_{0}$ for $\xi = 0.1$ (pink solid line) and $\xi = 0.05$ (pink dashed line), with $\lambda=0.065$.
 }}\label{fig:IndirectBound}
\end{figure*}
In figure~\ref{fig:IndirectBound} we show the bounds on $\langle \sigma v_{\rm rel}\rangle_{0}$ obtained using this procedure, considering both the KOL (left panel) and CON (right panel) propagation models. For comparison,
 we also plot the value of $\langle \sigma v_{\rm rel}\rangle_{0}$ using the two benchmark values $\xi=0.1$ and $\xi = 0.05$, with $\lambda = 0.065$. In both cases it is clear that the antiproton bound provides a stringent constraint on the annihilation cross-section. 
 Moreover, we repeat our analysis using the three different DM density profiles mentioned above. As expected, we find that the DM antiproton flux is larger for profile models in which the DM density is enhanced towards the Galactic center  while is smaller for 
 density distribution described by an isothermal sphere; as a consequence the bound in figure~\ref{fig:IndirectBound} is more (less) stringent for the Einasto (Isothermal) profile. Finally, notice that the difference between different DM  density profiles is less evident considering the CON propagation model; as already noticed in 
 ref.~\cite{Evoli:2011id}, in the convective model the antiproton flux from DM annihilations is dominated by local 
 contribution (i.e. from regions close to the Earth) where the three profiles are almost equivalent. For the KOL model the contribution from regions close
to the Galactic center is more important, and therefore the three profiles -- more or less peaked in this region -- give different bounds. 
 
In section~\ref{sec:results} we will present the antiproton bound as contour plot  in the plane $(m_{\eta},\lambda)$ considering both the KOL and CON propagation models. For definiteness, we will focus only on the NFW profile.

%
\section{Results}
\label{sec:results}

\begin{figure}[!htb!]
\vspace{-0.8cm}
\begin{center}
\fbox{\footnotesize $\xi = 0.1$} \\
\hspace*{-0.65cm} 
\begin{minipage}{0.5\linewidth}
\begin{center}
	\includegraphics[width=\linewidth]{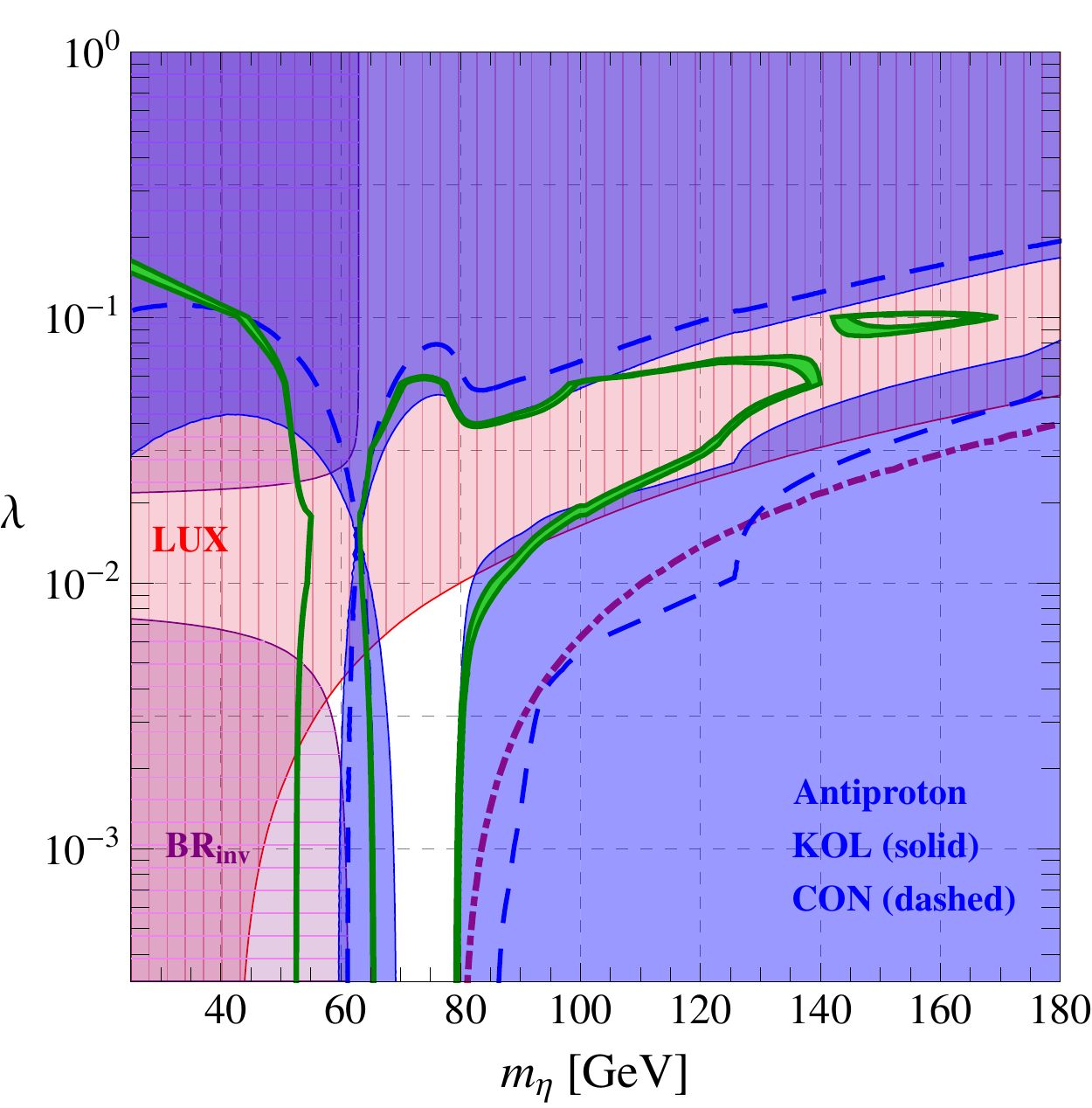}
\end{center}
\end{minipage}
\begin{minipage}{0.5\linewidth}
\begin{center}
	\includegraphics[width=\linewidth]{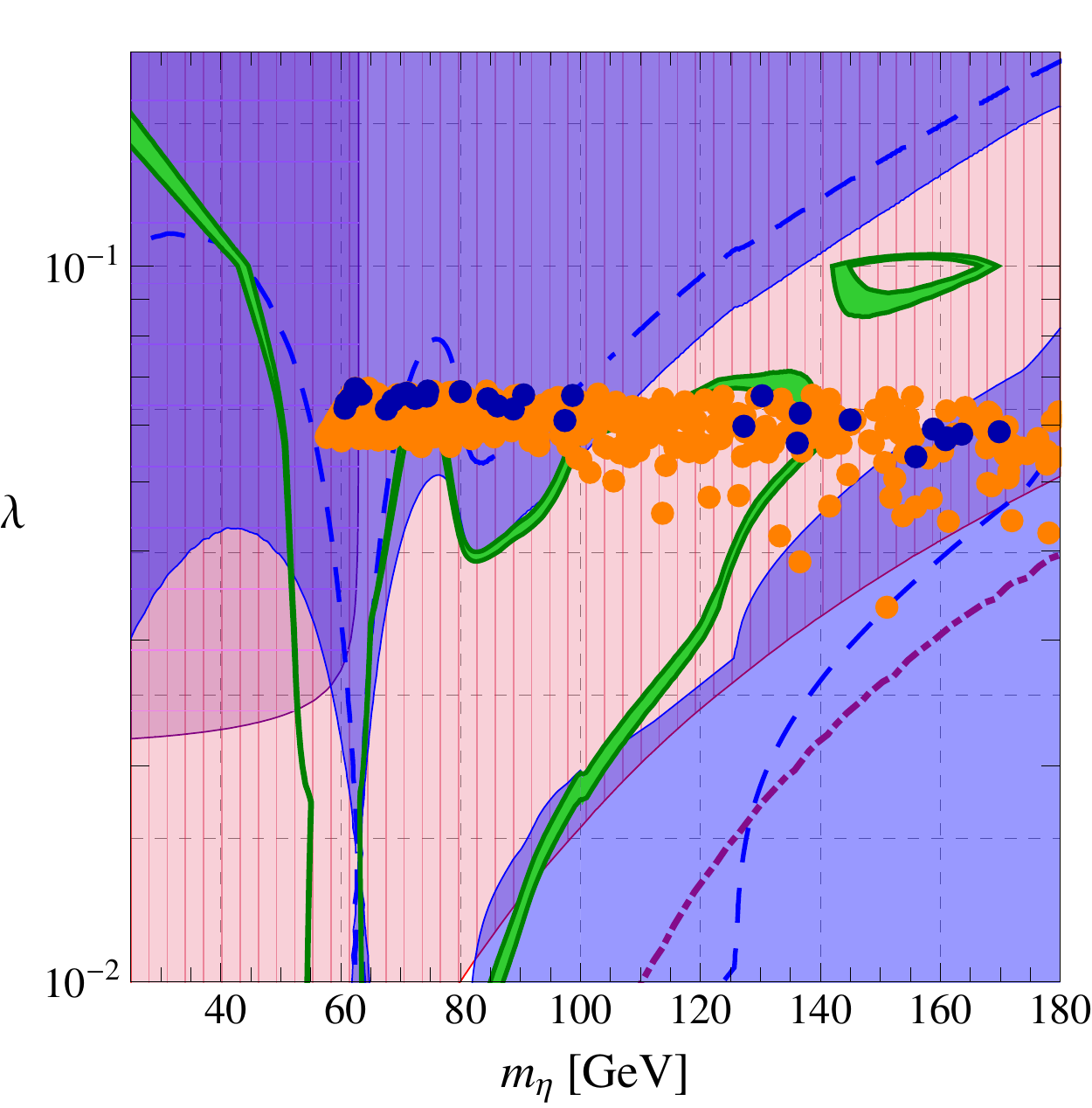}
\end{center}
\end{minipage}
\\[0.1cm]
\fbox{\footnotesize $\xi = 0.05$} \\
\hspace*{-0.65cm} 
\begin{minipage}{0.5\linewidth}
\begin{center}
	\includegraphics[width=\linewidth]{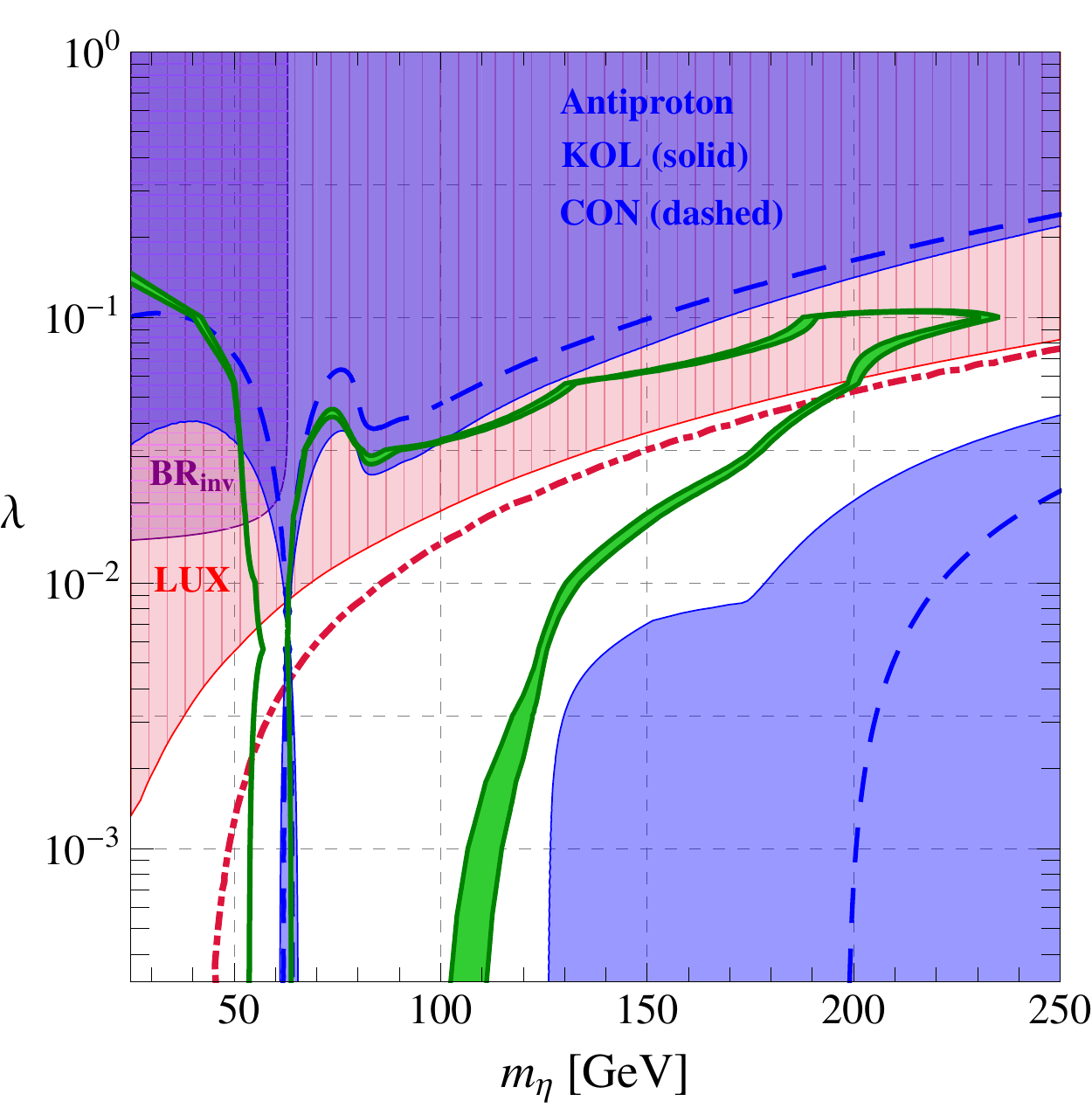}
\end{center}
\end{minipage}
\begin{minipage}{0.5\linewidth}
\begin{center}
	\includegraphics[width=\linewidth]{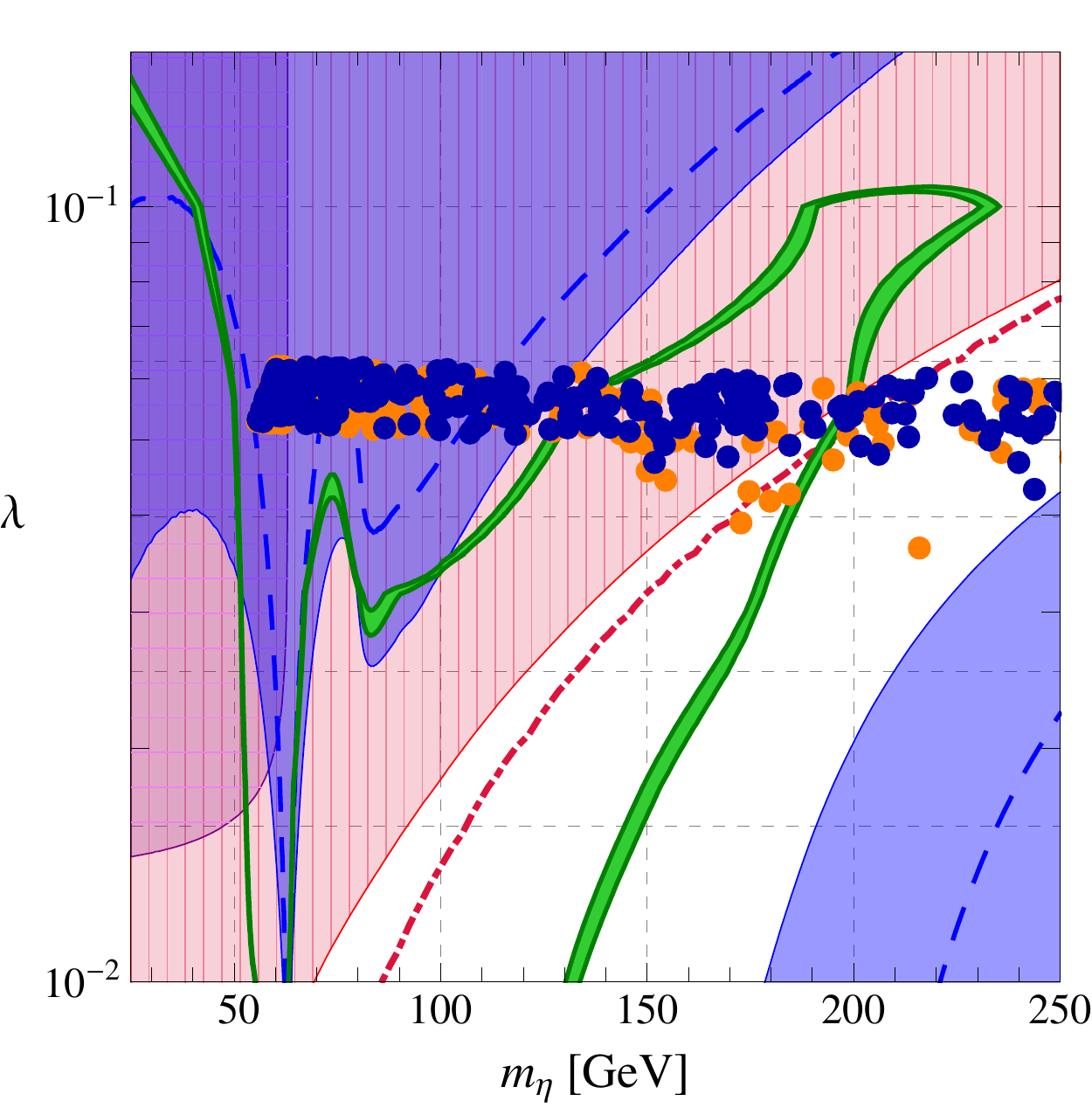}
\end{center}
\end{minipage}
\end{center}
\vspace*{-0.2cm}
\caption{\label{fig:CombinedPlot} 
\textit{
Green line: $3\sigma$ contour reproducing the correct DM relic abundance. Red region (vertical meshes): region excluded by the LUX experiment at 95\% C.L. assuming case 1 in eq.~(\ref{eq:DDCase1}) while the red dot-dashed line represents the bound assuming case 2 in eq.~(\ref{eq:DDCase2}). Purple region (horizontal meshes): region excluded by the LHC at $3\sigma$ considering the bound on the invisible Higgs branching ratio. Blue region (no meshes): region excluded at $3\sigma$ by the PAMELA measurement of the antiproton flux (solid line: KOL propagation model; dashed line: CON propagation models). In the upper (lower) plot we use $\xi = 0.1~(0.05)$.   
In the right panel we zoom on a specific window of values for $\lambda$, and we superimpose the result of the scan performed in section~\ref{sec:TopPartners}. All the points reproduce the correct top and Higgs masses; the orange points are excluded by direct searches of top partners and vector resonances, while the blue points pass the selection.}}
\end{figure}

Here we combine all the constraints obtained in our phenomenological analysis for the Composite DM model studied in this paper. 
We present our results in figure~\ref{fig:CombinedPlot} in the plane $(m_{\eta},\lambda)$. The green strip reproduces the correct amount of relic abundance as measured by the Planck collaboration \cite{Ade:2013zuv} (section~\ref{sec:RelicDensity}).
In the same plot we also show the bounds placed by the LUX experiment \cite{Akerib:2013tjd} in the context of direct detection of DM (section~\ref{sec:DirectDetection}), the PAMELA experiment \cite{Adriani:2010rc} in the context of indirect detection of DM (section~\ref{sec:IndirectDetection}) and the LHC experiment \cite{Aad:2014iia,CMS_invisible} considering the invisible decay width of the Higgs (section~\ref{sec:InvisibleWidth}).
On top of this, we superimpose the results of the scans performed in section~\ref{Sec:Potential} analyzing the effective potential, dividing the points among those which pass or not the bounds from direct searches of top partners and vector resonances at the LHC described in section~\ref{sec:TopPartners}.
We consider the two benchmark values $\xi = 0.1$ and $\xi = 0.05$.

Let us now describe in detail the features present in figure~\ref{fig:CombinedPlot}.
The region of the parameter space reproducing at 3$\sigma$ the correct value of the relic density is covered by the green strip. Considering DM annihilation, the interactions between $\eta$ and the Higgs boson described by the chiral Lagrangian in eq.~\ref{eq:ChiralLagrExpanded} grow with the DM mass and decrease with the scale $f$. For $\xi = 0.1~(0.05)$ and $m_{\eta}\gtrsim 180~(250)$ GeV  these annihilations become too efficient, thus leading to a value of relic density that is too small to match the observed one.\footnote{It is worth noting that this is a distinctive feature of the composite model. In the singlet scalar extension of the SM, in which the derivative interactions are absent, it is always possible to increase the value of $\lambda$ in order to reproduce the correct relic density  for large DM masses.} 
The funnel-shaped region that stretches towards this limit value $m_{\eta}\approx 180~(250)$ GeV corresponds to the condition $s- 2\lambda\xi(1-\xi)/v^2=0$ with $s=4m_{\eta}^2/(1-v_{\rm rel}^2/4)$ and $v_{\rm rel}\approx 1/2$, where an accidental cancellation between the derivative and the $\lambda$ contribution to the $\eta$-$\eta$-$h$ vertex partially counterbalances the growth of the cross-section discussed before.
On the basis of this observation, and in order to keep our discussion as clear as possible, let us divide the plane $(m_{\eta}, \lambda)$ in three parts: the low-mass region $m_{\eta}\lesssim m_h/2$, the resonant region $m_{\eta}\approx m_h/2$ and the funnel-shaped region defined above.

      For $\xi=0.1$, the region $m_{\eta}\lesssim m_h/2$ is ruled out by a combination of LHC and LUX bounds. On the one hand, as soon as the invisible decay channel $h\to \eta\eta$ is kinematically allowed, $\Gamma_{\rm inv}(h\to \eta\eta)$ easily dominates over the SM contribution $\Gamma_{\rm SM}^{~\xi =0.1}\approx 3$ MeV (eqs.~(\ref{eq:InvisibleWidth},~\ref{eq:BRinv})); on the other one, the LUX experiment reaches in this region its best sensitivity. Decreasing $\xi$, however, reduces the strength of the $\eta$-$\eta$-$h$ interaction for low values of $\lambda$. Therefore, for $\xi = 0.05$ a combination of LHC and LUX bound rules out only values of $\lambda \gtrsim 7\times 10^{-3}$ in the $m_{\eta}\lesssim m_h/2$ region; this bound can be further pushed towards lower values $\lambda \simeq 10^{-3}$ considering non-zero contact interactions between $\eta$ and light quarks (see section~\ref{sec:DirectDetection} and eq.~(\ref{eq:DDCase2})).
   
      The resonant region $m_{\eta}\simeq m_h/2$ cannot be ruled out by constraints on the invisible branching ratio or the spin-independent elastic DM-nucleon cross-section since in the first case ${\rm BR}_{\rm inv}\to 0$ if $m_{\eta}\to m_h/2$ while in the second one $-q^2 \ll m_h^2$.
      Around the Higgs resonance, however, DM particles mostly annihilate into $b\bar{b}$ pairs,  producing a large antiproton signal that is ruled out by the bound extracted from the local antiproton flux measured by the PAMELA experiment. This conclusion is still valid regardless the astrophysical uncertainties plaguing the propagation of charged particles in the Galaxy and the DM density profile and for both values of $\xi$ considered here.
      Note that for $\xi=0.1$ the antiproton bound, at least adopting the KOL propagation models, can also rule out the right boundary of the funnel-shaped region (i.e. the vertical line corresponding to $m_{\eta}\simeq 80$ GeV); this confirms the expected result that DM annihilation into $b\bar{b}$ with a cross-section of the order of the thermal value $\langle \sigma v_{\rm rel}\rangle \simeq 3\times 10^{-26}$ cm$^{3}$s$^{-1}$ is in tension with the limit extracted from the antiproton spectrum measured by the PAMELA experiment considering values of DM mass up to $\sim 100$ GeV \cite{Tavakoli:2013zva}.\footnote{The reader should keep in mind that, since in our model we combine different final states with different branching ratios, our result cannot be immediately linked to more general analyses that assume $100\%$ DM annihilation into one single channel. In particular if $m_{\eta}\simeq 80$ GeV we have, in addition to $b\bar{b}$, a sizable branching ratio into three-body $WW^*$ final states.} 
      
      As far as the bottleneck of the funnel-shaped region is concerned, the bound from antiproton cannot be applied since the accidental cancellation that characterizes this region also suppresses DM annihilations today ($v_{\rm rel}\approx 0$). On the contrary the spin-independent DM-nucleon elastic cross-section, relying on a different kinematic w.r.t. the annihilation process, does not suffer from the same cancellation and, as a consequence, the funnel-shaped region turns out to be ruled out by the LUX experiment for $\xi = 0.1$ and strongly constrained for $\xi=0.05$, in particular the upper half part of the region. For $\xi =0.05$ a viable candidate of DM, therefore, sits on the strip of the analyzed parameter space $(m_{\eta},\lambda)$ that spans values from $m_{\eta}\simeq 100$ GeV, $\lambda \simeq 3\times 10^{-4}$ up to  $m_{\eta}\simeq 200$ GeV, $\lambda \simeq 6\times 10^{-2}$.
      
      Finally, we also show in the right panels of figure~\ref{fig:CombinedPlot} the result of the numerical parameter scans performed in the next-to-minimal scenario discussed in section~\ref{sec:NTMmodel}. We do not show here the result for the minimal case since it predicts a very narrow region in this plane which is also contained in the next-to-minimal one. Both for $\xi = 0.1$ and $\xi=0.05$, the points reproducing the correct top and Higgs masses, as expected from eq.~(\ref{eq:Lambda2}), lie around the value $\lambda \simeq 0.065$ and vary between $m_{\eta} \sim m_h/2$ and $m_{\eta} \sim 700$ GeV; moreover the points with $m_{\eta} \lesssim 200$ GeV, shown in the plot, have the smaller amount of tuning, see figure~\ref{fig:ScanResultsNS2}.

      For $\xi=0.1$ all the points which provide the correct DM abundance lie in the region excluded by LUX or by the antiproton flux measurements. Moreover, most of the points are also disfavored by direct searches of top partners and vector resonances at the LHC. In conclusion we find that -- remarkably -- the entire region of the $(m_{\eta},\lambda)$ plane in which the model can accommodate a realistic DM candidate is ruled out by our phenomenological analysis.

      For the smaller value of $\xi$ considered here, $\xi = 0.05$, the constraints from direct searches at LHC are substantially alleviated. The favored region of the parameter space lies close to the bound imposed by DM direct detection experiments, $m_\eta \simeq 200$ GeV and $\lambda \simeq 6 \times 10^{-2}$.
      In this regard it should be noted that if we assume non-zero contact interactions between $\eta$ and light quarks the bound becomes even more stringent (red dot-dashed line in figure~\ref{fig:CombinedPlot}). In any case -- including or not this  theoretical uncertainty -- we expect that this region will be definitely covered in the near future by direct detection experiments.

\section{Conclusions}
\label{sec:conclusions}

In this paper we have analyzed the Composite DM model proposed in ref.~\cite{Frigerio:2012uc}.
The model assumes the existence of  a composite sector described by some new fundamental strongly-coupled theory and characterized by a global symmetry $SO(6)$ spontaneously broken to the subgroup $SO(5)$ by a condensate of the strong dynamics, at a scale $f$.
The NGBs arising from this breaking are the Higgs doublet $H$ and a real, gauge singlet, pseudo-scalar $\eta$. The former contains the physical Higgs boson $h$ while the latter plays the role of DM. The global $SO(6)$ symmetry is also explicitly broken by the linear mixing between the composite states and the elementary SM particles.
These terms induce, at one-loop, an effective potential for $h$ and $\eta$ which is assumed to be dominated by the contributions of  SM fields, spin-1/2 top partners and composite spin-1 resonances (i.e. the \textit{Minimal Higgs Potential} hypothesis proposed in ref.~\cite{Marzocca:2012zn}) and made calculable by imposing generalized Weinberg sum rules.

From a phenomenological viewpoint, the most important consequence of this theoretical construction is that the Higgs boson, the DM particle, the top quark and the 
composite resonances are inextricably linked by the effective potential. This fact allowed us to study the constraints imposed on the model considering both DM and collider searches. 
Combining the results from direct and indirect detection of DM, invisible Higgs decay width and direct searches of top partners and vector resonances at the LHC, 
we were able to show that the model can reproduce the observed value of relic density only if $\xi =0.05$ (or lower), corresponding to the value $f\simeq 1.1$ TeV. 
As far as the DM mass and the Higgs portal coupling are concerned, for  $\xi =0.05$ our phenomenological analysis predicts  $m_{\eta}\simeq 200$ GeV and $\lambda \simeq 6 \times 10^{-2}$.
Most importantly, we have shown that this prediction lies well within the reach of future DM direct detection experiments. We argue that the model presented in this paper, therefore, will be definitely ruled out -- or discovered -- in the near future.

\acknowledgments{
We thank Marco Serone for reading the manuscript and providing us with precious comments.
We also thank Riccardo Torre and Wei Xue for discussions.

The work of A.U. is supported by the ERC Advanced Grant n$^{\circ}$ $267985$, ``Electroweak Symmetry Breaking, Flavour and Dark Matter: One Solution for Three Mysteries" (DaMeSyFla).}

\begin{appendix}

\section{Parametrizing the $SO(6)/SO(5)$ coset and physical couplings}
\label{App:parametrizations}

In this appendix, after providing some definitions useful for our work, we present three different parametrizations of the physical $h$ and $\eta$ fields, used in previous literature, and the relations among them. In particular, we show how the couplings among the physical fields differ between the parametrizations: only physical observables are parametrization-independent.

Let us first define the broken and unbroken generators of $SO(6)/SO(5)$ in the fundamental representation of $SO(6)$. We classify them in the five broken ones of $SO(6)/SO(5)$ and the ten unbroken generators of the $SO(5)$ subgroup, which can be further divided into the six of the $SU(2)_L \otimes SU(2)_R \sim SO(4)\subset SO(5)$ subgroup and the four of the $SO(5)/SO(4)$ coset
\be \begin{split}
	T^{\hat{a}}_{ij} =&~ - \frac{i}{\sqrt{2}} \left( \delta^{\hat{a} i } \delta^{6 j} - \delta^{\hat{a} j } \delta^{6 i}\right)~, \\
	T^{a_{L,R}}_{ij} =&~ - \frac{i}{2} \left[ \frac{1}{2} \epsilon^{abc} (\delta^{bi} \delta^{cj} - \delta^{bj} \delta^{ci}) \pm (\delta^{ai} \delta^{4j} - \delta^{a j} \delta^{4 i}) \right]~, \\
	T^{\alpha}_{ij} =&~ - \frac{i}{\sqrt{2}} \left( \delta^{\alpha i } \delta^{5 j} - \delta^{\alpha j } \delta^{5 i} \right)~,
	\label{eq:SO6generators}
\end{split} \ee
where $\hat{a} = 1,\ldots,5$, $a_{L,R} = 1,2,3$ and $\alpha = 1,\ldots,4$.

The five NGBs can be parametrized, using the standard CCWZ formalism \cite{Coleman1,Coleman2}, by a $6\times 6$ unitary matrix obtained exponentiating a linear combination of the broken generators,
\be
	U(x) = \exp \left[ i \sqrt{2} \frac{\theta^\ahat(x)}{f} T^\ahat \right]~,
	\label{eq:NGBmatrix}
\ee
which transforms under a global $SO(6)$ transformation $g$ as $U(x) \rightarrow g \, U(x) \, k^\dagger(g, \theta^{\hat{a}}(x))$, where $k$ is a local transformation of the unbroken group $SO(5)$, which depends on $g$ and on the position via the NGB dependence.
From the NGB matrix $U$ one can define the standard CCWZ structures $d_\mu$ and $E_\mu$  as
\be
	d_\mu^{\hat{a}} T^{\hat{a}} + E_\mu^{a} T^a = - i (U^\dagger D_\mu U)~.
	\label{eq:CCWZ}
\ee
Defining $\Sigma(x) \equiv U(x) \Sigma_0$, with $\Sigma_0 = (0,0,0,0,0,1)^t$, one gets
\be \begin{split}
	\Sigma = & \sin \frac{\theta}{f} \left(
	\frac{\theta^{\hat{1}}}{\theta},
	\frac{\theta^{\hat{2}}}{\theta},
	\frac{\theta^{\hat{3}}}{\theta},
	\frac{\theta^{\hat{4}}}{\theta},
	\frac{\theta^{\hat{5}}}{\theta},
	\cot \frac{\theta}{f} \right) \\
	= & \frac{1}{f} \left(
	h_1, h_2, h_3, h_4, \eta,
	\sqrt{f^2 - h^2 - \eta^2 } \right)~,
	\label{eq:NGBsigmaGen}
\end{split} \ee
where $\theta^2 \equiv \sum_{\hat{a}=1}^5(\theta^{\hat{a})^2}$ and $h^2 \equiv \sum_{i=1}^4 h_i^2$.
The usual Higgs doublet can can be constructed as $H = \frac{1}{\sqrt{2}} ( h_1 + i h_2, h_3 + i h_4 )^t$.
The fields $h_i(x)$ and $\eta(x)$ live in the region $\sqrt{h^2 + \eta^2} \leq f$.
In the unitary gauge $h_1(x) = h_2(x) = h_4(x) = 0$ and $h(x) \equiv h_3(x)$
\be \begin{split}
	\Sigma_{unitary} = & \sin \frac{\theta}{f} \left(0, 0, \frac{\theta^{\hat{3}}}{\theta}, 0, \frac{\theta^{\hat{5}}}{\theta}, \cot \frac{\theta}{f} \right)  \\
	= & \frac{1}{f} \left(
	0, 0, h, 0, \eta,
	\sqrt{f^2 - h^2 - \eta^2 } \right)\\
	= & \left(
	0, 0, \sin \frac{\phi}{f} \cos \frac{\psi}{f}, 0, \sin \frac{\phi}{f} \sin \frac{\psi}{f},
	\cos \frac{\phi}{f} \right)~,
\end{split} \ee
where in the third line we introduced another parametrization \cite{Redi:2012ha}, in terms of two angles, which is related to the previous two as
\be \begin{split}
	\phi = \sqrt{ (\theta^{\hat{3}})^2 + (\theta^{\hat{5}})^2 }~, \qquad
	\tan \frac{\psi}{f} = \frac{\theta^{\hat{5}}}{\theta^{\hat{3}}}~,\\
	\sin \frac{\phi}{f} = \frac{1}{f}\sqrt{h^2 + \eta^2}~, \qquad
	\tan \frac{\psi}{f} = \frac{\eta}{h}~.\\
\end{split} \ee
Let us call the first parametrization, in terms of the $\theta^{\hat{a}}$ variables, \emph{Cartesian}, the one we use throughout the paper, in terms of $h$ and $\eta$, \emph{constrained} and the third one, in terms of the angles $\phi$ and $\psi$, \emph{polar}.
In the rest of this appendix we will show how the physical fields in the three parametrization have qualitatively different couplings, both from the chiral Lagrangian and from the effective potential. In the computation of physical quantities such as cross-sections or decay widths, these differences conspire and give the exact same result, as expected.

The leading-order chiral Lagrangian, eq.~\eqref{eq:LeadingLagr}, can be written in a compact form in both the constrained and in the polar parametrization, it reads
\be \begin{split}
	\Lag_{chiral} &= \frac{f^2}{4} \text{Tr}\left[ d_\mu d^\mu \right] = \frac{f^2}{2} (D_\mu \Sigma)^t D^\mu \Sigma = \\
		&= \frac{1}{2} \left[ \sin^2 \frac{\phi}{f} (\partial_\mu \psi)^2 + (\partial_\mu \phi)^2 \right] + \frac{f^2}{8} \sin^2 \frac{\phi}{f} \cos^2 \frac{\psi}{f} ~ ( \tilde g^2 A_\mu A^\mu) \\
		&= \frac{1}{2} \left[ (\partial_\mu h)^2 + (\partial_\mu \eta)^2 + \frac{( h \partial_\mu  h +  \eta \partial_\mu  \eta)^2 }{f^2 - h^2 - \eta^2 } \right] + \frac{h^2}{8} ( \tilde g^2 A_\mu A^\mu)~,
	\label{eq:ChirLagrParam}
\end{split} \ee
where, for convenience, we defined $~\tilde g^2 A_\mu A^\mu \equiv  g_0^2 [ (W^1_\mu)^2 + (W^2_\mu)^2 ] + (g_0^\prime B_\mu - g_0 W_\mu^3)^2$.
In the three parametrizations, the EWSB vacuum can be identified as ($\langle \theta^{\hat{3}}\rangle = f \sin^{-1}\sqrt{\xi}$, $\langle \theta^{\hat{5}} \rangle = 0$), ($\sin \langle \phi \rangle = \sqrt{\xi}$, $\langle \psi \rangle = 0$) or ($\langle h \rangle = v = f \sqrt{\xi}$, $\langle \eta \rangle = 0$), where $\xi = v^2 / f^2$.
It is then straightforward to identify the physical Higgs and DM fields in the three parametrizations
\be \begin{split}
	\theta^{\hat{3}} = f \sin^{-1}\sqrt{\xi} + h_{Cart}~, \quad & \quad
	\theta^{\hat{5}} = f \frac{\sin^{-1}\sqrt{\xi}}{\sqrt{\xi}} + \eta_{Cart}~; ~\\
	\phi = f \sin^{-1}\sqrt{\xi} + h_{pol}~, \quad & \quad
	\psi = \frac{1}{\sqrt{\xi}} ~ \eta_{pol}~; \\
	h = v + \sqrt{1 - \xi} ~ h_{con}~, \quad & \quad
	\eta = \eta_{con}~.
	\label{eq:PhysFieldsParam}
\end{split}\ee

Let us now look at the effective potential. With a simple spurionic analysis it is possible to obtain the possible functional dependence of the potential on the pNGBs. The gauge contribution to the potential depends only on $h^2 = f^2 \sin^2 \frac{\phi}{f} \cos^2 \frac{\psi}{f}$, instead the functional dependence of the fermion contribution depend on the particular embedding of the SM fermions in $SO(6)$ representations. In our models, that is embedding the third generation quarks in fundamentals as in eq.~\eqref{eq:SMfermEmbedding6}, the functional dependences are $h^2 = f^2 \sin^2 \frac{\phi}{f} \cos^2 \frac{\psi}{f}$ and $(h^2 + \eta^2) = f^2 \sin^2 \frac{\phi}{f}$.
Expanding for small values of $h^2, \eta^2$ and keeping terms up to quartic order, the effective potential can thus be parametrized as
\bea
	V_{\rm eff} &=& \frac{\mu^2_h}{2} h^2 + \frac{\lambda_h}{4}  h^4 + \frac{\mu^2_\eta}{2} \eta^2 + \frac{\lambda}{2}  h^2 \eta^2 + \frac{\lambda_\eta}{4}  \eta^4 + \ldots ~ \\
	&=& - \gamma \sin^2 \frac{\phi}{f} \cos^2 \frac{\psi}{f} + \beta \sin^4 \frac{\phi}{f} \cos^4 \frac{\psi}{f} + \delta \sin^2 \frac{\phi}{f} + \sigma \sin^4 \frac{\phi}{f} \cos^2 \frac{\psi}{f} + \chi \sin^4 \frac{\phi}{f} + \ldots. \nonumber
\label{eq:EffPotentialParam}
\eea
The relation between the coefficients in the two formalisms, at this order, is
\be \begin{split}
	\mu^2_h f^2 = - 2 (\gamma - \delta)~, \quad & \quad \mu^2_\eta f^2 = 2 \delta~, \\
	\lambda_h f^4 = 4 (\beta + \sigma + \chi)~, \qquad \lambda f^4 &= 2 (\sigma + 2 \chi)~ , \qquad\lambda_\eta f^4 = 4 \chi ~.
\end{split}\ee
The EWSB minimum is given by
\be
	\xi = \frac{v^2}{f^2} = -\frac{\mu^2_h}{\lambda_h \, f^2} =\frac{\gamma - \delta}{2(\beta + \sigma + \chi)}~.
\ee
The mass matrix for physical fields defined in eq.~\eqref{eq:PhysFieldsParam}, in all three parametrizations, is the same
\begin{eqnarray}
	m_h^2 = \left. \frac{\partial^2 V(h_{phys},\eta_{phys})}{\partial h_{phys}^2} \right|_{min} &=& 2 \lambda_h v^2 (1-\xi) = \frac{8 (\beta + \sigma + \chi)}{f^2} \xi (1 - \xi)~, \label{eq:HiggsMassParam} \\
	m_\eta^2 = \left. \frac{\partial^2 V(h_{phys},\eta_{phys})}{\partial \eta_{phys}^2} \right|_{min} &=& \mu_\eta^2 + \lambda v^2 = \frac{2 \delta}{f^2} + \frac{2 (\sigma + 2 \chi)}{f^2} \xi~, \label{eq:EtaMassParam}\\
	m_{h\eta}^2 = \left. \frac{\partial^2 V(h_{phys},\eta_{phys})}{\partial h_{phys} \partial \eta_{phys}} \right|_{min} &=& 0~.
\end{eqnarray}
Which confirms that the physical fields defined above are indeed mass eigenstates.

Let us now move to study the couplings of the physical fields in the three parametrizations arising from the Lagrangian of eq.~\eqref{eq:ChirLagrParam} and the potential in eq.~\eqref{eq:EffPotentialParam}. We parametrize the generic couplings of the physical fields following, and adapting, the formalism of ref.~\cite{Contino:2010mh}. Up to four-particle interaction terms and assuming custodial invariance and parity under $\eta \rightarrow - \eta$, (from now on we neglect the subscript ``$phys$''), we write the phenomenological Lagrangian
\be\begin{split}
	\hspace{-0.3cm} \Lag_{\text{pheno}} &=\; \frac{1}{2} (\partial_\mu h)^2 \left( 1 + 2 a_{hh} \frac{h}{v} + b_{hh} \frac{h^2}{v^2} + b_{h \eta} \frac{\eta^2}{v^2} + \ldots \right)  \\
	&\; + \frac{1}{2} (\partial_\mu \eta)^2 \left( 1 + 2 a_{\eta h} \frac{h}{v} + b_{\eta h} \frac{h^2}{v^2} + b_{\eta \eta} \frac{\eta^2}{v^2} + \ldots \right)  \\
	&\; + (\partial_\mu \eta \partial^\mu h) \left( c_{\eta} \frac{\eta}{v} + d_{\eta h} \frac{\eta h}{v^2} + \ldots \right) - V_\text{eff}(h,\eta)~ \\
	&\; + \left[ M_W^2 W_\mu^+ W^{- \mu} + \frac{M_Z^2}{2} Z_\mu Z^\mu \right] \left( 1 + 2 a_{Vh} \frac{h}{v} + b_{Vh} \frac{h^2}{v^2} + b_{V \eta} \frac{\eta^2}{v^2} + \ldots \right)  \\
	&\; - m_f \bar{\psi}_f \psi \left( 1 + c_{f h} \frac{h}{v} + b_{f h} \frac{h^2}{v^2} + b_{f \eta} \frac{\eta^2}{v^2} + \ldots \right)~,
	\label{eq:CouplingParam}
\end{split}\ee
where $f = u^i, d^i, e^i$ represents any SM fermion and
\be
	V_\text{eff}(h,\eta) = \frac{m_h^2}{2} h^2 + \frac{m_\eta^2}{2} \eta^2 + \frac{\lambda_{h^3}}{2} h^3 v + \frac{\lambda_{h^4}}{4} h^4 + \frac{\lambda_{\eta^2 h}}{2} \eta^2 h + \frac{\lambda_{\eta^2 h^2}}{4} \eta^2 h^2 + \frac{\lambda_{\eta^4}}{4} \eta^4~.
	\label{eq:CouplingPotParam}
\ee
We report the expression of the couplings in the three parametrizations, as functions of $\xi$, in table~\ref{table:CouplingsParam}. It can be noticed that the constrained parametrization offers the cleanest expressions for the physical couplings. For this reason, and for its intuitive relation with the physical Higgs and DM fields, we decided to use this parametrization throughout the work.

In table~\ref{table:CouplingsParam} it can be noted that the couplings of the physical fields differ also qualitatively among the three parametrizations. It can be checked that, however, when computing physical observables (for example cross-sections) they all give the same result. As an example it can be easily checked that the NGB scattering amplitudes for high energies, $E^2 \gg m_h^2, m_\eta^2, M_{W,Z}^2$, go like $|\mathcal{A}|^2 \sim E^4 / f^4$ in all three parametrizations. In order to check that also the couplings from the potential provide the same physical results (which can not be tested from the previous check), we explicitly computed the unpolarized cross-section $\sum_{pol} \sigma(\eta \eta \rightarrow W^+ W^-)$ in all parametrizations and for all energies above threshold and confirmed that the result is indeed the same in all three cases.

\begin{landscape}
\begin{table}[p]
\small
\begin{center}
\begin{tabular}{|c|c|c|c|} \hline
\textbf{Coupling} & \emph{Constrained} & \emph{Polar} & \emph{Cartesian}  \\
\hline \hline
	$a_{V h}$ & $\sqrt{1-\xi}$ & $\sqrt{1-\xi}$ & $\sqrt{1-\xi}$ \\  
	$b_{V h}$ & $1-\xi$ & $1- 2 \xi$ & $1- 2 \xi$ \\
	$b_{V \eta}$ & $0$ & $-1$ & $-1 + \frac{\sqrt{1-\xi}}{\sqrt{\xi}} \sin^{-1} \sqrt{\xi}$ \\ \hline
	$a_{h h}$ & $\frac{\xi}{\sqrt{1-\xi}}$ & $0$ & $0$ \\
	$b_{h h}$ & $\frac{\xi(1+3\xi)}{1-\xi}$ & $0$ & $0$ \\
	$b_{h \eta}$ & $\frac{\xi^2}{1-\xi}$ & $0$ & $-1 + \frac{\xi}{(\sin^{-1} \sqrt{\xi})^2}$ \\ \hline
	$a_{\eta h}$ & $0$ & $\sqrt{1-\xi}$ & $\sqrt{1-\xi} - \frac{\sqrt{\xi}}{\sin^{-1} \sqrt{\xi}}$ \\
	$b_{\eta h}$ & $0$ & $1- 2 \xi$ & $1 - 2\xi + \frac{3\xi}{(\sin^{-1} \sqrt{\xi})^2} - \frac{4 \sqrt{\xi} \sqrt{1-\xi}}{\sin^{-1} \sqrt{\xi}}$ \\
	$b_{\eta \eta}$ & $\frac{\xi}{1-\xi}$ & $0$ & $-2 + \sqrt{1-\xi}\frac{\sin^{-1} \sqrt{\xi}}{\sqrt{\xi}} + \frac{(\sin^{-1} \sqrt{\xi})^2}{\xi}$ \\ \hline
	$c_{\eta}$ & $\frac{\xi}{1-\xi}$ & $0$ & $\frac{\sin^{-1} \sqrt{\xi}}{\sqrt{\xi}} - \frac{\sqrt{\xi}}{\sin^{-1} \sqrt{\xi}}$ \\
	$d_{\eta h}$ & $\frac{\xi (1+\xi)}{1-\xi}$ & $0$ & $-1 + \frac{3\xi}{(\sin^{-1} \sqrt{\xi})^2} - \frac{2 \sqrt{\xi} \sqrt{1-\xi}}{\sin^{-1} \sqrt{\xi}}$ \\ \hline
	$c_{f h}$ & $\frac{1-2\xi}{\sqrt{1-\xi}}$ & $\frac{1-2\xi}{\sqrt{1-\xi}}$ & $\frac{1-2\xi}{\sqrt{1-\xi}}$ \\
	$b_{f h}$ & $-\frac{(3-2\xi)\xi}{2(1-\xi)}$ & $- 2 \xi$ & $- 2 \xi$ \\
	$b_{f \eta}$ & $-\frac{\xi}{2(1-\xi)}$ & $- \frac{1}{2}$ & $- \frac{1}{2} (1 - \frac{1-2\xi}{\sqrt{1-\xi}} \frac{\sin^{-1} \sqrt{\xi}}{\sqrt{\xi}})$ \\ \hline
	$\lambda_{h^3}$ & $2 \lambda_h (1-\xi)^{3/2}$ & $2 \lambda_h \sqrt{1-\xi}(1-2\xi)$ & $ 2 \lambda_h \sqrt{1-\xi}(1-2\xi)$ \\ 
	$\lambda_{h^4}$ & $\lambda_h (1-\xi)^2$ & $\lambda_h(1-\frac{28}{3}\xi (1-\xi))$ & $\lambda_h(1-\frac{28}{3}\xi (1-\xi))$ \\ 
	$\lambda_{\eta^2 h}$ & $2 \lambda \sqrt{1-\xi}$ & $2\frac{\mu_\eta^2}{v^2} + 2 (2\lambda - \lambda_h) \sqrt{1-\xi}$ & $2 \lambda - \frac{2}{3} \frac{\mu_\eta^2}{f^2} + \mathcal{O}(\xi)$ \\ 
	$\lambda_{\eta^2 h^2}$ & $2 \lambda (1-\xi)$ & $2\frac{\mu_\eta^2}{v^2}(1-2\xi) + 6\lambda(1-\frac{4}{3}\xi) - \lambda_h (5 - 6\xi)$ & $\lambda - \frac{1}{3} \frac{\mu_\eta^2}{f^2} + \mathcal{O}(\xi)$ \\
	$\lambda_{\eta^4}$ & $\lambda_\eta$ & $-\frac{2}{3} \frac{\mu_\eta^2}{v^2} + \lambda_\eta + \lambda_h - \frac{8}{3} \lambda$ & $\lambda_\eta - \frac{2}{3} \frac{\mu_\eta^2}{f^2} + \mathcal{O}(\xi)$  \\ \hline
\end{tabular}
\caption{
\textit{Expression of the couplings in eqs.~\eqref{eq:CouplingParam} and \eqref{eq:CouplingPotParam} in the three parametrizations considered. For the couplings to fermions we assumed the embedding of eq.~\eqref{eq:SMfermEmbedding6}. In the last three rows of the Cartesian parametrization we show the leading term for small $\xi$, since the whole expressions do not fit in the table.}}\label{table:CouplingsParam}
\end{center}
\end{table}
\end{landscape}


\section{Details on the effective potential}
\label{App:EffectivePotential}

The mixing terms between the elementary SM states and the heavy composite resonances, introduced in section~\ref{sec:fermionembedding}, break explicitly the $SO(6)$ symmetry. At one loop they generate a Coleman-Weinberg effective potential for the NGBs $h$ and $\eta$. This potential can be easily obtained from the mass matrix in each sector (gauge and fermionic), keeping $h$ and $\eta$ as background fields.
Let us parametrize the field-dependent mass terms for the spin-1 and spin-1/2 fields as
\be
	\Lag^{mass} = \frac{1}{2} V_\mu^i M^{2}_{V,ij}(h,\eta) V^{j \mu} - \left( \bar{\psi}_L^i M_{F,ij} (h,\eta) \psi_R^j + h.c. \right)~,
\ee
where $i,j$ run over all the fields in each sector and $M^{2}_V$ is a real symmetric matrix while $M_F$ is a generic complex matrix. From these matrices one can obtain the singular values with a $h,\eta$ background: $m_n(h,\eta)^2 > 0$, where $n$ runs over all the states with a spin $s_n = 1, \frac{1}{2}$.
These singular values can finally be used to obtain the one-loop effective potential. Regularizing the integral with dimensional regularization one has
\be \begin{split}
	V^{(1)}(h,\eta) &= \frac{1}{16 \pi^2} \sum_n \frac{(-1)^{2 s_n} (2 s_n + 1)}{4} m_n(h,\eta)^4 \left( \log \frac{m_n(h,\eta)^2}{Q^2} - k_{s_n} \right)  \\
		&= \frac{3}{64 \pi^2} \text{Tr} \left[ M_V^4(h,\eta) \left( \log \frac{M_V^2 (h,\eta)}{Q^2} - k_1 \right) \right]  \\
		& - \frac{2 N_c}{64 \pi^2} \text{Tr} \left[ (M^{\dagger}_F M_{F})^2(h,\eta) \left( \log \frac{(M^{\dagger}_F M_{F}) (h,\eta)}{Q^2} - k_{1/2} \right) \right] ~,
\end{split}\ee
where $Q$ is the sliding scale and $k_{s_n}$ are numerical factors which depend on the subtraction scheme used.
We see that, in general, the potential is scale-dependent as well as scheme-dependent, which would imply the necessity to fix some boundary conditions at some scale, for example by matching with the measured Higgs mass and vacuum expectation value. This, however, would imply our impossibility to predict those values from our explicit models. To avoid this, we impose a set of generalized Weinberg sum rules by asking that $\text{Tr}[M_V^4]$ and $\text{Tr}[(M^{\dagger}_F M_{F})^2]$ are independent on $h$ and $\eta$
\be
	\text{WSR: }\quad
	\text{Tr}\left[ M^{4}_V (h,\eta)\right] \equiv \text{const} \quad \text{and}\quad
	\text{Tr}\left[ \left(M^{\dagger}_F(h,\eta) M_{F} (h,\eta)\right)^2 \right] \equiv \text{const} ~.
\ee

Another, independent, method to obtain the one-loop effective potential is by integrating out the heavy resonances with $h$ and $\eta$ acting as background fields and writing an effective Lagrangian for the elementary SM fields with non-trivial form factors. Finally, by integrating out also the elementary fields one obtains the effective potential as an integral in momentum of these form factors, which can be performed, for example, with a cutoff regularization.
In general, the field-dependent terms of this potential are quadratically divergent in the UV, which would imply the need of fixing some boundary conditions and therefore a lack of predictability. In this formalism, the Weinberg sum rules are conditions imposed in order to cancel the quadratic and logarithmic divergencies, that is conditions on the UV behavior of the form factors.
In our numerical analysis we used both methods to derive the effective potential and checked that the results agree. To obtain the analytical results presented in this work we use the approach with the form factors, described in detail in the rest of this appendix.

\subsection{Vector contribution}
\label{App:GaugePotential}

Integrating out the heavy spin-1 fields one obtains a low-energy effective theory. The quadratic terms in the SM gauge bosons will be the relevant ones for deriving the one-loop Coleman-Weinberg potential.
In order to obtain the possible field-dependence of the gauge contributions to the scalar potential it is useful to embed the SM gauge fields in a spurionic complete representation of $SO(6)\otimes U(1)_X$, introducing spurionic gauge fields: $A_\mu = A_\mu^A T^A$ and $X_\mu$, where the only physical components are $A_\mu^{a_L} = W_\mu^a$, $A_\mu^{3R} = c_X B_\mu$ and $X_\mu = s_X B_\mu$, with $c_X = g_0^\prime / g_0$ and $s_X^2 = 1 - c_X^2$.
The effective Lagrangian for the SM gauge fields and NGBs can be parametrized, in momentum space, as
\be
	\Lag^{g,eff} = \frac{P_T^{\mu\nu}}{2} \left[ \Pi_0(q^2) \Tr\left[A_\mu A_\nu \right] + \Pi_1(q^2) \Sigma^t A_\mu A_\nu \Sigma + \Pi_0^X(q^2) X_\mu X_\nu\right]~.
\ee
Turning off the unphysical gauge fields we obtain
\begin{equation}
\begin{split}
	\mathcal{L}^{g,eff}=\frac{P_{t}^{\mu\nu}}{2} & \bigg[ \Pi_{0}W_{\mu}^{a}W_{\nu}^{a} + \Pi_{1} \frac{h^2}{4 f^2} \left(W_{\mu}^{1}W_{\nu}^{1} + W_{\mu}^{2}W_{\nu}^{2}\right) \\
 & + \Pi_{B} B_{\mu}B_{\nu} + \Pi_{1} \frac{h^2}{4 f^2} \left(\frac{g_0^\prime}{g_0} B_{\mu}-W_{\mu}^{3}\right)\left(\frac{g_0^\prime}{g_0} B_{\nu}-W_{\nu}^{3}\right) \bigg]~,
\end{split}\label{Lgauge}
\end{equation}
where $\Pi_B = (s_X^2 \Pi_0^X + c_X^2 \Pi_0)$ and where the form factors from the UV Lagrangian of eq.~\eqref{eq:spin1Lagr} are
\be \begin{split}
	\Pi_0 &= - p^2 + g_0^2 p^2 \frac{f_\rho^2}{p^2 - m_\rho^2}~, \\
	\Pi_1 &= g_0^2 f^2 + 2 g_0^2 p^2 \left[ \frac{f_a^2}{p^2 - m_a^2} - \frac{f_\rho^2}{p^2 - m_\rho^2} \right]~, \\
	\Pi_0^X & = - p^2~.
\end{split} \ee
From eq.~\eqref{Lgauge} we observe that the gauge sector contributes to the potential in eq.~\eqref{eq:ScalarPotential} only via the Higgs terms $\mu_h^2$, at the $g^2$ order, and to $\lambda_h$, at the $g^4$ order.
The gauge contribution to the Coleman-Weinberg potential for the NGBs is
\begin{equation}
V_g(h, \eta) = \frac{3}{2}\int\frac{d^{4}p_E}{(2\pi)^{4}}  \Big\{2\:\log \Pi_{WW}(-p_E^2)+\log\left[\Pi_{BB}(-p_E^2)\Pi_{WW}(-p_E^2)-\Pi_{W_3B}^2(-p_E^2)\right]\Big\}~,
\label{Vgauge}
\end{equation}
where
\be
	\Pi_{WW} = \Pi_0 +\frac{h^2}{4 f^2} \Pi_1~,\quad
	\Pi_{BB} = \Pi_B+ c_X^2 \frac{h^2}{4 f^2} \Pi_1~,\quad
	\Pi_{W_3B} = -c_X \frac{h^2}{4 f^2} \Pi_1~.
\label{eq:GaugeFormFactorHiggsDep}
\ee
The tree-level contribution from our models to the oblique $\hat{S}$ parameter \cite{Peskin:1990zt,Barbieri:2004qk} can be extracted from the last form factor in eq.~\eqref{eq:GaugeFormFactorHiggsDep} as \cite{Marzocca:2012zn}
\be
	\hat{S} = - \frac{g}{g^\prime} \Pi'_{W_3 B}(0) \simeq \frac{\langle h^2 \rangle }{4 f^2} \Pi'_1(0) = \frac{2 m_W^2}{f^2} \left( \frac{f_\rho^2}{m_\rho^2} - \frac{f_a^2}{m_a^2}\right) \stackrel{\text{WSRs}}{=} \frac{2 m_W^2}{m_\rho^2} \left( 1 - \frac{f^2}{4 f_\rho^2} \right)~,
	\label{eq:Sparam}
\ee
where the prime indicates a derivative with respect to $p^2$. In the second step we approximated $g \simeq g_0$ and $g' \simeq g'_0$ and in the last step we applied both Weinberg sum rules of eqs.(\ref{eq:WSR1gauge},\ref{eq:WSR2gauge}).

\subsection{Fermion contribution}
\label{App:FermionPotential}

After integrating out the composite resonances from eq.~\eqref{eq:LGeneralFermions}, the top quark effective Lagrangian in momentum space, up to quadratic order in the fermions and to any order in the scalar fields, can be written as
\begin{equation}
	\Lag^{f, eff} = \bar{t}_{L}{\slashed p} \,t_{L} \, \Pi_{t_{L}}(p^2, h, \eta) +\bar{t}_{R}{\slashed p}\,t_{R}\, \Pi_{t_{R}}(p^2, h, \eta) -(\bar{t}_{L}\, t_{R}\Pi_{t_{L}t_{R}}(p^2, h, \eta) +h.c.)~,
	\label{eq:EffLagrTop}
\end{equation}
resulting in the following contribution to the pNGB potential
\begin{equation}
V_f(h, \eta)=-2N_{c}\int\frac{d^{4}p_{E}}{(2\pi)^{4}}\log\left[p_E^2\Pi_{t_{L}}(-p_{E}^{2})\Pi_{t_{R}}(-p_{E}^{2})+\left|\Pi_{t_{L}t_{R}}(-p_{E}^{2})\right|^{2}\right]~.
\label{eq:ColWeinFermPot}
\end{equation}
With the embedding of the top in eq.~\eqref{eq:SMfermEmbedding6}, the pNGB dependence of these form factors can be made explicit as
\be \begin{split}
	&\Pi_{t_L} = \Pi_F +  \frac{h^2}{f^2} \Pi_{1F}~,~~~ \Pi_{t_R} = \Pi_S + \left(1 -  \frac{h^2}{f^2} - \frac{\eta^2}{f^2} \right) \Pi_{1S}~,\\
	&\Pi_{t_Lt_R} = \frac{h}{f} \sqrt{1 -  \frac{h^2}{f^2} - \frac{\eta^2}{f^2}} \Pi_{FS}~.
\end{split} \ee
Integrating out the fermion resonances $S$ and $F$ from the Lagrangian of eq.~\eqref{eq:LGeneralFermions}, we get the following expression for the form factors
\bea
\Pi_{F}(p^2)& = &  1-  \sum_{j=1}^{N_F} \frac{ |\epsilon_{qF}^j|^2}{p^2- m_{jF}^2}~,~~~~~~~~~~ 
\Pi_{1F}(p^2)  =  \frac{1}{2}  \left( \sum_{j=1}^{N_F} \frac{ |\epsilon_{qF}^j|^2}{p^2- m_{jF}^2} -  \sum_{i=1}^{N_S} \frac{ |\epsilon_{qS}^i|^2}{p^2- m_{iS}^2} \right) \,, \nn \\
\Pi_{S}(p^2)& = & 1- \sum_{j=1}^{N_F} \frac{ |\epsilon_{tF}^j|^2}{p^2- m_{jF}^2} ~,~~~~~~~~~~ 
\Pi_{1S}(p^2)  =   \sum_{j=1}^{N_F} \frac{ |\epsilon_{tF}^j|^2}{p^2- m_{jF}^2} - \sum_{i=1}^{N_S} \frac{ |\epsilon_{tS}^i|^2}{p^2- m_{iS}^2}
\,, \nn \\
\Pi_{FS}(p^2) & = &  \frac{1}{\sqrt{2}} \left( \sum_{j=1}^{N_F} \epsilon_{tF}^{j*}  \epsilon_{qF}^j   \frac{m_{jF}}{p^2- m_{jF}^2} + \sum_{i=1}^{N_S} \epsilon_{tS}^{i*}  \epsilon_{qS}^i   \frac{m_{iS}}{p^2- m_{iS}^2} \right) \,.
\eea
The top mass can be obtained either as the lightest singular value of the mass matrix of the $Q=2/3$ fields in eq.~\eqref{eq:LGeneralFermions}, or from eq.~\eqref{eq:EffLagrTop} by finding the pole of the propagator:
\be
	M_{top}^2 - \left. \frac{|\Pi_{t_L t_R}(M_{top}^2)|^2}{\Pi_{t_L}(M_{top}^2) \Pi_{t_R}(M_{top}^2)} \right|_{h=v, \eta=0} = 0~,
\ee
which, if the top is much lighter than the top partners, can be approximated as
\be
	M_{top} \simeq \left.\frac{|\Pi_{t_Lt_R}(0)|}{\sqrt{\Pi_{t_L} (0) \Pi_{t_R} (0)}} \right|_{h=v, \eta=0}~.
	\label{eq:GenericTopMass}
\ee

\end{appendix}


\bibliography{ref}
\bibliographystyle{jhep}

\end{document}